\documentclass[11pt]{article}
\usepackage[fleqn]{amsmath}
\usepackage[fleqn]{mathtools}
\usepackage{amssymb,mathrsfs}
\usepackage{graphicx,rotate,subfigure}
\usepackage{setspace}
\usepackage{cite}
\usepackage{float}
\usepackage[colorlinks=true,
            linkcolor=red,
            urlcolor=blue,
            citecolor=blue]{hyperref}
\usepackage[left=0.7in, right=0.5in, top=1in, bottom=1in]{geometry}
\usepackage{multirow}
\usepackage[utf8]{inputenc}

\usepackage{color}
\long\def\rpl#1!!#2!!{\textcolor{red}{#1} \textcolor{blue}{#2}}

\def \order(#1){{\cal O} \left(#1 \right)}

\allowdisplaybreaks
\begin{document}


\begin{center}
{\Large \bf Unitarity Constraints on non-minimal Universal Extra Dimensional Model} \\
\vspace*{1cm}  
{\sf  Tapoja Jha$^{a,b,c,}$\footnote{tapoja.phy@gmail.com}}
\\
\vspace{10pt} {\small }{\em $^a$Department of Physics, University of Calcutta,  
92 Acharya Prafulla Chandra Road,\\ Kolkata 700009, India}\\
\vspace{5pt} {\small}{\em $^b$Institute of Physics, 
	Sachivalaya Marg, 
	Bhubaneswar, Odisha 751005, India}\\
\vspace{5pt} {\small}{\em $^c$Homi Bhabha National Institute,
	Training School Complex,\\ Anushakti Nagar, 
	Mumbai 400085, India}\\

\normalsize
\end{center}
\setstretch{1.15}
\begin{abstract}
We examine the unitarity constraints in gauge and scalar sectors of non-minimal Universal Extra Dimensional model. We show that some of the tree-level two-body scattering amplitudes in gauge and scalar sectors do not respect partial wave unitarity. Unitarity analysis of this model leads to an upper bound on corresponding boundary-localized (BLT) parameter which depends on the maximum number of Kaluza-Klein (KK) mode considered in the analysis. This upper bound of the relevant BLT parameter decreases with the increasing KK-modes. The results are, in effect, independent of the inverse of compactifiaction radius. The upper bound on BLT parameter also results in a lower bound on gauge and scalar KK-masses.
\end{abstract}
\noindent Keyword: {Universal Extra Dimension, Boundary Localized Terms, Unitarity}

\bigskip
\section{Introduction}
Many unsolved puzzles of the Standard Model (SM) of particle physics would lead us to rely on the existence of extra spatial dimensions. These extra dimensional theories can provide a solution to gauge coupling unifications \cite{Dienes:1998vh, Dienes:1998vg, Bhattacharyya:2006ym}, and also can present a new perspective to address the issues on fermion mass hierarchy \cite{ArkaniHamed:1999dc}. Another interesting feature of extra dimensional theories is the possibility to provide a suitable dark matter candidate of the Universe \cite{Servant:2002aq, Majumdar:2003dj, Kong:2005hn, Burnell:2005hm}. At present, we are interested in a particular incarnation of extra dimensional theory proposed by Appelquist {\em et al.} \cite{Appelquist:2000nn} \footnote{Earlier, an alternate scenario for TeV scale extra dimensions was explored in Refs.~\cite{Antoniadis:1990ew, Antoniadis:1998ig}}. Here all the SM fields can propagate in five dimensional space-time. The extra spatial dimension (will be denoted as $y$) is compactified on a circle ($S^{1}$)  of radius $R$. This model is referred as Universal Extra Dimensional Model (UED). The five dimensional action consisting of the same fields of SM respect the same gauge symmetry as SM. After compactification, the four dimensional effective action of UED consists of SM particles and the towers of their Kaluza-Klein (KK) excitations. Members of this tower are specified by KK-number ($n$), which is nothing but the discretized momentum ($p_{5}$) in the direction of the extra spatial dimension. Masses of the KK-excitations in $n$th KK-state are proportional to $n^2/R^2$. The inverse of compactification radius ($R^{-1}$) is the specific energy scale at which the four dimensional effective theory would start to reflect the dynamics of KK-excitations of SM fields. 
  
To get zero-mode chiral fermions as in SM, one needs to orbifold the extra dimension by imposing a discrete $\mathbb Z_{2}$ symmetry : $y \leftrightarrow -y$. The fields having zero modes are even under this $\mathbb Z_{2}$ symmetry. The $n=0$ mode in this theory is identified as SM particles. Fields which are odd under $\mathbb Z_{2}$ transformation have only higher mode KK-excitations. The resulting manifold is called $S^{1}/\mathbb Z_{2}$ orbifold with effective domain of $y$ being from $0$ to $\pi R$. These two boundary points are called the {\em fixed points} of orbifold. Orbifolding breaks the translational invariance along $y$. Consequently the momentum $p_{5}$ in the fifth direction is no longer conserved. Thus KK-number is violated. There remains an additional discrete symmetry $(y \rightarrow y + \pi R)$ called KK-parity, which for $n$th KK-mode is $(-1)^{n}$.  The conserved KK-parity ensures the stability of the lightest Kaluza-Klein particle (LKP) resulting in a good dark matter candidate \cite{Servant:2002aq, Majumdar:2003dj, Kong:2005hn, Burnell:2005hm, Servant:2002hb}. The tree-level spectrum in this model (UED) is highly degenerate. Radiative corrections to KK-masses partially lift the degeneracy \cite{Cheng:2002iz}. The mass corrections are composed of two types: $(i)$ correction from the compactification, which is called bulk correction and $(ii)$ boundary correction which arises due to orbifolding. The mass shift coming from boundary corrections are not finite as bulk corrections, but are logarithmically divergent having a dependence on the unknown cut-off scale $\Lambda$. The counterterms  are localized at boundary points. In {\em minimal UED (mUED)} scenario boundary terms are assumed to be vanishing at the cut-off scale $\Lambda$, whereas in {\em non-minimal UED (nmUED)} this approximation is relaxed \cite{Dvali:2001gm, Carena:2002me, delAguila:2003bh, Haba:2009uu, Haba:2009wa}.  For unitarity analysis in scalar and gauge sectors of nmUED model, the kinetic terms of the scalar and gauge fields and also the mass and potential terms of scalar fields are added to their respective five dimensional actions at the boundary points. Coefficients of the boundary-localized terms (BLTs) are the free parameters of this model.

 Many theoretical and phenomenological aspects of nmUED have been discussed in \cite{delAguila:2003gv, Flacke:2008ne, Haba:2009uu, Haba:2009wa, Datta:2012xy}. Rigorous studies have been performed to constrain non-minimal parameters from the perspective of electroweak observables \cite{Flacke:2008ne, delAguila:2003gv, Flacke:2013pla}, relic density \cite{Datta:2013nua} and from LHC experiments \cite{Datta:2012tv, Datta:2013yaa, Datta:2013lja, Shaw:2014gba, Ghosh:2014uwa}. A lower limit on the inverse of compactification radius has been estimated from Higgs production and decay \cite{Dey:2013cqa}, $Z \rightarrow b \bar{b}$ decay \cite{Jha:2014faa} and from a study of $B_{s} \rightarrow \mu^{+} \mu^{-}$ \cite{Datta:2015aka}. Rare top decay channels  have been discussed elaborately in nmUED scenario in \cite{Dey:2016cve}. 

 Though various studies have been performed in nmUED scenario to constrain the lower limit of the inverse of compactification radius $R^{-1}$ and to constrain BLT parameters \cite{Datta:2012tv, Jha:2014faa, Dey:2016cve}, no studies have been done yet to set the upper limit on BLT parameters. The boundary terms which are generated by radiative corrections \cite{Georgi:2000ks, Cheng:2002iz} are evidently loop suppressed. But it is not evident what is the actual range of BLT parameters which are the coefficients of boundary terms, the new parameters of the theory. We do not know whether it should be very small as they are the coefficients of boundary terms which are originated from radiative corrections, or they might have some higher values. In nmUED the boundary terms can be viewed as some effective operators with unknown coefficients. So a study of unitarity is essential for determining their upper bound in four dimensional effective theory. In this article, a detailed study on unitarity has been performed in gauge and scalar sectors of nmUED scenario to set an upper bound on gauge and scalar BLT parameters. 

The paper is organized as follows. In the next section the basic idea to implement the unitarity constraint is discussed. In section 3, necessary Lagrangian and interactions in nmUED framework are given. In section 4, necessary processes are referred. The numerical results including the constraints on the upper bound of the nmUED parameters will be demonstrated in section 5. Finally in section 6, we will summarize the results and observations.

\section{Unitarity Constraints}
Any $2 \rightarrow 2$ scattering amplitudes $\mathcal{M}(\theta)$ can be expressed in terms of an infinite sum of partial waves as
\begin{flalign}
\label{unitarity_polynomial}
~~~~~~~~~~~~~~~~~~~~~~~~~~~~~~~~~\mathcal{M}(\theta) = 16\pi\sum_{J=0}^{\infty}a_J\,(2J+1)\,P_J(\cos\theta),
\end{flalign}
where $a_{J}$ is the scattering amplitude of $J$th partial wave, $\theta$ is the scattering angle and $P_{J}(\cos\theta)$ is
$J$th order Legendre polynomial. Several two-body scattering amplitudes have been analyzed in the context of SM in a seminal paper by Lee, Quigg and Thacker (LQT) \cite{Lee:1977eg}. Following Ref.~\cite{Lee:1977eg, SekharChivukula:2001hz}, one can translate the unitarity constraints of scattering amplitudes on partial-wave coefficients, in particular on the zeroth partial wave amplitudes $a_{0}$ as
\begin{flalign}
\label{unitarity_constraint}
~~~~~~~~~~~~~~~~~~~~~~~~~~~~~~~~~~~~~~~~~~~~~~~~~~~~~~~|{\rm Re}~(a_{0})|\leq \frac{1}{2}.
\end{flalign}
In the high energy limit, by virtue of equivalence theorem \cite{Lee:1977eg}, the unphysical scalars can be used instead of the original longitudinal components of the gauge bosons and the relevant $2 \rightarrow 2$ scatterings should get contributions from the quartic couplings. The contribution from trilinear couplings should safely be ignored due to the fact that the diagrams originating from the trilinear couplings will have an $E^{2}$-suppression coming from the intermediate propagators. A $t$-matrix, which is $t^{0}$ for $J=0$, can be constructed from different two-particle states represented as rows and columns \cite{Lee:1977eg}. Consequently each matrix element will give the scattering amplitude between the corresponding 2-particle state in the row and in the column. Evidently, the constraints on  the bounds on the eigenvalues of the $t^{0}$ matrix will imply
\begin{flalign}
\label{unitarity_constraint1}
~~~~~~~~~~~~~~~~~~~~~~~~~~~~~~~~~~~~~~~~~~~~~~~~~~~~~~~~|\mathcal{M}|\leq 8\pi.
\end{flalign}

However, the calculation of the eigenvalues of matrix stated above  can be extremely complicated in case of nmUED. In some cases, the trilinear couplings are effectively proportional to KK-masses. Thus in that case, one can not simply ignore the contributions from trilinear couplings, and evidently the matrix elements are not simple numbers but are functions of $s$, where $\sqrt{s}$ is the centre of mass energy of the respective processes. So one can end up with an intractable determinant \cite{DeCurtis:2003zt}. But single scattering channel is not enough to analyze the bad high energy behavior in five dimensional compactified theory, rather coupled channel analysis is suitable for the study of unitarity violation \cite{SekharChivukula:2001hz}. For an optimal implementation of the unitarity constraints we should consider all the processes involved in the analysis, i.e. we should consider the matrix elements $\mathcal{M}_{ij} = \mathcal{M}_{i \to j}$, where, $i$ and $j$ symbolically present all possible 2-particle states. For single channel analysis, or to be specific, in the present case, in $n, n \rightarrow n, n$ analysis, we actually omit some of the possible 2-particle states. Coupled channel analysis exhaustively includes all possible 2-particle states and consequently results in higher dimensional matrices. Those higher dimensional matrices in turn result in large number of eigenvalues (which is the immediate result of the inclusion of all possible states) implying much larger constraints on the parameter. Coupled channel analysis is performed by constructing $t^{0}$ matrix for all suitable channels and finding the eigenvalues as functions of model parameters and also demanding no eigenvalue exceeds $8 \pi$.  Thus, in the present scenario, at first we would like to find the expressions of $a_0$ for every possible $2 \rightarrow 2$ processes in the entire scalar sector of nmUED model, where the quartic couplings are not suppressed by KK-masses and can constrain the scalar and gauge BLKT parameters from Eq.~\ref{unitarity_constraint}. Then we can perform the coupled channel analysis with some selective channels which do not eventually fall with $s$ and can further constrain the parameters of the model using Eq.~\ref{unitarity_constraint1}.

\section{A Review of nmUED Framework}
\label{model_nmUED}
Now we will very briefly discuss the scalar and gauge sector of nmUED model. A more detailed analysis of this model can be found in Refs.  \cite{Dvali:2001gm, Carena:2002me, delAguila:2003bh, Flacke:2008ne, Datta:2012xy, Datta:2013nua, Datta:2012tv, Datta:2013yaa, Datta:2013lja, Shaw:2014gba, Dey:2013cqa, Jha:2014faa, Datta:2015aka, Dey:2016cve}.
The action of gauge fields and scalar fields are given as follows :

\begin{eqnarray}
\label{gauge}
\mathcal{S}_{G} &=& \int d^4 x \int_{0}^{\pi R} dy\Big[-\frac{1}{4}F^{MNa}F_{MN}^{a} -\frac{r_{g}}{4}\{ \delta(y) + \delta(y - \pi R)\}F^{\mu \nu a}F_{\mu \nu}^{a} \nonumber\\
 & & -\frac{1}{4}B^{MN}B_{MN} -\frac{r_{g}}{4}\{ \delta(y) + \delta(y - \pi R)\}B^{\mu \nu }B_{\mu \nu} \Big],\\
\label{higgs}
\mathcal{S}_{\Phi} &=& \int d^4 x \int_{0}^{\pi R} dy \Big[\left(D^{M}\Phi\right)^{\dagger}\left(D_{M}\Phi\right) +\mu_{5}^{2}\Phi^{\dagger}{\Phi}-\lambda_{5}{(\Phi^{\dagger}{\Phi})}^{2} \nonumber \\
 & & + \{ \delta(y) + \delta(y - \pi R)\}\{r_{\phi}\left(D^{\mu}\Phi\right)^{\dagger}\left(D_{\mu}\Phi\right) + \mu_{B}^{2}\Phi^{\dagger}{\Phi}-\lambda_{B}{(\Phi^{\dagger}{\Phi})}^{2} \} \Big],
\end{eqnarray}
 where, $a$ is the $SU(2)_L$ gauge index. The five dimensional Lorentz indices are given by $M,N=0,1,2,3,4$ with the metric convention $g_{MN} \equiv {\rm diag}(+1,-1,-1,-1,-1)$. The convention of covariant derivative $D_M$, field strength tensors $F_{MN}^a$ and $B_{MN}$ of respective $SU(2)_L$ and $U(1)_Y$ gauge groups and the convention of $\Gamma_M$ are the same as given in Refs. \cite{Jha:2014faa, Dey:2016cve}.

Here, $\Phi$ is the standard Higgs doublet. The symbols $\mu_{5}$ and $\lambda_{5}$ respectively represent the $5D$ bulk Higgs mass parameter and scalar self-coupling. The  BLKT parameters for the gauge and scalar fields are $r_{g}$ and $r_{\phi}$ respectively; $\mu_{B}$ and $\lambda_{B}$ are the boundary-localized Higgs mass parameter and the scalar quartic coupling respectively. 

For the zero-mode of Higgs to be flat \cite{Flacke:2013pla, Datta:2013yaa} the following conditions must hold \footnote{If the boundary parameters are unequal the mass term would involve KK-mode mixing and diagonalization of KK-mass matrix would  modify the wave functions implying a $y$-dependent zero mode \cite{Flacke:2013pla}.
}.
\begin{equation}
\label{higsblt}
~~~~~~~~~~~~~~~~~~~~~~~~~~~~~~~~~~~~~~~~~\mu_{B}^{2}= r_{\phi}\mu_{5}^{2}~~{\rm and}~~\lambda_{B} = r_{\phi}\lambda_{5}.
\end{equation}

In the limit, $r_{\phi}=r_g $  the scalar and gauge fields will have the same $y$-dependent profile given in Eqs. \ref{fn} and \ref{gn}. If  the two BLT parameters are taken to be different, the breakdown of electroweak symmetry results in a term proportional to $r_{\phi}$ in the differential equations governing the dynamics of gauge profile in $y$ direction \cite{Datta:2013yaa, Muck:2004zz, Datta:2014sha}. Consequently, the $y$-profile solutions of gauge field will be different from what is given below (Eqs. \ref{fn} and \ref{gn}). Throughout the analysis, the two BLKT parameters will be taken as equal to avoid the complications. Thus the $y$-dependent wave functions for scalar or gauge fields for $n$th KK-mode with appropriate boundary conditions are given by
\begin{eqnarray}
 \label{fn}
f_{\phi}^{n} = N_{\Phi n} \left\{ \begin{array}{rl}
                \displaystyle \frac{\cos(M_{\Phi_{n}} \left (y - \frac{\pi R}{2}\right))}{C_{\Phi_{n}}}  &\mbox{for $n$ even,}\\
                \displaystyle \frac{{-}\sin(M_{\Phi_{n}} \left (y - \frac{\pi R}{2}\right))}{S_{\Phi_{n}}} &\mbox{for $n$ odd.}
                \end{array} \right.
\end{eqnarray}
Since the fifth component of gauge field are projected out by $\mathbb Z_{2}$ odd condition, no zero-mode appears for $W_{5}^{\pm}$, and the $y$-profile for $n$th KK-mode is given by \cite{ Datta:2013yaa, Jha:2014faa, Datta:2015aka, Dey:2016cve, Muck:2004zz}
\begin{eqnarray}
\label{gn}
g_{\phi}^{n} = N_{\Phi n} \left\{ \begin{array}{rl}
                \displaystyle \frac{\sin(M_{\Phi_{n}} \left (y - \frac{\pi R}{2}\right))}{C_{\Phi_{n}}}  &\mbox{for $n$ even,}\\
                \displaystyle \frac{\cos(M_{\Phi_{n}} \left (y - \frac{\pi R}{2}\right))}{S_{\Phi_{n}}} &\mbox{for $n$ odd,}
                \end{array} \right.
\end{eqnarray}

with
\begin{equation}
~~~~~~~~~~~~~~~~~~~~~~~~~~~~~~~~C_{\Phi_{n}} = \cos\left( \frac{M_{\Phi_{n}} \pi R}{2} \right)\,\, , 
\,\,\,\,
S_{\Phi_{n}} = \sin\left( \frac{M_{\Phi_{n}} \pi R}{2} \right).
\end{equation}

These wave functions $f_{\phi}^{n}$ and $g_{\phi}^{n}$ satisfy the orthonormality conditions
\begin{equation}
~~~~~~~~~~~~~~~~\int dy \left[1 + r_{f}\{ \delta(y) + \delta(y - \pi R)\}
\right] ~f_{\phi}^n(y) ~f_{\phi}^m(y) = \delta^{n m} = \int dy ~g_{\phi}^n(y) ~g_{\phi}^m(y).
\end{equation}
 which give the normalization constant as
\begin{equation}
\label{normalisation}
~~~~~~~~~~~~~~~~~~~~~~~~~~~~~~~~~~~~~~~N_{\Phi n} = \sqrt{\frac{2}{\pi R}}\left[ \frac{1}{\sqrt{1 + \frac{r_\phi^2 M_{\Phi n}^{2}}{4} 
+ \frac{r_\phi}{\pi R}}}\right].
\end{equation}

The  mass $M_{\Phi n}$ of the $n$th KK-mode now satisfies the following transcendental equations 
\begin{eqnarray}
\label{scalar_masses}  
  r_{\phi} M_{\Phi n}= \left\{ \begin{array}{rl}
         -2\tan \left(\frac{M_{\Phi n}\pi R}{2}\right) &\mbox{for $n$ even,}\\
          2\cot \left(\frac{M_{\Phi n}\pi R}{2}\right) &\mbox{for $n$ odd,}
          \end{array} \right.       
 \end{eqnarray}
and the KK-masses are not equal to $n/R$ as in UED. Evidently, $M_{\Phi n}$ vanishes for zero modes $(n=0)$.

Here $\tilde g$ and $\tilde g'$, the five dimensional gauge coupling constants are related to their four dimensional counterparts $g$ and $g'$ by 
\begin{equation}
\label{g5_g4}
~~~~~~~~~~~~~~~~~~~~~~~~~~~~~~~~~~~~~~g \; (g') = N_{\Phi 0}~{\tilde g} \;(\tilde g')=\frac {\tilde g \;(\tilde g')}{\sqrt{r_{g} + \pi R}}.
\end{equation}

As we will do our calculations in 't-Hooft Feynman gauge, the gauge fixing actions are very essential for the calculation and the gauge-fixing actions are given by
\begin{eqnarray}
\label{gauge_fix}
\mathcal{S}_{\rm GF}^{A} &=& -\frac{1}{2 \xi _y}\int d^{4}x\int_{0}^{\pi R} dy \Big(\partial_{\mu}A^{\mu}+\xi_{y}\partial_{5}A^{5}\Big) ^2 , \\
\mathcal{S}_{\rm GF}^{Z} &=& -\frac{1}{2 \xi _y}\int d^{4}x\int_{0}^{\pi R} dy \Big\{\partial_{\mu}Z^{\mu}+\xi_{y}(\partial_{5}Z^{5}-iM_{Z}\chi\{1 + r_{\phi}\left( \delta(y) + \delta(y - \pi R)\right)\})\Big\} ^2 , \\
\mathcal{S}_{\rm GF}^{W} &=& -\frac{1}{\xi _y}\int d^{4}x\int_{0}^{\pi R} dy \Big\vert\partial_{\mu}W^{\mu +}+\xi_{y}(\partial_{5}W^{5+}-iM_{W}\phi^{+}\{1 + r_{\phi}\left( \delta(y) + \delta(y - \pi R)\right)\})\Big \vert ^2 .
\end{eqnarray}
In the above, $M_Z$ and $M_W$ are the respective masses of the $Z$ and $W$ boson; $\mathcal {S}_{\rm GF}^{A}$, $\mathcal {S}_{\rm GF}^{Z}$ and $\mathcal {S}_{\rm GF}^{W}$ are the gauge fixing actions for photon, $Z$ boson and $W$ boson respectively. The $y-$dependent gauge fixing parameter $\xi _y$ is related to the $y-$independent gauge fixing parameter $\xi$ (equal to $1$ in Feynman gauge, and $0$ in Landau gauge) by \cite{Jha:2014faa, Datta:2015aka, Dey:2016cve, Muck:2004zz}
\begin{equation}
~~~~~~~~~~~~~~~~~~~~~~~~~~~~~~~~~~\frac{1}{\xi_{y}}= \frac{1}{\xi}\{1 + r_{\phi}\left( \delta(y) + \delta(y - \pi R)\right)\}.
\end{equation}

While dealing with the four dimensional effective Lagrangian there exist  bilinear terms involving the KK-excitations (from first and higher KK-levels) of the $5th$ components of $Z$ bosons and the KK-excitations of $\chi^0$ of the Higgs doublet field; and similarly there are mixing terms between the $5th$ component of $W^{\pm}$ and the KK-excitations of $\phi^\pm$ of the Higgs doublet field \cite{Petriello:2002uu}. Mixing between $A_{\mu}^{n}$ and $A_{5}^{n}$ cancels by adding $\mathcal {S}_{\rm GF}^{A}$, and the new spectrum consists of a massless zero-mode photon, a tower of KK-modes with masses $M_{\Phi n}^{2}$ for both $A_{\mu}^{n}$ and $A_{5}^{n}$. Using the gauge fixing actions and appropriate mode functions of gauge and scalar fields (Eqs.~\ref{fn}, \ref{gn}) and finally integrating over $y$, the mass matrices for the mixing between KK-modes of $Z^{5}$ and $\chi^n$ and that for the mixing between the KK-modes of $W_5 ^{\pm n}$ and $\phi^{\pm n}$ are respectively given by
\begin{equation}
\label{mass_mat1}
~~~~~~~~~~~~~~~~~~~~~~~~~~~~~~\begin{pmatrix}
Z^{5 n} & \chi^{n}
\end{pmatrix}
\begin{pmatrix}
M_{Z}^{2}+\xi M_{\Phi n}^{2} & (1-\xi)M_{Z}M_{\Phi n} \\ (1-\xi)M_{Z}M_{\Phi n} & M_{\Phi n}^{2}+\xi M_{Z}^{2}
\end{pmatrix}
\begin{pmatrix}
Z^{5 n} \\ \chi^{n}
\end{pmatrix},
\end{equation}
and
\begin{equation}
\label{mass_mat2}
~~~~~~~~~~~~~~~~~~~~~~~~~~\begin{pmatrix}
W_{5}^{(n)-} & \phi^{(n)-}
\end{pmatrix}
\begin{pmatrix}
M_{W}^{2}+\xi M_{\Phi n}^{2} & -i(1-\xi)M_{W}M_{\Phi n} \\ i(1-\xi)M_{W}M_{\Phi n} & M_{\Phi n}^{2}+\xi M_{W}^{2}
\end{pmatrix}
\begin{pmatrix}
W_{5}^{(n)+} \\ \phi^{(n)+}
\end{pmatrix} + {\rm h.c. }.
\end{equation}
Diagonalization of the mass matrices~\ref{mass_mat1} leads to a tower of Goldstone modes of $Z$ ($G_{Z}^{n}$ with mass squared $\xi  (M_{\Phi n} ^2 + M_Z^2) $) and a physical CP-odd scalars ($A^{n}$ with mass squared $M_{\Phi n} ^2 + M_Z^2$) respectively given as
$$G_{Z}^{n} = \frac{1}{M_{Z_{n}}}\left(-M_{\Phi n}Z^{5 n} + M_{Z}\chi^{n}\right),$$
$$A^{n} = \frac{1}{M_{Z_{n}}}\left(M_{\Phi n}\chi^{n} + M_{Z} Z^{5 n}\right).$$
Diagonalization of another set of matrices~\ref{mass_mat2} also generates KK-tower of charged Goldstone bosons (with mass squared $\xi  (M_{\Phi n} ^2 + M_W^2) $) and a physical charged Higgs pair (with mass squared $M_{\Phi n} ^2 + M_W^2$) given by
$$ G^{\pm (n)} = \frac{1}{M_{W_{n}}}\left(M_{\Phi n}W^{\pm5(n)}\mp  iM_{W}\phi^{\pm(n)}\right), $$
$$H^{\pm (n)} = \frac{1}{M_{W_{n}}}\left(M_{\Phi n}\phi^{\pm(n)}\mp iM_{W}W^{\pm5(n)}\right).$$
The fields $Z^{\mu n}$, $G_{Z}^{n}$ and $A^{n}$ all posses the common mass eigenvalue as  $M_{Z n} \equiv \sqrt {M_{\Phi n} ^2 + M_Z^2}$. Similarly $W^{\mu (n)\pm}$, $G^{(n)\pm}$ and $H^{(n)\pm}$ share the same mass eigenvalue $M_{W n} \equiv \sqrt {M_{\Phi n} ^2 + M_W^2}$ in 't-Hooft Feynman gauge ($\xi  = 1$).
The above combinations of charged Higgs and charged Goldstone ensure the vanishing coupling of $A^{\mu 0} H^{n \pm}W_{\nu}^{n \mp}$.

Substituting Eq.~\ref{higsblt} in Eq.~\ref{higgs}, we can have the form of five dimensional Lagrangian of scalar field $ \mathcal L_{\phi}$ as
\begin{eqnarray}
\label{Lphi}
\mathcal{L}_{\phi} &=& \int_{0}^{\pi R} dy \Big[\{1+ r_{\phi}\left( \delta(y) + \delta(y - \pi R)\right)\}\left(D^{\mu}\Phi\right)^{\dagger}\left(D_{\mu}\Phi\right) \nonumber \\ 
 & & + \{1+ r_{\phi}\left( \delta(y) + \delta(y - \pi R)\right)\}(\mu_{5}^{2}\Phi^{\dagger}{\Phi}-\lambda_{5}{(\Phi^{\dagger}{\Phi})}^{2}) -(D_{5} \Phi)^\dagger(D_{5} \Phi)\Big].
\end{eqnarray}
The terms $ \int_{0}^{\pi R} dy \{1+ r_{\phi}\left( \delta(y) + \delta(y - \pi R)\right)\}(\mu_{5}^{2}\Phi^{\dagger}{\Phi}-\lambda_{5}{(\Phi^{\dagger}{\Phi})}^{2})$ and $ \int_{0}^{\pi R} dy \{-(D_{5} \Phi)^\dagger(D_{5} \Phi)\}$ in the above equation give the required scalar interactions. Only scalar interactions are required, as in high energy limit, we can replace all the longitudinal modes of gauge bosons by their corresponding unphysical scalars, i.e., Goldstone modes by exploiting equivalence theorem. Higgs doublet $\Phi$ can be expanded~\cite{CorderoCid:2011ja, Garcia-Jimenez:2016cqe} in terms of zero-mode and its KK-tower as
$$\Phi = \frac {1}{\sqrt{r_{\phi} + \pi R}}\Phi^{0} + \Phi^{n} f_{\phi}^{n},$$ 
and $D_{5}\Phi$ can be written as
$$D_{5}\Phi=-M_{\Phi n}\Phi^{n} g_{\phi}^{n} -i \mathcal{\widetilde{X}}^{n}g_{\phi}^{n}\frac{\Phi^{0}}{\sqrt{r_{\phi}+\pi R}}-i\mathcal{\widetilde{X}}^{p}g_{\phi}^{p}\Phi^{n}f_{\phi}^{n},$$
where,
\begin{equation}
~~~~~~~~~~~~~~~~~~~~~~~~~~~~~~~~~~\mathcal{\widetilde{X}}^{n}= \frac{1}{2}\begin{pmatrix}
\tilde{g} W_{5}^{n3}+ \tilde g'B_{5}^{n} & \sqrt{2} \tilde{g} W_{5}^{n +} \\ \sqrt{2} \tilde{g} W_{5}^{n -} & -\tilde{g} W_{5}^{n3}+ \tilde g'B_{5}^{n}
\end{pmatrix}
\end{equation}
Substituting all the required $y$-profile in the above and integrating over $y$, the final form of the last two parts of Eq.~\ref{Lphi} (denoted as $ \mathcal L_{1}~{\rm and}~\mathcal L_{2}$) can be represented as
\begin{eqnarray}
\label{finalLag}
\mathcal{L}_{1} &=& \mu^2(\Phi^{0 \dagger}{\Phi ^0})-\lambda (\Phi^{0 \dagger}{\Phi})^2 +  \mu^2(\Phi^{n \dagger}{\Phi^n}) - 2 \lambda (\Phi^{0 \dagger}{\Phi ^0})(\Phi^{n \dagger}{\Phi^n})\nonumber \\
 & & -\lambda (\Phi ^{0 \dagger} \Phi^{n} + \Phi ^{n \dagger} \Phi^{0})(\Phi ^{0 \dagger} \Phi^{n} + \Phi ^{n \dagger} \Phi^{0}) -2 \lambda~ \mathcal{I}^{npq}(\Phi ^{0 \dagger} \Phi^{n} + \Phi ^{n \dagger} \Phi^{0})(\Phi^{p \dagger}{\Phi^q})\nonumber \\
 & & -\lambda ~\mathcal{I}^{npqr}(\Phi^{n \dagger}{\Phi^p})(\Phi^{q \dagger}{\Phi^r}),\\
\label{finalLag1}
\mathcal{L}_{2} &=& -M_{\Phi n}^2 \Phi^{n \dagger}{\Phi^n} - iM_{\Phi n}\Phi^{n \dagger}\mathcal{X}^{n}{\Phi ^0} + i M_{\Phi n}\Phi^{0 \dagger}\mathcal{X}^{\dagger n}{\Phi^ n} - \Phi^{0 \dagger}\mathcal{X}^{\dagger n}\mathcal{X}^{n}{\Phi^0}\nonumber \\
 & & -i M_{\Phi n}~\mathcal{I_{\rm 1}}^{npq}\Phi^{n \dagger}\mathcal{X}^{p}{\Phi ^q} + i M_{\Phi q}~\mathcal{I_{\rm 1}}^{pqn}\Phi^{n \dagger}\mathcal{X}^{p \dagger}{\Phi ^q}- \mathcal{I_{\rm 1}}^{npq}\Phi^{0 \dagger}\mathcal{X}^{n \dagger}\mathcal{X}^{p}{\Phi ^q}- \mathcal{I_{\rm 1}}^{pqn}\Phi^{n \dagger}\mathcal{X}^{p \dagger}\mathcal{X}^{q}{\Phi ^0}\nonumber \\
 & & - \mathcal{I_{\rm 1}}^{pqn}\Phi^{n \dagger}\mathcal{X}^{p \dagger}\mathcal{X}^{q}{\Phi ^0}- \mathcal{I_{\rm 1}}^{prnq}\Phi^{n \dagger}\mathcal{X}^{p \dagger}\mathcal{X}^{r}{\Phi ^q},
\end{eqnarray}
where, the sum over all possible KK-indices are implied. In the above, $\lambda$ is the four dimensional counterpart of $\lambda_{5}$ given as
$$ \lambda=\frac{\lambda_{5}}{r_{\phi} + \pi R}. $$
$\mathcal X$s are the matrices given as
\begin{equation}
~~~~~~~~~~~~~~~~~~~~~~~~~~~~~~~~~~\mathcal{X}^{n}= \frac{1}{2}\begin{pmatrix}
g W_{5}^{n3}+ g'B_{5}^{n} & \sqrt{2} g W_{5}^{n +} \\ \sqrt{2} g W_{5}^{n -} & -g W_{5}^{n3}+ g'B_{5}^{n}
\end{pmatrix}
\end{equation}
which are related to its five dimensional counterpart as 
\begin{equation}
~~~~~~~~~~~~~~~~~~~~~~~~~~~~~~~~~~~~~~~~~\mathcal{X}^{n}= \frac{\mathcal{\widetilde{X}}^{n}}{\sqrt{r_{g} + \pi R}}.
\end{equation}
Overlap integrals which arise from the integrations of $y$-profiles (which are actually not present in UED as the wave functions are of simple form like $\sin (\frac{ny}{R})$ or $\cos (\frac{ny}{R})$) are given as
\begin{eqnarray}
\label{Is}
\mathcal{I}^{npq} &=& \sqrt{r_{\phi} +\pi R}\int_{0}^{\pi R} dy \;[1+r_{\phi}\{\delta(y)+\delta(y-\pi R)\}]f_{\phi}^{n}f_{\phi}^{p}f_{\phi}^{q},\\
\label{In}
\mathcal{I}^{npqr} &=& (r_{\phi} +\pi R)\int_{0}^{\pi R} dy \;[1+r_{\phi}\{\delta(y)+\delta(y-\pi R)\}]f_{\phi}^{n}f_{\phi}^{p}f_{\phi}^{q}f_{\phi}^{r},\\
\label{Is1}
\mathcal{I_{\rm 1}}^{npq} &=& \sqrt{r_{\phi} +\pi R}\int_{0}^{\pi R} dy \;g_{\phi}^{n}g_{\phi}^{p}f_{\phi}^{q},\\
\label{Is2}
\mathcal{I_{\rm 1}}^{pqn} &=& \sqrt{r_{\phi} +\pi R}\int_{0}^{\pi R} dy \;g_{\phi}^{p}g_{\phi}^{q}f_{\phi}^{n},\\
\label{In1}
\mathcal{I_{\rm 1}}^{prnq} &=& (r_{\phi} +\pi R)\int_{0}^{\pi R} dy \;g_{\phi}^{p}g_{\phi}^{r}f_{\phi}^{n}f_{\phi}^{q}.
\end{eqnarray}

From Eqs.~\ref{finalLag} and \ref{finalLag1}, we can get the mass of $n$th mode Higgs as $m_{h n} \equiv \sqrt {M_{\Phi n} ^2 + m_h^2}$, where $m_{h}$ denotes the mass of zero-mode Higgs. The overlap integrals are nonzero when the sum of all indices ($n+p+q+r$) are even and zero when the sum is odd as a consequence of conservation of KK-parity. Substituting all the expressions in terms of $A^{n}$, $G_{Z}^{n}$, $H^{\pm n}$ and $G^{\pm n}$ in Eqs.~\ref{finalLag} and \ref{finalLag1} all the couplings can be calculated. We list all the necessary Feynman rules in APPENDIX B.
\section{Relevant Scattering Processes}
In this section, all necessary processes are given from which we can set an upper bound on gauge and scalar BLT parameters using Eq.~\ref{unitarity_constraint}. The calculations will be restricted to $n, n \rightarrow n, n$ processes, that is KK-numbers of initial and final states are the same. We will consider only those processes arising from quartic couplings which are not suppressed by KK-masses\footnote{For example, let us consider the process $G^{n+}G^{n-} \rightarrow G^{n+}G^{n-}$. The corresponding quartic coupling is given by $(-2\frac{m_{h}^{2}}{v^{2}}\frac{M_{W}^{4}}{M_{Wn}^{4}}I^{n})$, which is suppressed by $M_{Wn}^{4}$; i.e., the coupling $\sim$ $1/({\rm KK-mass})^{4}$ and hence are numerically insignificant. Therefore, we have ignored those kind of processes which are suppressed by KK-masses (e.g., the processes corresponding to $G^{n+}G^{n-}$ in initial or in final state).}. Thus we have altogether 13 quartic couplings in scalar sector satisfying the above conditions. In $2 \rightarrow 2$ processes, there are neutral two-particle states and charged two-particle states.
 The bases of neutral two-particle states are given by 
$$\Big\{ \frac{h^{n}h^{n}}{\sqrt{2}}, \frac{A^{n}A^{n}}{\sqrt{2}}, \frac{G_{Z}^{n}G_{Z}^{n}}{\sqrt{2}},G_{Z}^{n}A^{n},H^{n+}H^{n-}, H^{n \pm}G^{n \mp}\Big\}~{\rm and}~\Big\{h^{n}A^{n}, h^{n}G_{Z}^{n}\Big\},$$
and the bases of charged two-particle states are 
$$\Big\{ H^{n \pm }h^{n}, G^{n \pm} h^{n}, H^{n \pm}G_{Z}^{n}\Big\}~{\rm and}~\Big\{H^{n \pm}A^{n}, G^{n \pm}A^{n}\Big\}.$$
 The above bases show that there are two different neutral two-particle states and two types of charged two-particle states. Since we are working in CP-conserving scenario, $h^{n}$ being CP-even and $A^{n}, G_{Z}^{n}$ being CP-odd, there will be no mutual interactions among these two different kinds of states. The diagrams for the required processes are given in Figs.~\ref{fig1}-\ref{fig6}, and their corresponding expressions of $a_{0}$ are given in APPENDIX A. In this analysis, radiative correction of Weinberg angle ($\theta_{W}$)~\cite{Cheng:2002iz} in KK-mode has not been included.

The quartic couplings in Fig.~\ref{fig3} (a) can also generate the processes $h^{n}h^{n}\rightarrow A^{n}A^{n}$, $h^{n}h^{n}\rightarrow G_{Z}^{n}G_{Z}^{n}$. But the amplitudes in these cases will be further suppressed by a factor of $\frac{1}{2}$ as compared to the amplitudes of the  processes given in Fig.~\ref{fig3} corresponding to the same set of parameters. This suppression occurs due to the normalization factor $\frac{1}{\sqrt{2}}$ for the presence of same bosonic state in both the initial and the final state.  Same argument will be used for the process $A^{n}A^{n}\rightarrow G_{Z}^{n}G_{Z}^{n}$ arising from the quartic coupling in Fig.~\ref{fig1} (a). This amplitude will be also suppressed by a factor of $\frac{1}{2}$ as compared to the process $A^{n}G_{Z}^{n}\rightarrow A^{n}G_{Z}^{n}$ mentioned in Fig.~\ref{fig1}. The quartic coupling in Fig.~\ref{fig5} (a) can also gives rise to the processes $h^{n}h^{n} \rightarrow H^{n +}H^{n -}$, $h^{n}h^{n} \rightarrow G^{n +}G^{n-}$, $A^{n}A^{n} \rightarrow H^{n +}H^{n -}$, $A^{n}A^{n} \rightarrow G^{n +}G^{n -}$, $G_{Z}^{n}G_{Z}^{n} \rightarrow H^{n +}H^{n -}$, which will be relatively suppressed by a factor of $\frac{1}{\sqrt{2}}$ for the same value of BLT parameter as compared to the processes given in Fig.~\ref{fig5}, due to the presence of identical bosonic state in the initial state. Though these processes will contribute to unitarity breaking at some relatively larger value of $r_{\phi}$ and will cancel the effect of the factor $\frac{1}{2}$ or $\frac{1}{\sqrt{2}}$ at this larger value, but the main motive of  $n, n \rightarrow n, n$ single channel analysis in this section and also in the immediate next section~\ref{nn_nn} is to zero in on the relevant channels, or bases for coupled channel analysis, which will give more stringent constraints on the parameter space. For example, both $h^{n}h^{n}\rightarrow A^{n}A^{n}$ and $h^{n}A^{n}\rightarrow h^{n}A^{n}$ involve the same quartic coupling but the amplitude of one process carries an extra factor of 1/2 and thus the bound on the BLKT parameter from unitarity violation will be different for these two different channels. Therefore we are considering only those processes which give much stringent constraint on BLT parameter\footnote{We shall see shortly, in section \ref{coupled_chnl}, in case of coupled channel analysis, all the bases will exhaustively include all the required quartic couplings (even the dibosonic states).}. Similar arguments hold for the processes $h^{n}h^{n} \rightarrow H^{n +}H^{n -}$, $h^{n}h^{n} \rightarrow G^{n +}G^{n-}$, $A^{n}A^{n} \rightarrow H^{n +}H^{n -}$, $A^{n}A^{n} \rightarrow G^{n +}G^{n -}$, $G_{Z}^{n}G_{Z}^{n} \rightarrow H^{n +}H^{n -}$ that are not being considered in this section and also in section~\ref{nn_nn} since the expressions of $a_{0}$ in that case will result in a unitarity violation at relatively larger value of $r_{\phi}$ and therefore will provide a relatively relaxed bound as compared to the processes shown in Figs.~\ref{fig1}, \ref{fig3} and \ref{fig5}.

\begin{figure}[!htbp]
\centering
 \subfigure[]{
   \includegraphics[scale=0.4]{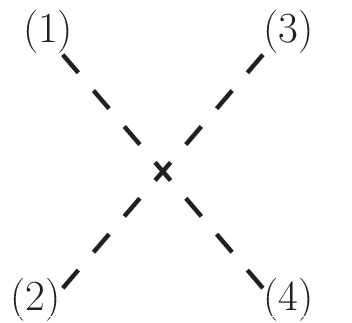}    
    }
    \subfigure[]{
   \includegraphics[scale=0.4]{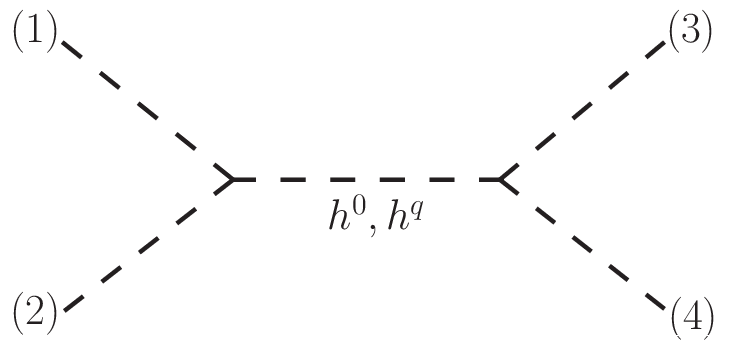}
   }
  \subfigure[]{
   \includegraphics[scale=0.38]{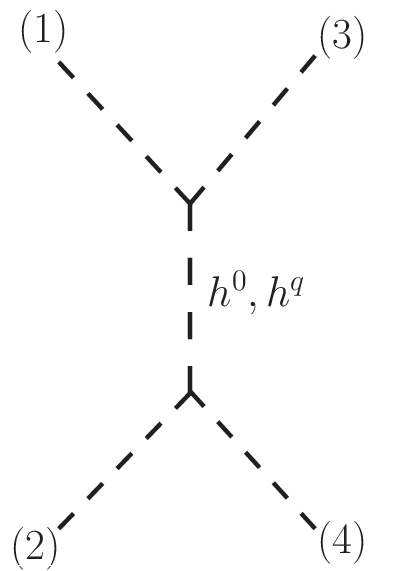}
   }
  \subfigure[]{
   \includegraphics[scale=0.38]{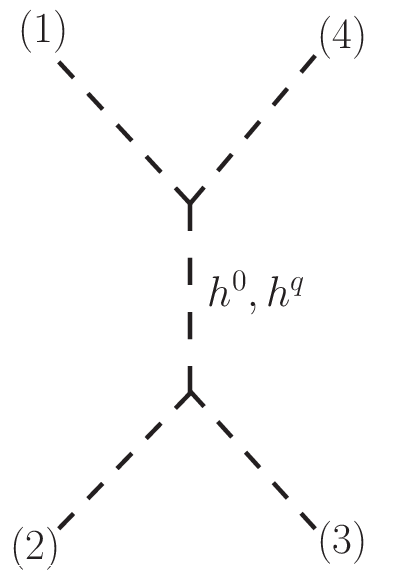}
    }
    \caption[]{Diagrams for the processes $h^{n}(1)h^{n}(2) \rightarrow h^{n}(3)h^{n}(4)$, $A^{n}(1)A^{n}(2) \rightarrow A^{n}(3)A^{n}(4)$, $G_{Z}^{n}(1)A^{n}(2) \rightarrow G_{Z}^{n}(3)A^{n}(4)$.}
\label{fig1}
\end{figure}
\begin{figure}[!htbp]
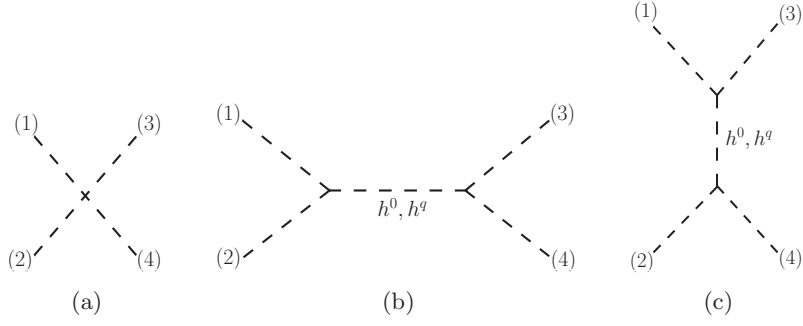

\centering
 \subfigure[]{
   \includegraphics[scale=0.4]{a02.eps}    
   }
   \subfigure[]{
   \includegraphics[scale=0.4]{a0s2.eps}
   }
  \subfigure[]{
   \includegraphics[scale=0.38]{a0t2.eps}
   }
    \caption[]{Diagrams for the process involving $H^{n +}(1)H^{n -}(2) \rightarrow H^{n +}(3)H^{n -}(4)$.}
\label{fig2}
\end{figure}
\begin{figure}[!htbp]
\centering
 \subfigure[]{
   \includegraphics[scale=0.4]{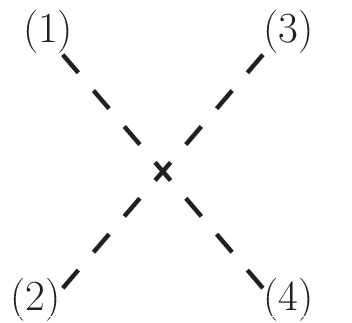}    
    }
    \subfigure[]{
   \includegraphics[scale=0.4]{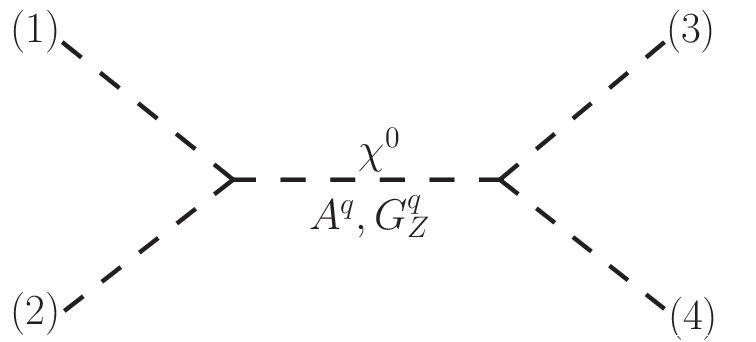}
   }
  \subfigure[]{
   \includegraphics[scale=0.38]{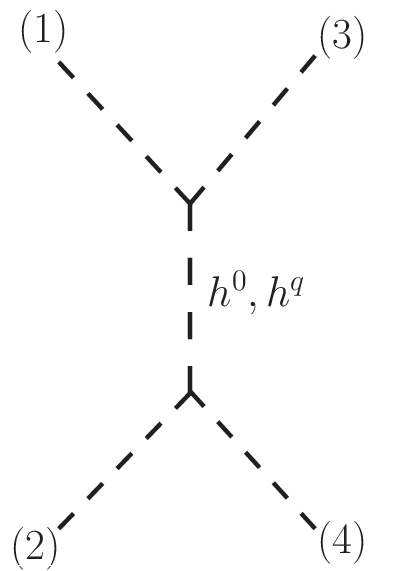}
   }
  \subfigure[]{
   \includegraphics[scale=0.38]{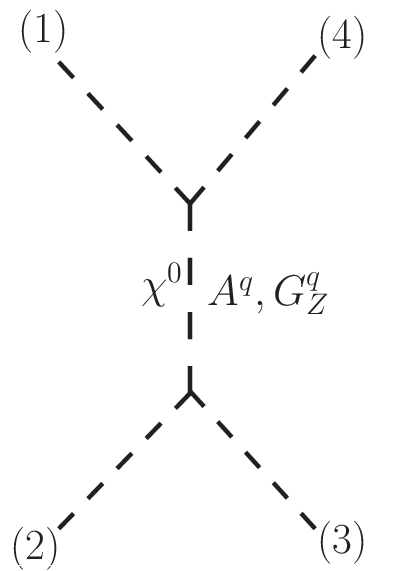}
    }
    \caption[]{Diagrams for the processes $h^{n}(1)A^{n}(2) \rightarrow h^{n}(3)A^{n}(4)$, $h^{n}(1)G_{Z}^{n}(2) \rightarrow h^{n}(3)G_{Z}^{n}(4)$.}
\label{fig3}
\end{figure}
\begin{figure}[!htbp]
\centering
 \subfigure[]{
   \includegraphics[scale=0.4]{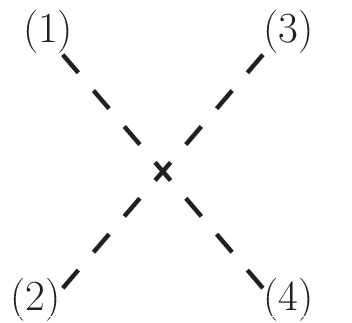}    
    }
    \subfigure[]{
   \includegraphics[scale=0.4]{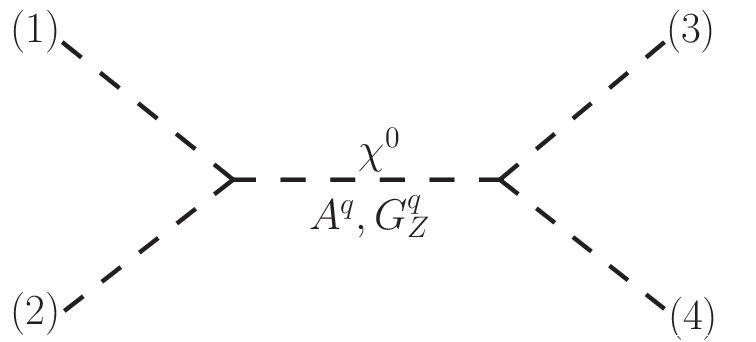}
   }
  \subfigure[]{
   \includegraphics[scale=0.38]{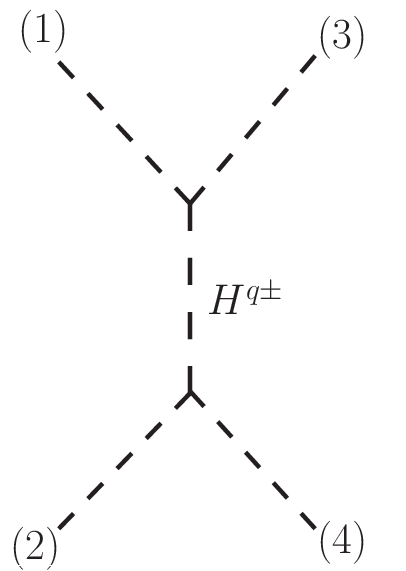}
   }
  \subfigure[]{
   \includegraphics[scale=0.38]{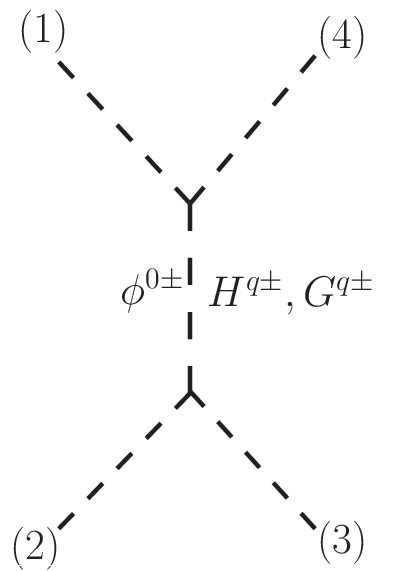}
    }
    \caption[]{Diagrams for the process $h^{n}(1)G_{Z}^{n}(2) \rightarrow H^{n \pm}(3)G^{n \mp}(4)$.}
\label{fig4}
\end{figure}
\begin{figure}[!htbp]
\centering
 \subfigure[]{
   \includegraphics[scale=0.4]{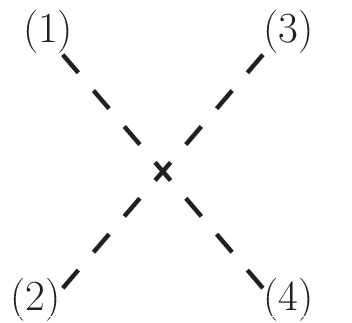}    
    }
    \subfigure[]{
   \includegraphics[scale=0.4]{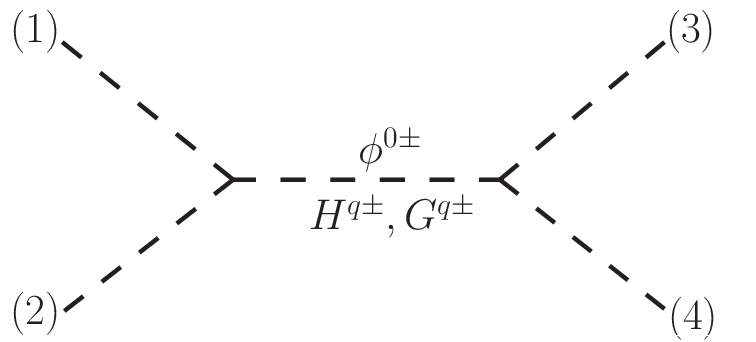}
   }
  \subfigure[]{
   \includegraphics[scale=0.38]{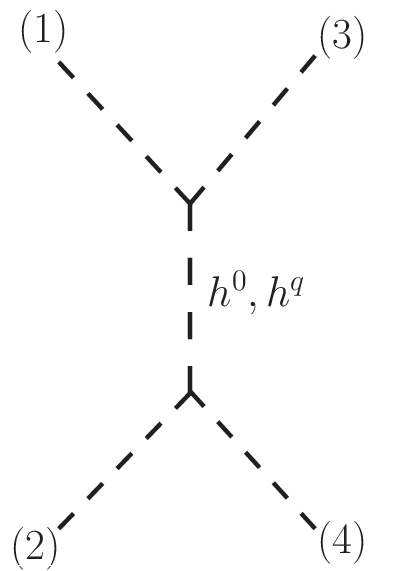}
   }
  \subfigure[]{
   \includegraphics[scale=0.38]{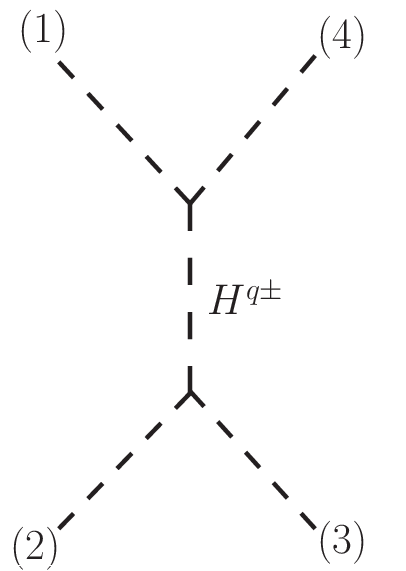}
    }
    \caption[]{Diagrams for the processes $H^{n \pm}(1)h^{n}(2) \rightarrow H^{n \pm}(3)h^{n}(4)$, $G^{n \pm}(1)h^{n}(2) \rightarrow G^{n \pm}(3)h^{n}(4)$, $H^{n \pm}(1)A^{n}(2) \rightarrow H^{n \pm}(3)A^{n}(4)$, $G^{n \pm}(1)A^{n}(2) \rightarrow G^{n \pm}(3)A^{n}(4)$, $H^{n \pm}(1)G_{Z}^{n}(2) \rightarrow H^{n \pm}(3)G_{Z}^{n}(4)$.}
\label{fig5}
\end{figure}
\begin{figure}[!htbp]
\centering
 \subfigure[]{
   \includegraphics[scale=0.4]{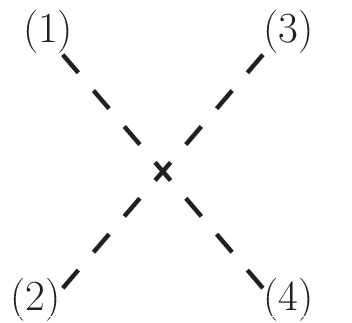}    
    }
    \subfigure[]{
   \includegraphics[scale=0.4]{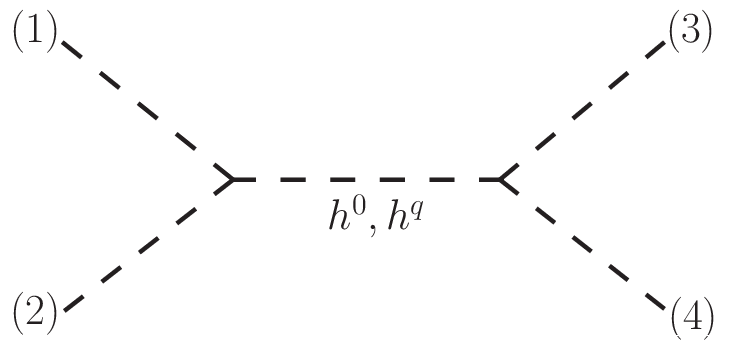}
   }
  \subfigure[]{
   \includegraphics[scale=0.38]{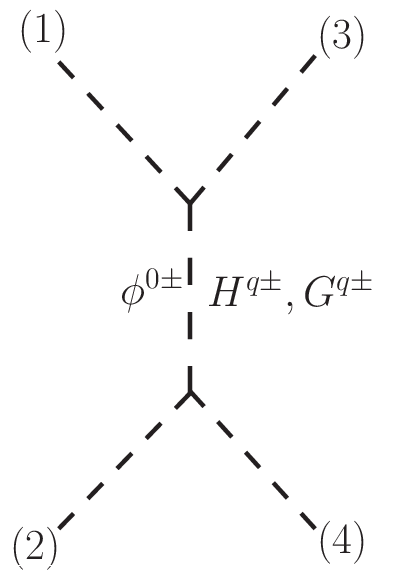}
   }
  \subfigure[]{
   \includegraphics[scale=0.38]{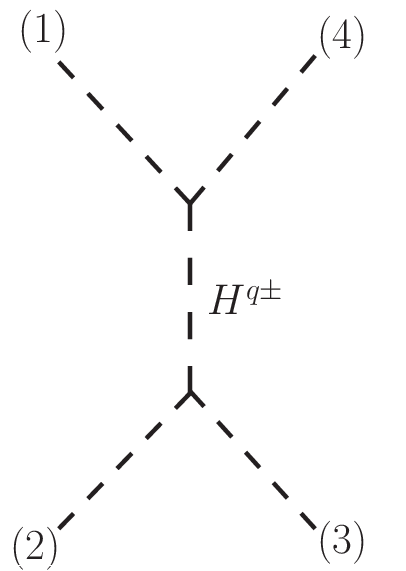}
    }
    \caption[]{Diagrams for the process $A^{n}(1)G_{Z}^{n}(2) \rightarrow G^{n \mp}(3)H^{n \pm}(4)$.}
\label{fig6}
\end{figure}

The expressions for $a_{0}$ for each processes can be studied as function of $s$ ($\sqrt{s}$ being the centre of mass energy for respective processes) for different values of $r_{\phi}$ and one can set an upper bound on BLT parameters using the Eq.~\ref{unitarity_constraint}. This $a_{0}$ can also be studied as function of BLT parameter $r_{\phi}$ for a fixed value of $s$. The value of $r_{\phi}$ for which $|{\rm Re}~a_{0}|$ will be greater than half even at large limit of $s$, would give us the required upper bound on the BLT parameter. 

 Now, the coupled channel analysis (which was mentioned in Sec. 2) can be performed for suitable set of processes to get further constraint on the upper bound on BLT parameters \cite{SekharChivukula:2001hz}. However, it is not possible to obtain the channels for this purpose before getting the results of single channel scattering analysis. The procedure of the formation of the $t^{0}$ matrix with the appropriate basis as well as the results will be analyzed elaborately in the next section after showing the results of $n, n \rightarrow n, n$ scattering.

%

%
%
%
%
%
%
%
\section{Results}
\subsection{$n, n \rightarrow n, n$ processes}
\label{nn_nn}
 In this section, we will discuss the variations of $a_{0}$ for different processes as function of $s$ for a fixed value of BLT parameter $r_{\phi}$ and vice versa. Since we are dealing only with $n, n \rightarrow n, n$ processes, the variation of $a_{0}$ will be analyzed for specific KK-modes ($n=1-4$). In this single channel scattering analysis, we restrict ourselves to the KK-number up to 4. It would be clear in the later part of this section that to obtain suitable channels for the coupled channel analysis it is sufficient to study the single channel analysis with KK-mode up to 4.
 
\begin{figure}[!htbp]

 \centering
 \subfigure[]{
   \includegraphics[scale=0.4]{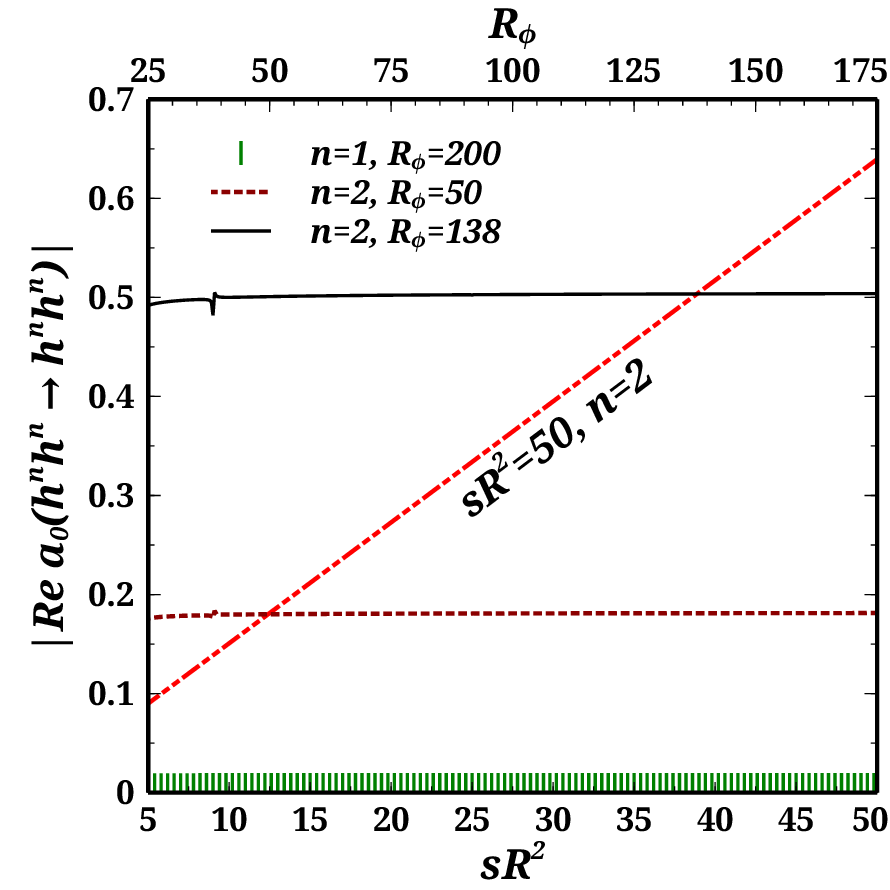}    
    }
    \subfigure[]{
   \includegraphics[scale=0.4]{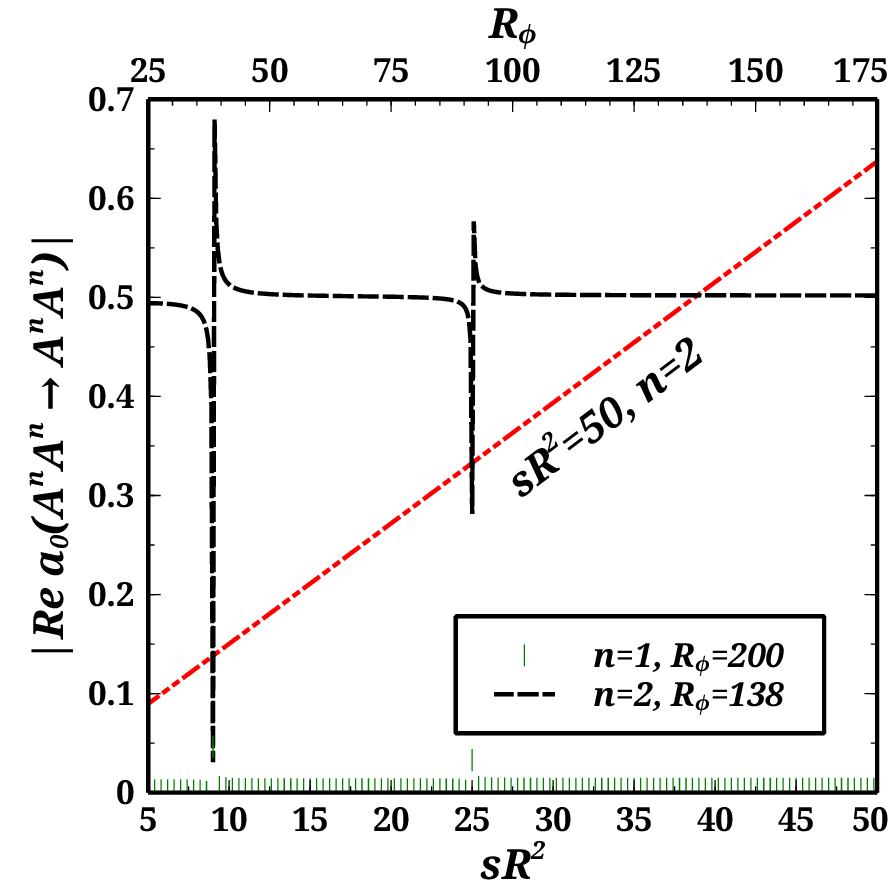}
   } \\
  \subfigure[]{
   \includegraphics[scale=0.4]{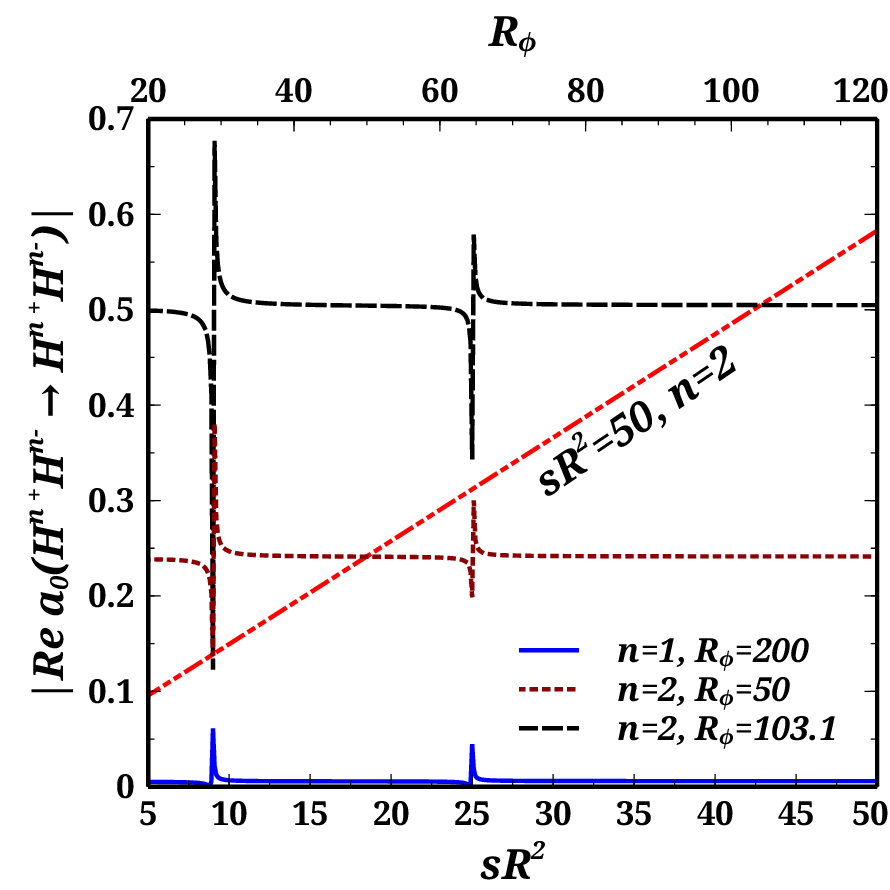}
   } 
  \subfigure[]{
   \includegraphics[scale=0.4]{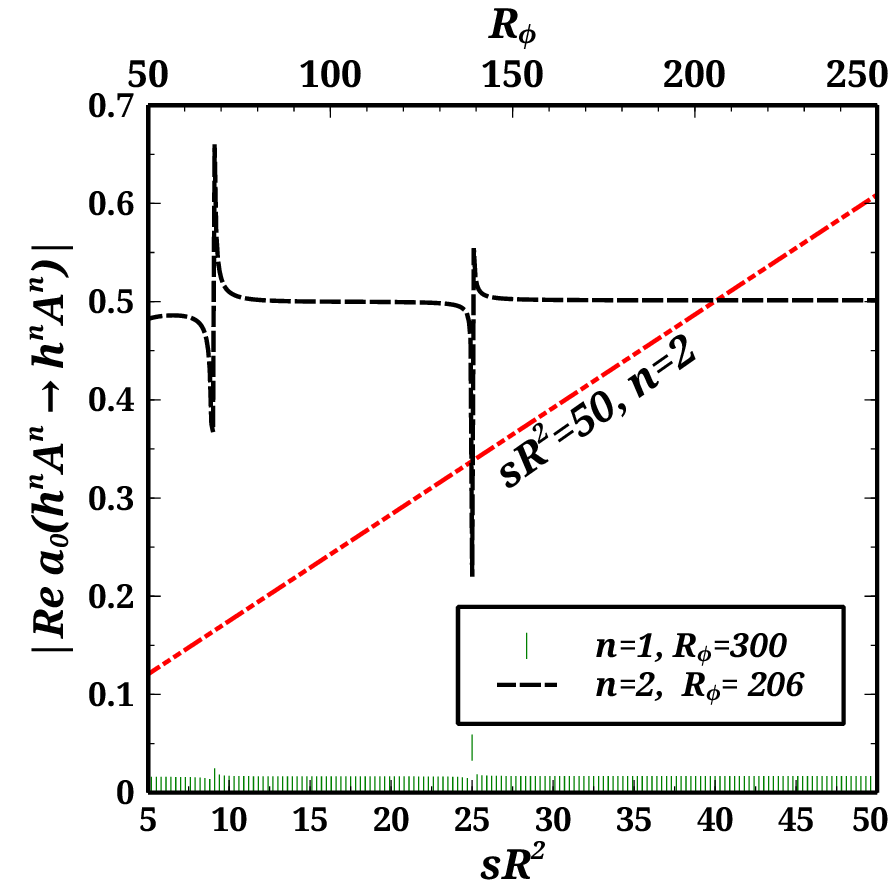}
    } \\
 \subfigure[]{
   \includegraphics[scale=0.4]{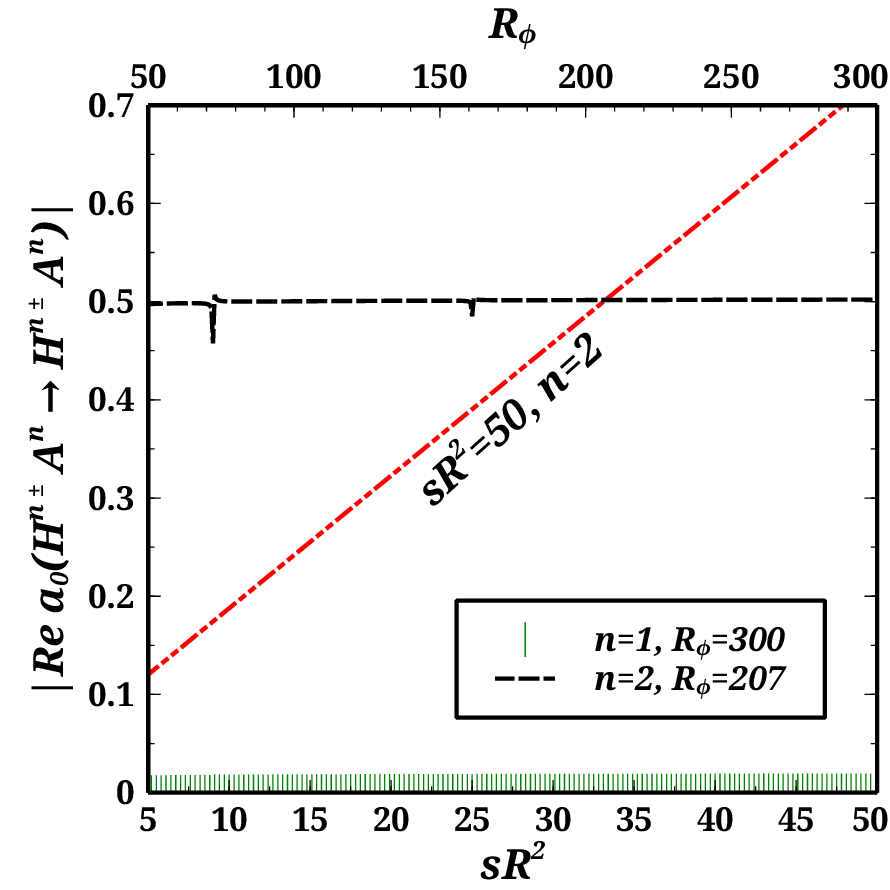}
   } 
  \subfigure[]{
   \includegraphics[scale=0.4]{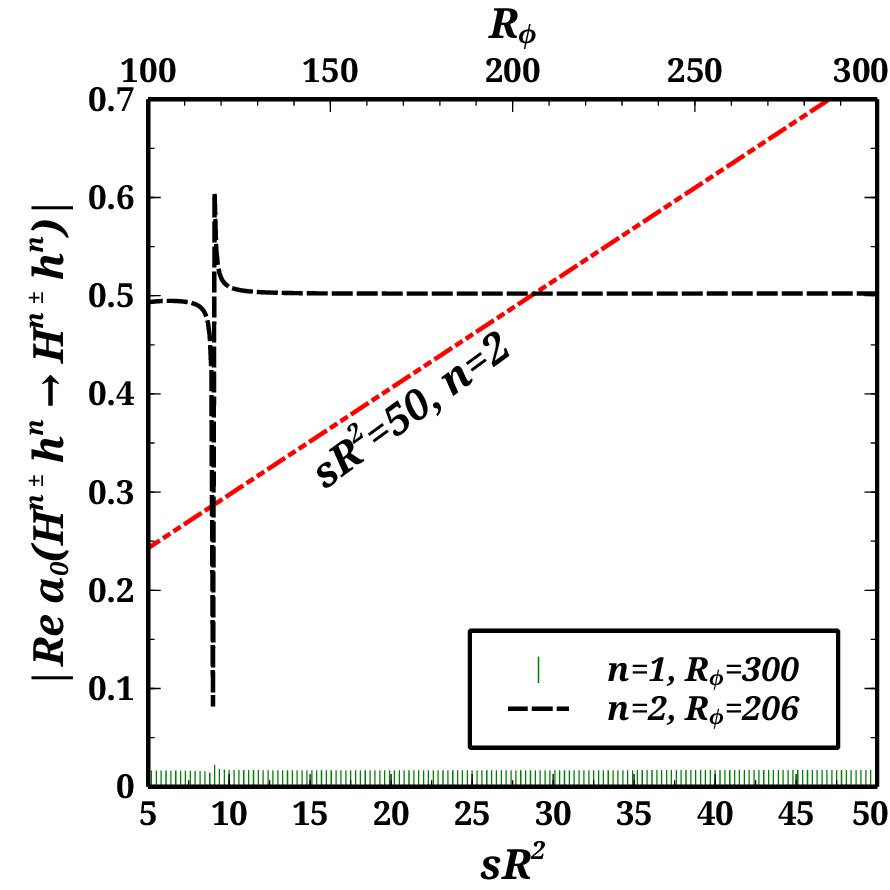}
    }

    \caption[]{Variation of $a_{0}$ as a function of $sR^{2}$ for different KK-mode with different values of $R_{\phi}$, and also as a function of $R_{\phi}$ for second KK-mode with $sR^{2}=50$. There are two horizontal axes in each plot. The lowest one corresponds to $sR^{2}$ for different values of $R_{\phi}$ and the upper one gives the variation of $a_{0}$ as a function of $R_{\phi}$ for fixed value of $sR^{2}$. Both dependences have been shown for specific KK-modes. Variations of $a_{0}$ for the processes $h^{n}h^{n} \rightarrow h^{n}h^{n}$, $A^{n}A^{n} \rightarrow A^{n}A^{n}$, $H^{n+}H^{n-} \rightarrow H^{n+}H^{n-}$, $h^{n}A^{n} \rightarrow h^{n}A^{n}$, $H^{n\pm}A^{n} \rightarrow H^{n\pm}A^{n}$, $H^{n\pm}h^{n} \rightarrow H^{n\pm}h^{n}$ are shown. Here $R^{-1}$ is taken to be 1500 GeV.}
    \label{figa0}
\end{figure}

 In the Fig.~\ref{figa0}, the variation of $a_{0}$ for these six processes $h^{n}h^{n} \rightarrow h^{n}h^{n}$, $A^{n}A^{n} \rightarrow A^{n}A^{n}$, $H^{n+}H^{n-} \rightarrow H^{n+}H^{n-}$, $h^{n}A^{n} \rightarrow h^{n}A^{n}$, $H^{n\pm}A^{n} \rightarrow H^{n\pm}A^{n}$, $H^{n\pm}h^{n} \rightarrow H^{n\pm}h^{n}$ have been presented. As the BLKT parameter $r_{\phi}$ is a dimensionful parameter, we will use scaled BLKT parameter $R_{\phi}\equiv r_{\phi}/R$ while presenting our results. There are two horizontal axes for each plot, lower one corresponds to $sR^{2}$ and the other corresponds to $R_{\phi}$. The vertical axis gives the values of $a_{0}$ for different values of $s$ and $R_{\phi}$. Here we have taken $R^{-1}$ as 1500 GeV. From these figures, we can see, for $n=1$ the $|{\rm Re}~a_{0}|$ is much less than half with the variation of $s$ even at very large value of $R_{\phi}$. These figures also reflects the fact that $|{\rm Re}~a_{0}|$ is almost independent of $s$ for $n=1$. There is no unitarity violation for these processes for $n=1$.

From the Fig.~\ref{figa0}, it is evident that the variation of $a_{0}$ is quite different for $n=2$ from the variation of $a_{0}$ for $n=1$. The value of $|{\rm Re}~a_{0}|$ can be greater than half for some specific value of $R_{\phi}$ for a given value of $R^{-1}$. As example, $|{\rm Re}~a_{0}|$ for the processes  
$h^{n}h^{n} \rightarrow h^{n}h^{n}$, $A^{n}A^{n} \rightarrow A^{n}A^{n}$, $H^{n+}H^{n-} \rightarrow H^{n+}H^{n-}$, $h^{n}A^{n} \rightarrow h^{n}A^{n}$, $H^{n\pm}A^{n} \rightarrow H^{n\pm}A^{n}$, $H^{n\pm}h^{n} \rightarrow H^{n\pm}h^{n}$ at $R^{-1} = 1500$ GeV becomes greater than half when $R_{\phi}$ is 138, 138, 104, 206, 207 and 206 respectively even at large value of $sR^{2}$. So among all the processes mentioned in Fig.~\ref{figa0}, the process $H^{n+}H^{n-} \rightarrow H^{n+}H^{n-}$ gives the most stringent upper limit on the value of $R_{\phi}$ (Fig.~\ref{figa0} (c)) for $n=2$. One can see at $R_{\phi} = 104$, $|{\rm Re}~a_{0}|$ becomes greater than half signaling the breakdown of unitarity.

  The discontinuity along the curves corresponds to different values of pole masses of the propagators. The conservation of KK-parity ensures that, whether $n=1$ or $n=2$, only even KK-modes can arise along the propagators (the KK index along the propagator is denoted by $q$). When $a_{0}$ is considered as a function of $R_{\phi}$ for a fixed value of $sR^{2}$ for specific KK-mode (in Fig.~\ref{figa0}, $sR^{2}=50$ and $n=2$), variation of $|{\rm Re}~a_{0}|$ is a straight line and will be greater than half at the same value of $R_{\phi}$ at which the violation occurs with the variation of $sR^{2}$ even at large $s$ limit.

For $n=3,~{\rm and}~n=4$ and with $R^{-1}=1500$ GeV, the values of $R_{\phi}$ at which $|{\rm Re}~a_{0}|> 1/2 $ are given in Table~\ref{table}. For higher values of KK-modes violation occurs at relatively lower values of $R_{\phi}$ reducing the allowed upper value of BLT parameters. The data in the table reflects the fact that the process $H^{n+}H^{n-} \rightarrow H^{n+}H^{n-}$ gives the tightest upper bound on $R_{\phi}$, for $n=3$ and $n=4$ the bounds are slightly different, $R_{\phi}$ should be less than $99.9$ for $n=3$ and $99.4$ for $n=4$.

\begin{table}[!htbp]
\begin{center}
\begin{tabular}{|c||c|c|} 
\hline 
Processes & \begin{tabular}[c]{@{}c@{}}Value of $R_{\phi}$\\ ($n=3$)\end{tabular} & \begin{tabular}[c]{@{}c@{}}Value of $R_{\phi}$\\ ($n=4$)\end{tabular} \\ \hline
$h^{n}h^{n} \rightarrow h^{n}h^{n}$        & 134.2                                                     & 133.9                                                               \\ \hline
$A^{n}A^{n} \rightarrow A^{n}A^{n}$          & 134.2                                                     & 133.7                                                               \\ \hline
$H^{n+}H^{n-} \rightarrow H^{n+}H^{n-}$          & 99.9                                                      & 99.4                                                                \\ \hline
$h^{n}A^{n} \rightarrow h^{n}A^{n}$          & 202.1                                                      & 202                                                               \\ \hline
$H^{n\pm}A^{n} \rightarrow H^{n\pm}A^{n}$         & 202.9                                                      & 202.2                                                                \\ \hline
$H^{n\pm}h^{n} \rightarrow H^{n\pm}h^{n}$         & 202                                                      & 201.3                                                                \\ \hline
\end{tabular}
\caption{Values of $R_{\phi}$ for KK-mode $n=3$ and $n=4$ for different processes $h^{n}h^{n} \rightarrow h^{n}h^{n}$, $A^{n}A^{n} \rightarrow A^{n}A^{n}$, $H^{n+}H^{n-} \rightarrow H^{n+}H^{n-}$, $h^{n}A^{n} \rightarrow h^{n}A^{n}$, $H^{n\pm}A^{n} \rightarrow H^{n\pm}A^{n}$, $H^{n\pm}h^{n} \rightarrow H^{n\pm}h^{n}$, at which unitarity violation occurs i.e., $|{\rm Re}~a_{0}|> 1/2 $. Here, $R^{-1}$ is taken as $1500$ GeV.}
\label{table}
\end{center}
\end{table}

In Table~\ref{table1}, the values of $R_{\phi}$  at which the unitarity  violation occurs for different values of $R^{-1}$ have been given for the process $H^{n+}H^{n-} \rightarrow H^{n+}H^{n-}$. The values are given for different KK-modes. In this table, the $a_{0}$ coming only from  quartic coupling contributions and from total amplitude (i.e., quartic coupling contributions along with contributions coming from trilinear coupling) are separately analyzed for different values of $R^{-1}$ and for different KK-modes. We can see that, the $R^{-1}$ has a nominal effect on the bounds and slightly shifts the bounds to a lower value  for all KK-modes when $R^{-1}$ is increased. For $R^{-1}=1.5$ TeV and $n=2$, the upper bounds on $R_{\phi}$ coming from the contributions from quartic interactions and from total amplitude are differed by $0.1$ only; $a_{0}$ coming from only the quartic interactions gives the upper bound as $R_{\phi}< 103$ for $n=2$. This small discrepancy vanishes for higher values of $R^{-1}$ which results in a nominal shift in bound as $R_{\phi}< 102.6$ for same KK-mode. This value is same for all $R^{-1}$ from 5 TeV onwards. At $R^{-1}=10$ TeV and for $n=3$, $R_{\phi}$ should be less than $99.8$, and for $n=4$ the value will be $99.3$.
 So for sufficiently large values of $R^{-1}$, contributions coming from trilinear couplings are fully suppressed by $E^{2}$ and $a_{0}$ solely depends on $R_{\phi}$; the contributions are mostly determined by quartic couplings. Clearly the sum over KK-modes along the propagators which has been taken up to $q=4$ does not affect the result significantly. Since the quartic couplings in these processes mentioned in Fig.~\ref{figa0} are not suppressed by KK-masses and also the overlap integrals in these couplings are independent of $R^{-1}$, the results have nominal dependence on the value of $R^{-1}$. 
\begin{table}[]
\centering
\begin{tabular}{|c||c|c||c|c||c|c|}
\hline
\multirow{3}{*}{\begin{tabular}[c]{@{}c@{}}The\\  Value of\\ $R^{-1}$\\ in GeV\end{tabular}} & \multicolumn{6}{c|}{\begin{tabular}[c]{@{}c@{}}The value of $R_{\phi}$ for $|{\rm Re}~a_{0}| > \frac{1}{2}$\\                   for different KK-modes\end{tabular}}                                                                                                                                                                                                                                                                                                                                                               \\ \cline{2-7} 
                                                                                             & \multicolumn{2}{c|}{$n=2$}                                                                                                                                           & \multicolumn{2}{c|}{$n=3$}                                                                                                                                           & \multicolumn{2}{c|}{$n=4$}                                                                                                                                           \\ \cline{2-7} 
                                                                                             & \begin{tabular}[c]{@{}c@{}}From quartic \\ coupling\end{tabular} & \begin{tabular}[c]{@{}c@{}}From total \\ amplitude\end{tabular} & \begin{tabular}[c]{@{}c@{}}From quartic \\ coupling\end{tabular} & \begin{tabular}[c]{@{}c@{}}From total \\ amplitude\end{tabular} & \begin{tabular}[c]{@{}c@{}}From quartic \\ coupling\end{tabular} & \begin{tabular}[c]{@{}c@{}}From total\\  amplitude\end{tabular} \\ \hline \hline
1500                                                                                         & 103                                                                             & 103.1                                                                            & 99.9                                                                               & 99.9                                                                              & 99.3                                                                             & 99.4                                                                              \\ \hline
2500                                                                                         & 102.8                                                                             & 102.8                                                                            &99.9                                                                                   &99.9                                                                                 &99.3                                                                                   &99.3                                                                                  \\ \hline
5000                                                                                         & 102.6                                                                             & 102.6                                                                            &99.8                                                                                   &99.8                                                                                  &99.3                                                                                     &99.3                                                                                  \\ \hline
7500                                                                                         & 102.6                                                                             & 102.6                                                                            &99.8                                                                                   &99.8                                                                                  &99.3                                                                                     &99.3                                                                                  \\ \hline
10000                                                                                        & 102.6                                                                                                                                                   &102.6                                                                                  &99.8                                                                                   &99.8                                                                                  &99.3                                                                                     &99.3                                                                                  \\ \hline
\end{tabular}
\caption{Value of $R_{\phi}$ for the process $H^{n+}H^{n-} \rightarrow H^{n+}H^{n-}$ for different KK-modes at which unitarity violation occurs for different values of $R^{-1}$ (GeV). Here contributions to $a_{0}$ from quartic coupling and from total amplitude have been presented separately. The center of mass energy is taken as $\sqrt{s}=10$ TeV.}
\label{table1}
\end{table}

\begin{figure}[H]

 \centering
 \subfigure[]{
   \includegraphics[scale=0.4]{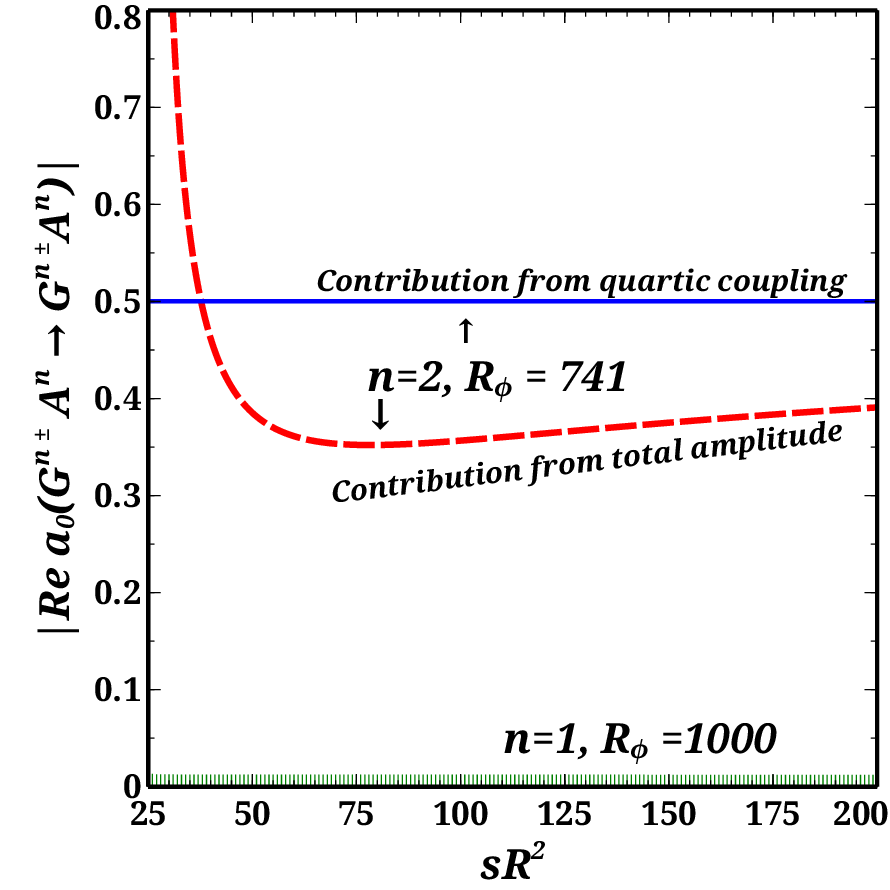}    
    }
    \subfigure[]{
   \includegraphics[scale=0.4]{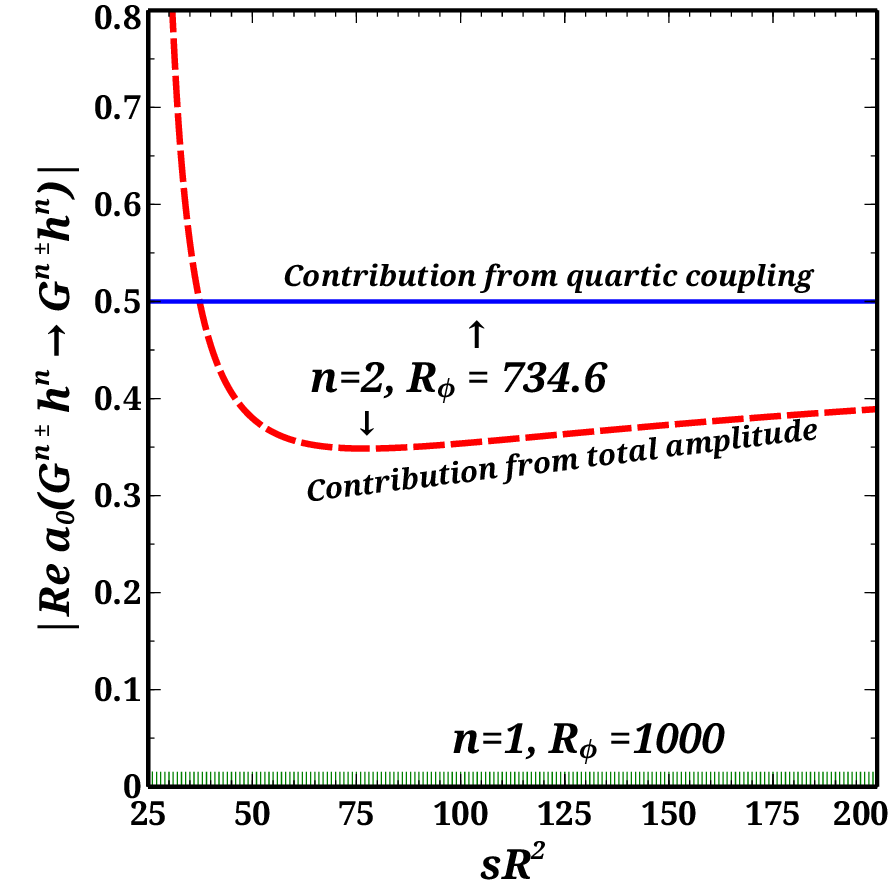}
   } \\
  \subfigure[]{
   \includegraphics[scale=0.4]{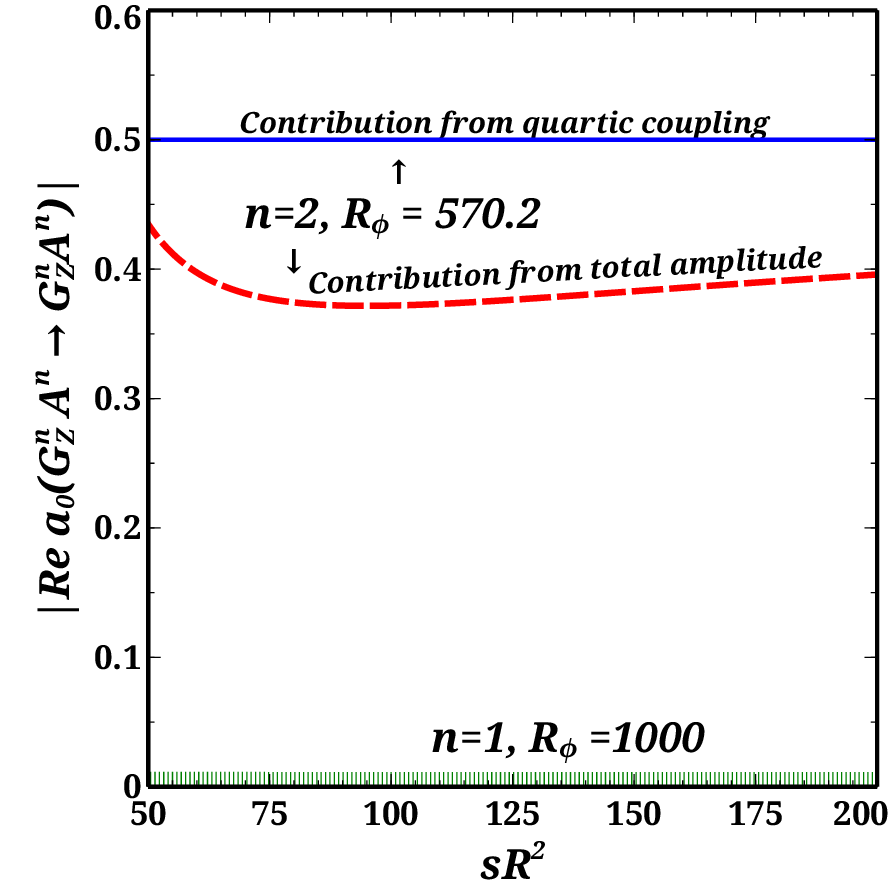}
   } 
  \subfigure[]{
   \includegraphics[scale=0.4]{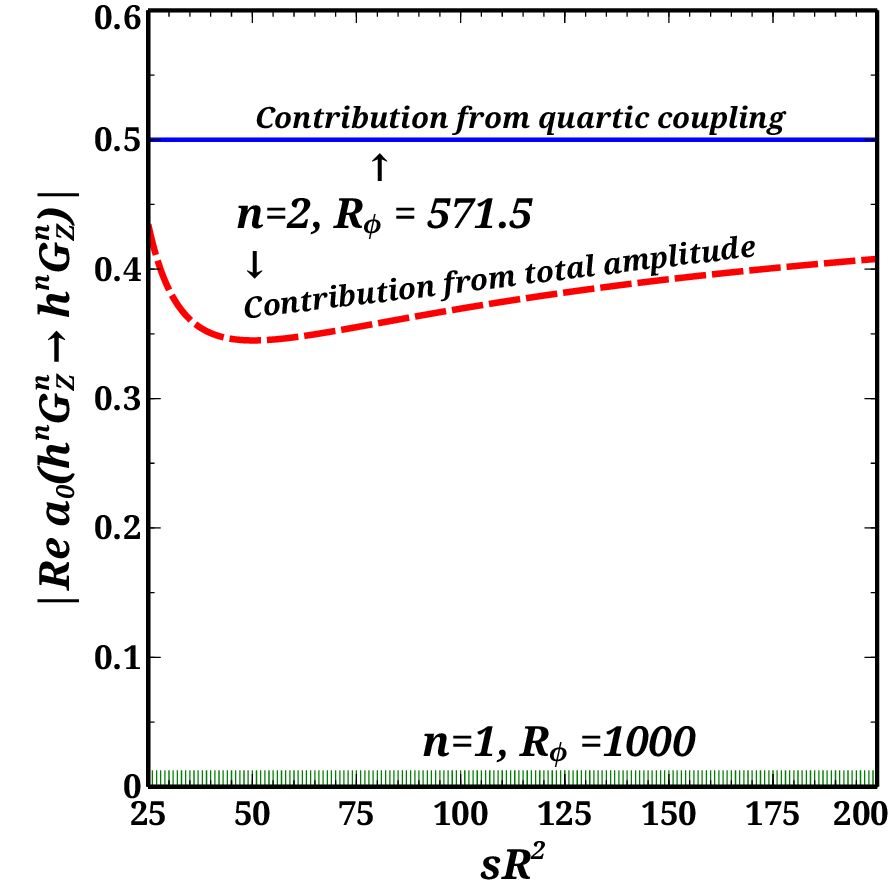}
    } 

    \caption[]{Variation of $a_{0}$ for processes $G^{n \pm}A^{n} \rightarrow G^{n \pm}A^{n}$, $G^{n \pm}h^{n} \rightarrow G^{n \pm}h^{n}$, $G_{Z}^{n}A^{n} \rightarrow G_{Z}^{n}A^{n}$, $h^{n}G_{Z}^{n} \rightarrow h^{n}G_{Z}^{n}$ as a function of $sR^{2}$ for different KK-mode with different values of $R_{\phi}$. Here $R^{-1}$ is taken to be 1500 GeV.}
    \label{figa0_1}
\end{figure}

      In the Fig.~\ref{figa0_1}, the variation of $a_{0}$ for the processes  $G^{n \pm}A^{n} \rightarrow G^{n \pm}A^{n}$, $G^{n \pm}h^{n} \rightarrow G^{n \pm}h^{n}$, $G_{Z}^{n}A^{n} \rightarrow G_{Z}^{n}A^{n}$, $h^{n}G_{Z}^{n} \rightarrow h^{n}G_{Z}^{n}$ as a function of $sR^{2}$ has been shown. For $n=1$ there is no unitarity violation. For $n=2$, the specific nature of $a_{0}$ due the contributions of quartic coupling and that from the total amplitude have been separately shown for a particular value of $R_{\phi}$. As an example, for the process $G^{n \pm}A^{n} \rightarrow G^{n \pm}A^{n}$, $|{\rm Re}~a_{0}|$ will become $\frac{1}{2}$ for $R_{\phi}= 741$, but contributions coming from total amplitude are much less than half. As the trilinear coupling in this case is effectively proportional to KK-masses, the numerator in the terms generated from trilinear couplings is effectively proportional to the square of KK-masses. Thus $a_{0}$ coming from the contributions of trilinear interactions falls from higher value than $1/2$, resulting initially a falling nature of $a_{0}$ with variation of $s$. In this case, $R^{-1}$ is taken as 1500 GeV. Evidently higher value of $R^{-1}$ will result in higher rate of falling of $a_{0}$ with $sR^{2}$. As $s$ increases the $E^{2}$-suppression increases, that is evident from the plots in Fig.~\ref{figa0_1}. Uniratity violation will occur at either very large value of $R_{\phi}$ or at very large value of $s$. So, the contributions coming from trilinear coupling can not be ignored when the couplings are effectively proportional to KK-masses. The same explanations will hold good for higher KK-modes. Here also, the sum over KK-modes along the propagators has been taken up to $q=4$. Further increase in $q$ does not change the result significantly as the contributions from higher modes will decouple.

   The other processes $A^{n}G_{Z}^{n} \rightarrow H^{n \pm}G^{n \mp}$, $h^{n}G_{Z}^{n} \rightarrow H^{n \pm}G^{n \mp}$, $H^{n \pm}G_{Z}^{n} \rightarrow H^{n \pm}G_{Z}^{n}$ give unitarity violation at very large value of $R_{\phi}$ and are irrelevant to our discussions. The uniratity violation with some specific value of $R_{\phi}$ actually occurs due to the presence of overlap integral (Eq.~\ref{In}) in quartic coupling, which for $n=p=q=r$ is denoted as $I^{n}$ in APPENDIX A. But the value of this overlap integral is very small for $n=1$ even at very large value of $R_{\phi}$. A Table~\ref{table2} for this overlap integral $I^{n}$ as function of $R_{\phi}$ is shown for different KK-modes ($n = 1 - 4$), which reflects the fact why there is no unitarity violation at $n=1$. Rather $I^{n}$ decreases with increasing value of $R_{\phi}$ for $n=1$. 

We would also like to make some additional remarks on the underlying effects of the overlap integrals on unitarity violation. The overlap integral $I^{n}$ enhances the couplings of the respective cross-sections of different $n, n \rightarrow n, n$ channels.  This enhancement makes the value of $|{\rm Re}~(a_{0})|\geq 1/2$ indicating the breakdown of unitarity. It is also noteworthy that this scenario is totally different from the case of basic five-dimensional UED where in the absence of Kaluza-Klein (KK) Higgs sector, gauge bosons do not respect partial wave unitarity if other KK-modes are involved. With the inclusion of higher modes of Higgs boson the unitarity is completely preserved~\cite{Belyaev:2012ai}. But this is the case of the five-dimensional UED, where there is no BLT and the theory is effectively one parameter theory. In this case the only parameter $R^{-1}$ does not play any role in unitarity violation. Things are certainly different in the case of nmUED scenario where the BLTs are present. Due to the presence of BLTs in nmUED, the $y$-profile solutions are different from that of the case of basic UED and in effective four-dimensional theory the integrations of $y$-profiles give rise to overlap integrals which play the crucial role in unitarity violation.

  A sanity check of the above considerations would be to take the $n=0$ limit and see if the well-known unitarity bound~\cite{Lee:1977eg, Dawson:1998yi} on the SM Higgs can be obtained or not. Note that in the $n=0$ limit, every coupling will reproduce the corresponding SM coupling. Let us take the example of the quartic interaction of the field $h^{n}$. In the $n=0$ limit the overlap integral in Eq.~\ref{In} will be unity and $n=p=q=r=0$ will give exactly the SM coupling which is $(-3i\frac{m_{h}^{2}}{v^{2}})$. Another example is the quartic interaction of the field $H^{n \pm}$. For $n=0$ it will reproduce the quartic coupling of the SM longitudinal field $W_{L}$ as $(-2i\frac{m_{h}^{2}}{v^{2}})$\footnote{See, the expression of $H^{n\pm}$ in the section~\ref{model_nmUED}.}. The similar thing holds for trilinear couplings too. We have checked explicitly that in the $n=0$ limit all the KK-mode couplings will reproduce the respective SM couplings and we can definitely say that in the limit $n=0$ the SM processes corresponding to the bases $Z_{L}Z_{L}$, $W_{L}^{+}W_{L}^{-}$, $Z_{L}W_{L}^{+}$, $hh$, $hW_{L}$, $hZ_{L}$ would reproduce the exact upper bound on the Higgs mass as 870 GeV~\cite{Lee:1977eg, Dawson:1998yi}.

\begin{table}[!htbp]
\begin{center}
\begin{tabular}{|c||c|c|c|c|} 
\hline 
Value of $R_{\phi}$ & \begin{tabular}[c]{@{}c@{}}\\ Value of $I^{n}$\\($n=1$)\end{tabular} & \begin{tabular}[c]{@{}c@{}}\\ Value of $I^{n}$\\ ($n=2$)\end{tabular} & \begin{tabular}[c]{@{}c@{}}\\ Value of $I^{n}$\\($n=3$)\end{tabular} & \begin{tabular}[c]{@{}c@{}}\\ Value of $I^{n}$\\($n=4$)\end{tabular}\\ \hline
 50      & 1.03  &  23.60  & 24.90  & 25.16    \\ \hline
100      & 1.02  & 47.45   &  48.78 & 49.04   \\ \hline
150      & 1.01  & 71.31   &  72.66 & 72.91  \\ \hline
200      & 1.01  & 95.18   & 96.53  & 96.79   \\ \hline
250      & 1.01  & 119.05   & 120.41  & 120.66   \\ \hline
300      & 1.01  & 142.92   & 144.28  & 144.53   \\ \hline
350      & 1.00  & 166.80   & 168.15  & 168.41   \\ \hline
400      & 1.00  & 190.67    & 192.03  & 192.28    \\ \hline
450      & 1.00  & 214.54   & 215.90  & 216.15    \\ \hline
500      & 1.00   & 238.41   & 239.77   & 240.03    \\ \hline
550      & 1.00  & 262.29   & 263.65  & 263.90   \\ \hline
600      & 1.00  & 286.16   & 287.52  & 287.77   \\ \hline
650      & 1.00  & 310.03   & 311.39  & 311.65    \\ \hline
700      & 1.00  & 333.91   & 335.27  & 335.52    \\ \hline
750      & 1.00   & 357.78   & 359.14  & 359.40    \\ \hline
800      & 1.00  & 381.65   & 383.01  & 383.27   \\ \hline
850      & 1.00  & 405.52   & 406.89  & 407.14    \\ \hline
900      & 1.00  & 429.40   & 430.76   & 431.02    \\ \hline
950      & 1.00  & 453.27   & 454.63  & 454.89    \\ \hline
1000     & 1.00  & 477.14   & 478.51  & 478.76    \\ \hline
\end{tabular}
\caption{Values of $I^{n}$ as a function of $R_{\phi}$ for different KK-modes.}
\label{table2}
\end{center}
\end{table}


\subsection{Coupled Channel Analysis}
\label{coupled_chnl}
In the previous section, we have shown a detailed analysis of unitarity violation from $n, n \rightarrow n, n$ channels. Coupled channel analysis should be taken into account to get more stringent constraint on the BLT parameter $R_{\phi}$. This analysis includes the construction of $t^{0}$ matrix generated by different two-body channels as rows and columns. Restoration of unitarity leads to the restrictions on each of the eigenvalues of this $t^{0}$ matrix to lie below $8 \pi$ (Eq.~\ref{unitarity_constraint1}). From the previous discussion of scattering amplitudes of $n, n \rightarrow n, n$ channels, we can construct the matrix from the processes given in Fig.~\ref{figa0}, as only for those processes the unitarity violation takes place at relatively much lower values of $R_{\phi}$. It is also convenient to analyze the formation of matrix with these processes as they are $s$ independent, i.e., contributions coming from trilinear couplings can safely be ignored. We will first start coupled channel unitarity analysis considered KK-mode up to $4$th level.

 Fig.~\ref{figa0} shows that there can be neutral two-particle states and charged two-particle states in case of $t^{0}$ construction. We first consider construction of $t^{0}$ matrix from  neutral two-particle states. This $t^{0}$ will be a $70 \times 70$ matrix in neutral scenario. The states are given by 

\begin{flalign*}
&\left \lbrace \frac{h^{0}h^{0}}{\sqrt{2}}, \frac{h^{1}h^{1}}{\sqrt{2}}, \frac{h^{2}h^{2}}{\sqrt{2}}, \frac{h^{3}h^{3}}{\sqrt{2}}, \frac{h^{4}h^{4}}{\sqrt{2}}, h^{0}h^{1}, h^{0}h^{2}, h^{0}h^{3}, h^{0}h^{4}, h^{1}h^{2}, h^{1}h^{3}, h^{1}h^{4}, h^{2}h^{3}, h^{2}h^{4}, h^{3}h^{4}, \frac{A^{1}A^{1}}{\sqrt{2}},  \frac{A^{2}A^{2}}{\sqrt{2}},  \right. & \\
&\left.   \frac{A^{3}A^{3}}{\sqrt{2}}, \frac{A^{4}A^{4}}{\sqrt{2}}, A^{1}A^{2}, A^{1}A^{3}, A^{1}A^{4}, A^{2}A^{3}, A^{2}A^{4}, A^{3}A^{4}, \phi^{+}\phi^{-}, H^{1+}H^{1-},  H^{2+}H^{2-},  H^{3+}H^{3-},  H^{4+}H^{4-},\right. & \\ 
&\left.  \phi^{+}H^{1-}, \phi^{-}H^{1+}, \phi^{+}H^{2-}, \phi^{-}H^{2+}, \phi^{+}H^{3-}, \phi^{-}H^{3+}, \phi^{+}H^{4-}, \phi^{-}H^{4+}, H^{1+}H^{2-}, H^{2+}H^{1-}, H^{1+}H^{3-}, \right. & \\ 
&\left. H^{3+}H^{1-}, H^{1+}H^{4-}, H^{4+}H^{1-}, H^{2+}H^{3-}, H^{3+}H^{2-}, H^{2+}H^{4-}, H^{4+}H^{2-}, H^{3+}H^{4-}, H^{4+}H^{3-} \right \rbrace ,&
\end{flalign*}
and,
\begin{flalign*}
&\left \lbrace h^{0}A^{1}, h^{0}A^{2}, h^{0}A^{3}, h^{0}A^{4}, h^{1}A^{1}, h^{1}A^{2}, h^{1}A^{3}, h^{1}A^{4}, h^{2}A^{1}, h^{2}A^{2}, h^{2}A^{3}, h^{2}A^{4}, h^{3}A^{1}, h^{3}A^{2}, h^{3}A^{3}, h^{3}A^{4},  h^{4}A^{1}, \right. & \\
&\left. h^{4}A^{2}, h^{4}A^{3}, h^{4}A^{4}  \right \rbrace .&
\end{flalign*}

Due to CP conservation (which has been explained previously) the $70 \times 70$ matrix will have $50 \times 50$ and $20 \times 20$ block diagonal form. So the eigenvalues of these matrices can be separately analyzed as functions of BLT parameters. The $50 \times 50$ charge neutral matrix can be written as 
\begin{equation}
~~~~~~~~~~~~~~~~~~~~~~~~~~\mathcal{M}^{(1)}_{NC, 50\times 50 }= \begin{pmatrix}
\mathcal{A}_{15\times 15} & \mathcal{B}_{15\times 10} & \mathcal{C}_{15\times 25} \\ \mathcal{B}^{T}_{10\times 15} & \mathcal{D}_{10\times 10} & \mathcal{E}_{10\times 25}  \\ \mathcal{C}^{T}_{25\times 15}  & \mathcal{E}^{T}_{25\times 10}  &  \mathcal{F}_{25\times 25}    
\end{pmatrix} \,,
\label{mat}
\end{equation} 
where,  $\mathcal{M}^{(1)}_{NC, 50\times 50 }$ matrix can have eigenvalues $\lambda 1_{la}~(la=1,\ldots,50)$. Other charge neutral matrix $\mathcal{M}^{(2)}_{NC, 20\times 20 }$ has eigenvalues $\lambda 2_{lb}~(lb=1,\ldots,20)$. General form of matrix elements are given in APPENDIX E.

 In a similar manner, with $h^{n}$ being CP-even whereas $A^{n}$ being CP-odd, charged two-particle states lead to $45 \times 45$ matrix having $20 \times 20$ and $25 \times 25$ block diagonal form as
\begin{equation}
~~~~~~~~~~~~~~~~~~~~~~~~~~~~~\mathcal{M}_{CC, 45\times 45 }= \begin{pmatrix}
\mathcal{G}_{20\times 20} & \mathbf{0}_{20\times 25} \\ \mathbf{0}_{25\times 20} & \mathcal{H}_{25\times 25}
\end{pmatrix} \,.
\label{matt}
\end{equation}
Here, $\mathcal{G}_{20\times 20}$ and $\mathcal{H}_{25\times 25}$ have eigenvalues denoted by $\lambda 3_{lb}~(lb=1,\ldots,20)$ and $\lambda 4_{lc}~(lc=1,\ldots,25)$ respectively. The set of two charge two-particle states are given as
\begin{flalign*}
&\left \lbrace \phi^{+}A^{1}, \phi^{+}A^{2}, \phi^{+}A^{3}, \phi^{+}A^{4}, H^{1+}A^{1}, H^{1+}A^{2}, H^{1+}A^{3}, H^{1+}A^{4},  H^{2+}A^{1}, H^{2+}A^{2}, H^{2+}A^{3}, H^{2+}A^{4}, \right. & \\
&\left.   H^{3+}A^{1}, H^{3+}A^{2}, H^{3+}A^{3}, H^{3+}A^{4}, H^{4+}A^{1}, H^{4+}A^{2}, H^{4+}A^{3}, H^{4+}A^{4} \right \rbrace ,&
\end{flalign*}
and,
\begin{flalign*}
&\left \lbrace \phi^{+}h^{0}, \phi^{+}h^{1}, \phi^{+}h^{2}, \phi^{+}h^{3}, \phi^{+}h^{4}, H^{1+}h^{0}, H^{1+}h^{1}, H^{1+}h^{2}, H^{1+}h^{3}, H^{1+}h^{4},  H^{2+}h^{0}, H^{2+}h^{1}, H^{2+}h^{2}, \right. & \\
&\left. H^{2+}h^{3}, H^{2+}h^{4}, H^{3+}h^{0}, H^{3+}h^{1}, H^{3+}h^{2}, H^{3+}h^{3}, H^{3+}h^{4}, H^{4+}h^{0}, H^{4+}h^{1}, H^{4+}h^{2}, H^{4+}h^{3}, H^{4+}h^{4} \right \rbrace .&
\end{flalign*}
 Note that due to CP-conservation, there will be no mutual interactions among the above two bases and the corresponding couplings will eventually be zero. 

\begin{figure}[H]

 \centering
 \subfigure[]{
   \includegraphics[scale=0.4]{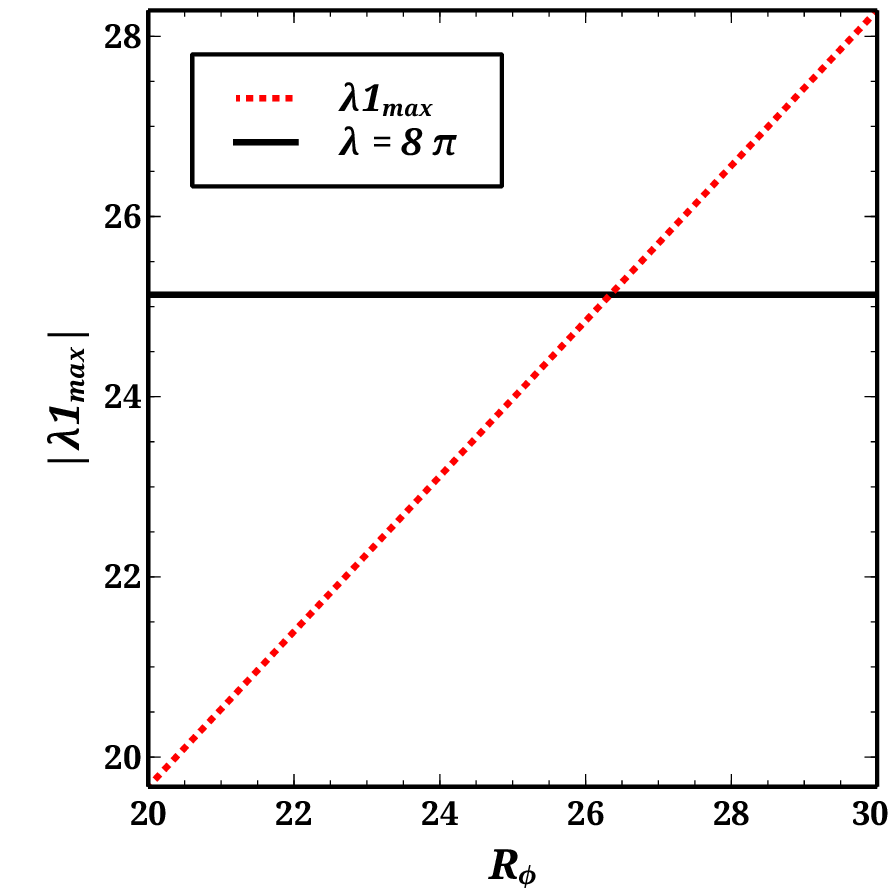}    
    }
    \subfigure[]{
   \includegraphics[scale=0.4]{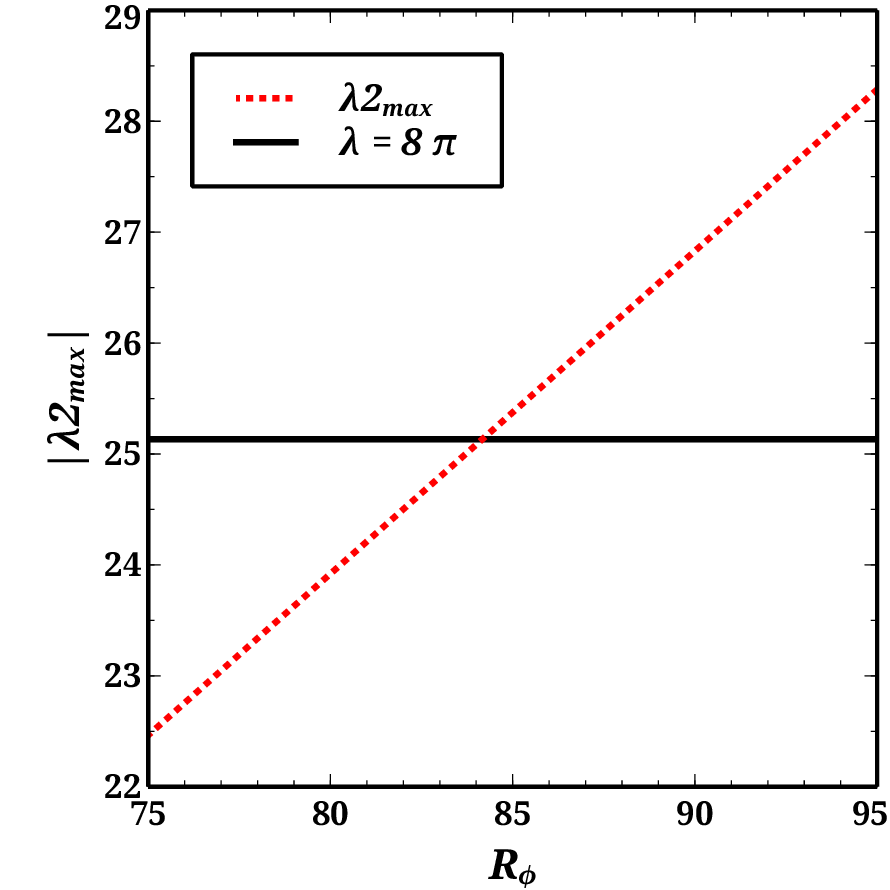}
   } \\
  \subfigure[]{
   \includegraphics[scale=0.4]{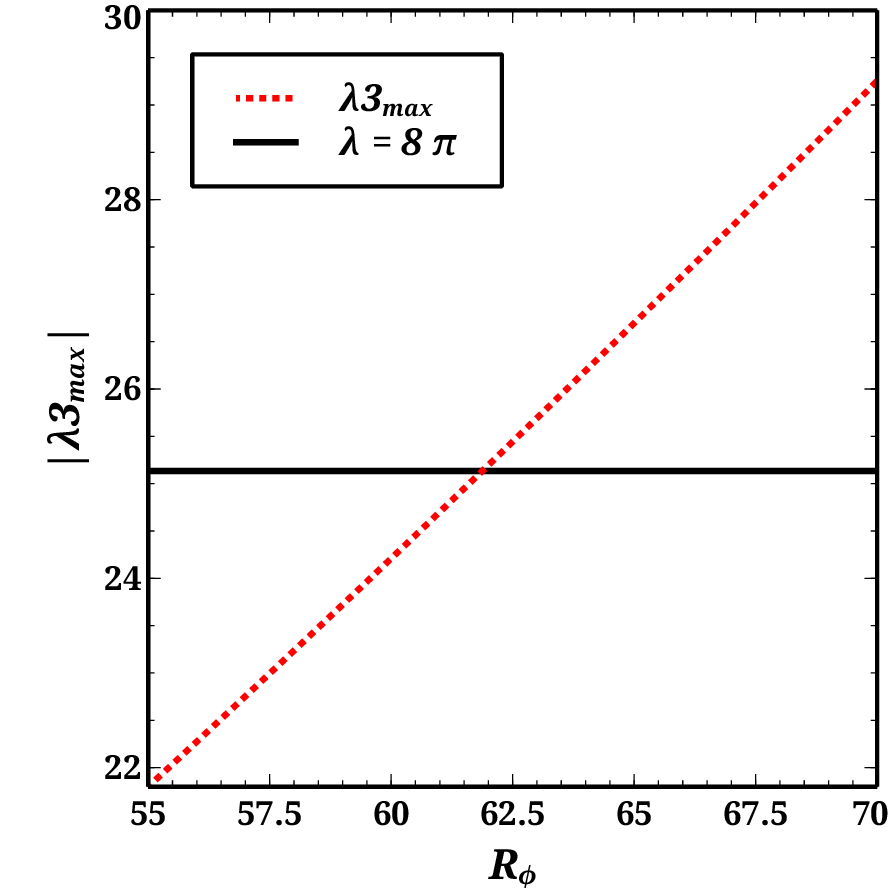}
   } 
  \subfigure[]{
   \includegraphics[scale=0.4]{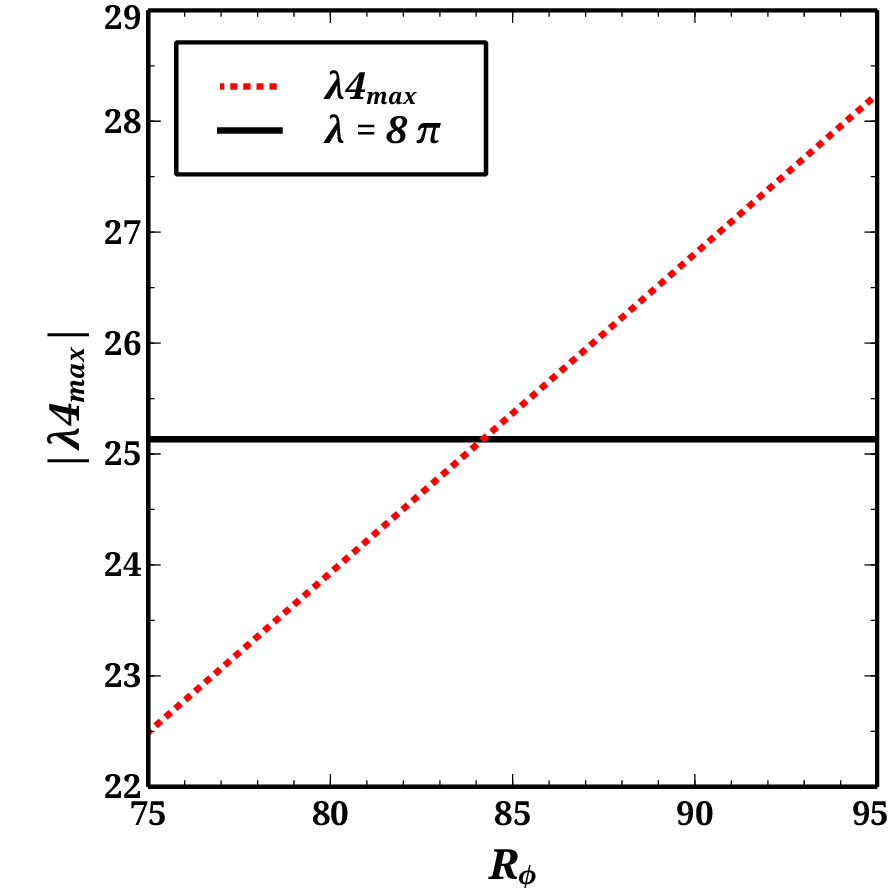}
    } 

    \caption[]{The variation of largest eigenvalues as function of $R_{\phi}$.}
    \label{lamda_max}
\end{figure}
In Fig.~\ref{lamda_max}, the variation of maximum eigenvalues corresponding to different $R_{\phi}$ have been shown. While calculating the eigenvalues we have neglected all the masses $m_{h}, M_{Z}, M_{W}$ with respect to KK-masses. The channels have negligible contributions from non-abelian Lagrangian part, since they are suppressed by KK-masses or higher power of KK-masses. We have neglected that part also. So, the result is evidently $R^{-1}$ independent (APPENDIX E). Also the Table~\ref{table1} reflects the fact, $R^{-1}$ does not play any important role in unitarity violation. So these simplifications would not affect the actual result. Now, from unitarity bound every eigenvalue of matrix should lie below $8\pi$. Consequently, the analysis of largest eigenvalue ($\lambda_{\rm max}$) from each set of 50, 20, or 25 number of eigenvalues will give the desired result. From Fig.~\ref{lamda_max}~(a), it can be seen  $\mathcal{M}^{(1)}_{NC, 50\times 50 }$ gives more stringent upper bound on $R_{\phi}$; at $R_{\phi}=26.4$ maximum value of $\lambda 1$ is greater than $8 \pi$.
The upper bound on $R_{\phi}$ implies a lower bound on KK-masses. In Refs. \cite{Datta:2012tv, Dey:2016cve}, the dependence of KK-masses as a function of scaled BLT parameters has been shown. KK-mass decreases with increase in $R_{\phi}$. Here, $R_{\phi}$ should not exceed the value $26$ implies a lower bound on KK-masses which for scalars and gauge field is given by $0.22~R^{-1}$, $1.05~R^{-1}$, $2.02~R^{-1}$ and $3.02~R^{-1}$ for $n=1-4$ respectively. Since, the upper bound on $R_{\phi}$ is , in effect, independent of $R^{-1}$, the results on the lower bound on KK-masses are true for any $R^{-1}$.

\begin{figure}[!htbp]
\centering
 {
   \includegraphics[scale=0.6]{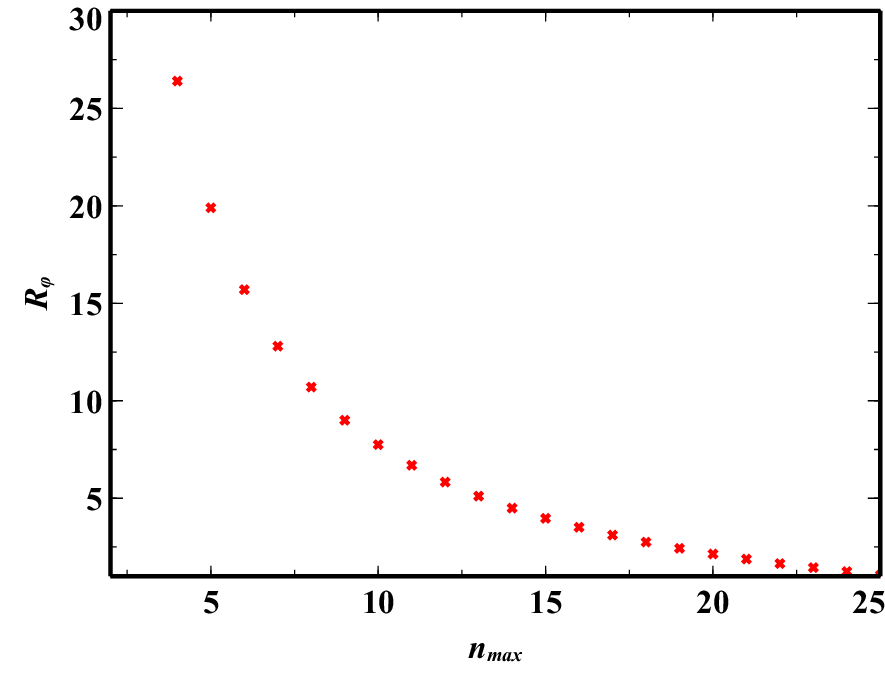}    
    }
\caption[]{The variation of $R_{\phi}$ at which the unitarity violation occurs as function of maximum KK-number considered  ($n_{\rm max}$) in the analysis.}
    \label{Rphi_vs_n}
\end{figure}

 In this case, apart from the overlap integral $I^{n}$, additionally there are other overlap integrals arising from the different combinations of KK-numbers given in Eq.~\ref{In} (e.g. $I^{nnmm}$ of Eq.~\ref{Innmm}) that result in the breakdown of unitarity. Here also, we have taken KK-modes up to 4 and the inclusion of higher KK-modes will definitely lead to higher dimensional matrices.  These higher dimensional matrices would result in the unitarity violation at relatively lower value of $R_{\phi}$. In the above analysis, since Eq.~\ref{mat} gives the most stringent upper bound on $R_{\phi}$, we will extend our analysis with higher KK-modes with its corresponding basis only. Fig.~\ref{Rphi_vs_n} shows that the upper bound on $R_{\phi}$ decreases with increasing KK-modes. If $n_{max}$ be the maximum KK-number taken, the dimension of the matrix would be $\{2(n_{max}+1)^{2} \times 2(n_{max}+1)^{2}\}$. To generate the $t^{0}$ matrices for different KK-modes and to obtain their corresponding eigenvalues numerically, we have used in-house \textsc{Mathematica} and \texttt{Python} codes. Fig.~\ref{Rphi_vs_n} shows, if $n_{max}$ is $25$ the upper bound on $R_{\phi}$ falls down to nearly one. It also exhibits the fact that KK-number from 18 onwards the values of $R_{\phi}$ resulting the breakdown of unitarity are more closely spaced. Clearly, inclusion of more higher modes, i.e.  KK-number from 26 onwards, will not change significantly the upper bound on $R_{\phi}$.

However, in Refs.~\cite{Jha:2014faa, Dey:2016cve} it has been shown explicitly that in radiative analysis higher modes from 5 or 6 onwards will not change the physical amplitude significantly. In those cases, the upper bound on $R_{\phi}$ can be taken as $19$, as for $n_{max}=5$ the violation would occur at $R_{\phi} \sim 20$. But allover, it is not possible to determine the upper bound on $n$ from unitarity analysis which can be done in some other five dimensional theories. In Ref.~\cite{SekharChivukula:2001hz}, it has been shown that for a fixed value of $R^{-1}$ one can find a lower bound on the number of KK-mode $n$ as 
\begin{flalign}
\label{bound_n}
~~~~~~~~~~~~~~~~~~~~~~~~~~~~~~~~~~~~~~~~~~~~~~~~~~~~~\frac{n}{R} \leq \frac{8 \pi^{2}}{N} \frac{1}{\tilde{g}^{2}},
\end{flalign}
for a five dimensional $SU(N)$ Yang-Mills theory. In our case, i.e., in nmUED, the scenario is somewhat more nontrivial as there are also BLKT parameters which were absent in simple universal extra dimensional theories. In that case, the normalized four-dimensional gauge-singlet $s$-wave amplitude $a_{0}(n,n \rightarrow m,m)$ of Ref.~\cite{SekharChivukula:2001hz}, will be modified by some overlap integrals as
\begin{flalign}
\label{T00}
~~~~~~~~~~~~~~~~~~~~~~~~~~~~~~~~~~~~~~~~~~~~~a_{0}(n,n \rightarrow m,m)= I^{nnmm}\frac{n}{R}\frac{N \tilde{g}^{2}}{16 \pi^{2}},
\end{flalign}
where, $I^{nnmm}$ is the overlap integral of Eq.~\ref{In}. Consequently, Eq.~\ref{bound_n} will be modified as
\begin{flalign}
\label{bound1_n}
~~~~~~~~~~~~~~~~~~~~~~~~~~~~~~~~~~~~~~~~~~~~~~~~~~~~~\frac{n}{R} \leq  \frac{1}{I^{nnmm}}\frac{8 \pi^{2}}{N\tilde{g}^{2}}.
\end{flalign}
The expression of overlap integral $I^{nnmm}$ is given as
\begin{eqnarray}
\label{Innmm}
I^{nnmm}&=&\frac{1}{\left(1 + \frac{(R_{\phi}m_{\Phi n})^{2}}{4} + \frac{R_{\phi}}{\pi}\right)\left(1 + \frac{(R_{\phi}m_{\Phi m})^{2}}{4} + \frac{R_{\phi}}{\pi}\right)}\left\lbrace 1 + \frac{2 R_{\phi}}{\pi} + \frac{R_{\phi}^{2}}{\pi^{2}}+ \frac{1}{4} (R_{\phi}m_{\Phi n})^{2} + \frac{1}{4} (R_{\phi}m_{\Phi m})^{2}\right. \nonumber \\
& & \left.+ \frac{1}{16}(R_{\phi}^{2}m_{\Phi n}m_{\Phi m})^{2} - \frac{R_{\phi}^{2}}{4\pi^{2}} (R_{\phi}m_{\Phi n})^{2} - \frac{R_{\phi}^{2}}{4\pi^{2}} (R_{\phi}m_{\Phi m})^{2}+ \frac{R_{\phi}}{16\pi}(R_{\phi}^{2}m_{\Phi n}m_{\Phi m})^{2}\right\rbrace.
\end{eqnarray}
If instead of $a_{0}(n,n \rightarrow m,m)$ in Eq.~\ref{T00}, we consider $a_{0}(n,n \rightarrow n,n)$, the overlap integral will be $I^{n}$ instead of $I^{nnmm}$ and given by
 \begin{eqnarray}
I^{n}&=&\frac{3}{\left(1 + \frac{(R_{\phi}m_{\Phi n})^{2}}{4} + \frac{R_{\phi}}{\pi}\right)^{2}}\left\lbrace \frac{1}{2}+ \frac{R_{\phi}}{\pi} + \frac{R_{\phi}^{2}}{2\pi^{2}}+ \frac{1}{4} (R_{\phi}m_{\Phi n})^{2} + \frac{R_{\phi}}{8\pi} (R_{\phi}m_{\Phi n})^{2} \right. \nonumber \\
& & \left. - \frac{1}{8\pi^{2}} (R_{\phi}^{2}m_{\Phi n})^{2} +\frac{1}{32} (R_{\phi}m_{\Phi n})^{4} + \frac{R_{\phi}}{32\pi} (R_{\phi}m_{\Phi n})^{4}\right\rbrace.
\end{eqnarray}
Clearly, the overlap integrals are not directly proportional to $R_{\phi}$. However, these overlap integrals are explicit function of $R_{\phi}$ as well as of $m_{\Phi n} \equiv M_{\Phi n} R$. On the other hand, $m_{\Phi n}$ has an implicit dependence on $n$. Overall, there would exist one possibility to find out the bound on $n$ through unitarity analysis that at some $n_{max}$ unitarity violation would occur at every possible value of $R_{\phi}$ ($R_{\phi} > (-\pi)$, as evident from Eq.~\ref{g5_g4}). Though the Fig.~\ref{Rphi_vs_n} shows that inclusion of higher KK-modes would result in unitarity violation at much lower value of $R_{\phi}$, it also reflects that from KK-number 18 onwards the values of $R_{\phi}$ at which the unitarity violation occurs are more closely spaced. Even the difference between the values of $R_{\phi}$ which violate the unitarity at KK-number  24 and 25 is less than 0.2. Therefore, after KK-number 25 the result will not change considerably.

\begin{figure}[!htbp]  
\centering
 {
   \includegraphics[scale=0.5]{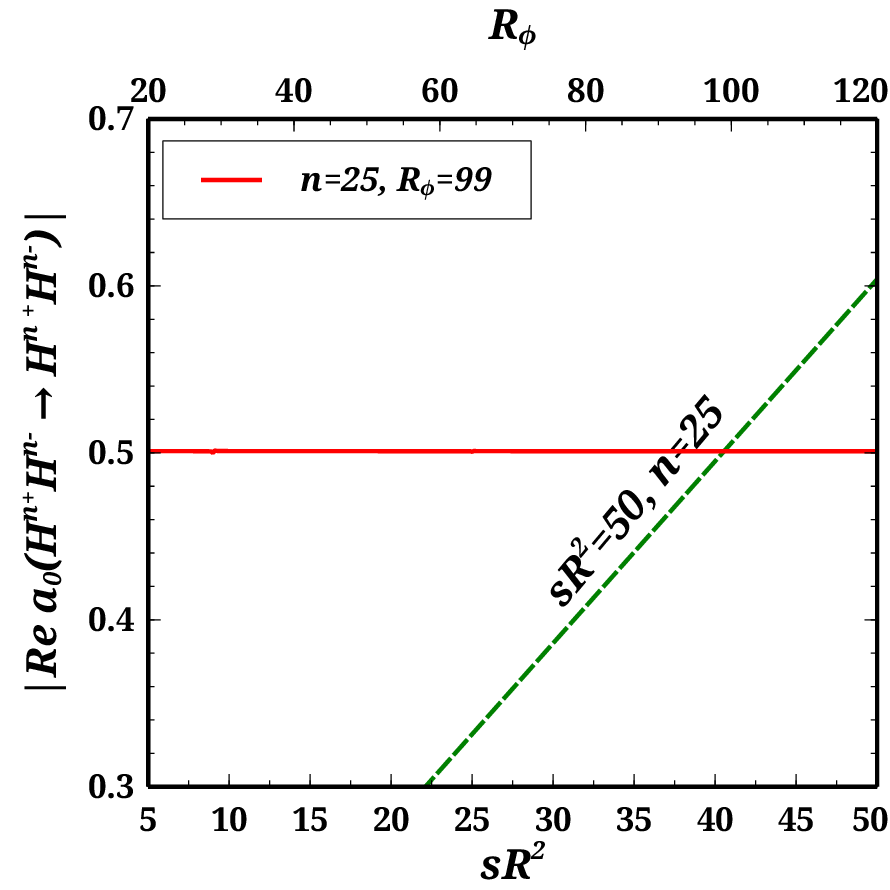}    
    }
\caption[]{Variation of $a_{0}$ as a function of $sR^{2}$ for KK-mode $n=25$, with $R_{\phi}=99$, and also as a function of $R_{\phi}$ for the same KK-mode with $sR^{2}=50$ (for the process $H^{n+}H^{n-} \rightarrow H^{n+}H^{n-}$). There are two horizontal axes in each plot. The lowest one corresponds to $sR^{2}$ for different values of $R_{\phi}$ and the upper one gives the variation of $a_{0}$ as a function of $R_{\phi}$ for fixed value of $sR^{2}$. Here $R^{-1}$ is taken to be 5000 GeV.}
    \label{n25}
\end{figure}

In passing, we would like to comment on the justification of ignoring the effect of trilinear couplings for higher KK-modes. 
The basis considered in the results shown in Fig.~\ref{Rphi_vs_n}, are independent of trilinear contributions. To put it in another way, for higher KK-modes (e.g. for $n=25$), the results are independent of $s$. For example, the Fig.~\ref{n25} shows the variation of $|{\rm Re}~a_{0}|$ for the process $H^{n+}H^{n-} \rightarrow H^{n+}H^{n-}$ as function of $sR^{2}$, for $n=25$. This clearly shows that the results for higher KK-modes are $s$ independent. It has been mentioned in the previous sections that we consider only those processes whose quartic couplings are not suppressed by the KK-masses. For example, in Sec.~\ref{nn_nn}, we have not neglected KK-masses anywhere in the entire single channel analysis. Similarly, in Sec.~\ref{coupled_chnl}, we neglect only the masses $m_{h}, M_{W}$ with respect to KK-masses, and since we have considered the quartic couplings independent of KK-masses, or the power of KK-masses in the  numerator is exactly same as that in the denominator, neglecting $m_{h}$ and $M_{W}$ eventually leads to quartic interactions independent of KK-masses.

In Fig.~\ref{n25}, $a_{0}$ is also considered as a function of $R_{\phi}$ for $sR^{2}=50$ and for KK-mode $n=25$; variation of $|{\rm Re}~a_{0}|$ is also a straight line and will be greater than half at the same value of $R_{\phi}$ at which the violation occurs with the variation of $sR^{2}$ even at large $s$ limit. This clearly justifies that the contributions of trilinear couplings for the basis considered in Fig.~\ref{Rphi_vs_n}, can be ignored for higher KK-modes too. 



 Besides, the bound on $R_{\phi}$ obtained from unitarity analysis yields the values of quartic coupling lower than the value $4\pi$. For example, at $R_{\phi}=15$, the value of the quartic coupling of $h^{n}$ will be $\sim$ 0.85, 5.43 and 6.71 for $n=1, 2~{\rm and}~25$ respectively; for $R_{\phi}=26$, the corresponding values will be $\sim$ 0.82, 9.44 and 10.77. So the results are consistent with the perturbativity limit of the theory.

%
%
%
%
%
%
%
%
%

%
%
%
%
%
%
%
%
%
%
\section{Summary and Conclusions}
We have done simple partial wave unitarity analysis as well as coupled channel analysis in gauge and scalar sectors in non-minimal Universal Extra Dimensional model where kinetic terms involving fields as well as mass and potential terms of the scalar fields are added to their respective five dimensional actions at the fixed boundary points. By virtue of equivalence theorem, we have used all the Goldstone modes instead of the longitudinal modes of vector bosons. First, all the necessary two-body tree level scattering amplitudes have been calculated to study the upper bound on scalar BLT parameter by the simple method of partial wave analysis. After that coupled channel analysis has been performed for some selective channels to get further constraint.

 All scattering amplitudes can be expressed in terms of an infinite sum of partial waves. For a process to maintain unitarity the zeroth partial wave amplitude $a_{0}$ should respect the condition $|{\rm Re}~a_{0}|\leq \frac{1}{2}$. Initially the construction of $t$-matrix has not been considered as in many cases, the contributions coming from trilinear coupling can not be ignored where the interaction is effectively proportional to KK-masses and hence the contribution is not $E^{2}$-suppressed. Therefore, some entries of the matrix elements are not simple numbers but are also the functions of centre of mass energy $\sqrt{s}$ of respective processes. So initially we have considered the single channel analysis for $n, n \rightarrow n, n$ processes to get the suitable channels for $t^{0}$ construction.

 While dealing with single channel scattering processes we have restricted our calculations to two-body scattering processes for specific KK-modes, i.e. KK-numbers for all initial and final particles are the same. We have considered only those processes whose quartic interactions are not suppressed by KK-masses. Thus we have altogether thirteen quartic interactions in scalar sectors satisfying the conditions. Quartic interactions where there are two same neutral particles in initial states and another same two neutral particles in final states, can generate two kinds of processes. Also the processes involving two same or different charged particles in initial states and another kind of same two neutral particles in final states can give rise to two kinds of processes. In that case only those processes have been preferred where the amplitudes are not suppressed by the factor of $1/2$ or by $1/\sqrt{2}$ arising from normalization factors due to the identical bosonic states, as that suppression would result in unitarity violation at some larger value of $R_{\phi}$ and thus giving a relatively less tight bound. Here, $R_{\phi}$ is the scaled scalar boundary-localized parameter given by $\equiv \frac{r_{\phi}}{R}$.

 Among all thirteen processes, $H^{n+}H^{n-} \rightarrow H^{n+}H^{n-}$ gives the most stringent constraint on the bound on $R_{\phi}$. A detailed analysis on $n,n \rightarrow n,n$ shows that the channels involving the processes $h^{n}h^{n} \rightarrow h^{n}h^{n}$, $A^{n}A^{n} \rightarrow A^{n}A^{n}$, $H^{n+}H^{n-} \rightarrow H^{n+}H^{n-}$, $h^{n}A^{n} \rightarrow h^{n}A^{n}$, $H^{n\pm}A^{n} \rightarrow H^{n\pm}A^{n}$, $H^{n\pm}h^{n} \rightarrow H^{n\pm}h^{n}$ are preferable for coupled channel analysis, i.e., construction of $t^{0}$ matrix generated by two-body states as rows and columns. Consequently each matrix element refers to the amplitude of respective processes. Each eigenvalue of the matrices should lie below $8\pi$. Coupled channel analysis leads to upper bound on BLT parameter $R_{\phi}$, the value of $R_{\phi}$ should be less than $26.4$ for complete restoration of unitarity if KK mode up to four is considered. The results are independent of $R^{-1}$. Since KK-mass decreases with increasing $R_{\phi}$, the upper bound on $R_{\phi}$ results in a lower bound on the masses of scalar and gauge fields. For example, the upper bound on $R_{\phi}$ as $26$ implies a lower bound on gauge or scalar KK-masses as $0.22~R^{-1}$, $1.05~R^{-1}$, $2.02~R^{-1}$ and $3.02~R^{-1}$ for $n=1-4$ respectively. The upper bound on $R_{\phi}$ does not depend on $R^{-1}$, so the results on lower bound on KK-masses are true for any value of $R^{-1}$. Besides, the upper bound on $R_{\phi}$ decreases with increasing KK-modes. For $n=25$, the upper bound on $R_{\phi}$ falls down to nearly one.  From KK mode 18 onwards, the constraints on $R_{\phi}$ change very slowly and will not change significantly for KK-modes higher than 25.   

 In nmUED, the boundary terms are generated due to radiative corrections. So those terms are loop suppressed. The coefficients of boundary terms are the free parameters of the theory which we call the BLT parameters. Though the terms are loop suppressed and should be small, but we do not know the effective range of BLT parameters, or how do they behave in the four dimensional effective theory. Unitarity analysis in gauge and scalar sectors will give the range of BLT parameters which maintains perturbativity of the model. We have performed this analysis in gauge and scalar sectors exploiting the equivalence theorem to get the upper bound on gauge and scalar BLT parameters (which are same in our analysis), and have found that the scaled scalar or gauge BLT parameters signaling the breakdown of unitarity decrease with increasing higher modes and saturated to nearly unity with KK-number 25. We hope to return to this issue in fermion sectors and to study the perturbativity in future work.
\section*{Acknowledgements :} TJ thanks Anindya Datta for many useful comments and advice at different stages of the work. TJ also thanks Ayon Patra for collaborating in the initial stages of the work and Ujjal Kumar Dey for many useful discussions. TJ acknowledges the financial support from CSIR in terms of an SRF.

\section{APPENDIX}
\begin{center}
{\bf \large APPENDIX A}
\end{center}
\section*{Expressions for $a_{0}$ :}
\setlength{\mathindent}{0pt}
\begin{flalign}
&a_{0}(h^{n}h^{n} \rightarrow h^{n}h^{n}) = - \frac{3}{16\pi}\frac{m_{h}^{2}}{v^{2}}\left[I^{n}+3m_{h}^{2}\left \lbrace\frac{1}{s-m_{h}^{2}}-\frac{2}{s-4m_{hn}^{2}}\ln\left(\frac{s-3m_{h}^{2}-4M_{\Phi n}^{2}}{m_{h}^{2}}\right)\right \rbrace \right. \nonumber \\
&\left.~~~~~~~~~~~~~~~~~~~~~~~~~~+3m_{h}^{2}\sum\limits_{q={\rm even}}^{\infty}{I^{nnq}}^{2}\left \lbrace\frac{1}{s-m_{hq}^{2}}-\frac{2}{s-4m_{hn}^{2}}\ln\left(\frac{s-3m_{h}^{2}-4M_{\Phi n}^{2}+M_{\Phi q}^{2}}{m_{hq}^{2}}\right)\right \rbrace\right],&\\
&a_{0}(A^{n}A^{n} \rightarrow A^{n}A^{n})  = - \frac{1}{16\pi}\frac{1}{v^{2} M_{Zn}^{4}}\left[3(m_{h}^{2}M_{\Phi n}^{4}I^{n}+4M_{Z}^{4}M_{\Phi n}^{2}I^{\prime n})+(m_{h}^{2}M_{\Phi n}^{2}+2M_{Z}^{2}M_{Zn}^{2})^{2}\left \lbrace\frac{1}{s-m_{h}^{2}}\right.\right. \nonumber \\
&\left.~~~~~~~~~~~~~~~~~~~~~~~~~~~\left.-\frac{2}{s-4M_{Zn}^{2}}\ln\left(\frac{s-4M_{Zn}^{2}+m_{h}^{2}}{m_{h}^{2}}\right)\right \rbrace +\sum\limits_{q={\rm even}}^{\infty}c1_{nnq}^{2}\left \lbrace\frac{1}{s-m_{hq}^{2}}\right.\right. \nonumber \\
&\left.~~~~~~~~~~~~~~~~~~~~~~~~~~~\left.-\frac{2}{s-4M_{Zn}^{2}}\ln\left(\frac{s-4M_{Zn}^{2}+m_{hq}^{2}}{m_{hq}^{2}}\right)\right \rbrace\right], &\\
&a_{0}(H^{n+}H^{n-} \rightarrow H^{n+}H^{n-})  = - \frac{1}{16\pi}\frac{1}{v^{2} M_{Wn}^{4}}\left[2(m_{h}^{2}M_{\Phi n}^{4}I^{n}+4M_{W}^{4}M_{\Phi n}^{2}I^{\prime n})+(m_{h}^{2}M_{\Phi n}^{2}+2M_{W}^{2}M_{Wn}^{2})^{2}\right. \nonumber \\
&\left.~~~~~~~~~~~~~~~~~~~~~~~~~~~~~~~~~~~~~\times \left \lbrace \frac{1}{s-m_{h}^{2}}-\frac{2}{s-4M_{Wn}^{2}}\ln\left(\frac{s-4M_{Wn}^{2}+m_{h}^{2}}{m_{h}^{2}}\right)\right \rbrace \right. \nonumber \\
&\left.~~~~~~~~~~~~~~~~~~~~~~~~~~~~~~~~~~~~~+ \sum _{q={\rm even}}^{\infty}c2_{nnq}^{2}\left \lbrace \frac{1}{s-m_{hq}^{2}}-\frac{2}{s-4M_{Wn}^{2}}\ln\left(\frac{s-4M_{Wn}^{2}+m_{hq}^{2}}{m_{hq}^{2}}\right)\right \rbrace \right], &\\
&a_{0}(G_{Z}^{n}A^{n}\rightarrow G_{Z}^{n}A^{n}) = - \frac{1}{16\pi}\frac{M_{Z}^{2}}{v^{2} M_{Zn}^{4}}\left[q1_{n}+M_{\Phi n}^{2}\left(m_{h}^{2}-M_{Zn}^{2}\right)^{2}\right. \nonumber \\
&\left.~~~~~~~~~~~~~~~~~~~~~~~~~~~~~~~\times\left \lbrace \frac{1}{s-m_{h}^{2}}-\frac{1}{\left(s-4M_{Zn}^{2}\right)}\ln\left(\frac{s-4M_{Zn}^{2}+m_{h}^{2}}{m_{h}^{2}}\right)\right \rbrace \right. \nonumber \\
&\left.~~~~~~~~~~~~~~~~~~~~~~~~~~~~-\frac{m_{h}^{2}(m_{h}^{2}M_{\Phi n}^{2}+2M_{Z}^{2}M_{Zn}^{2})}{\left(s-4M_{Zn}^{2}\right)}\ln\left(\frac{s-4M_{Zn}^{2}+m_{h}^{2}}{m_{h}^{2}}\right) \right. \nonumber \\
&\left.~~~~~~~~~~~~~~~~~~~~~~~~~~~~+ \sum_{q={\rm even}}^{\infty}c7_{nnq}^{2}\left \lbrace \frac{1}{s-m_{hq}^{2}}-\frac{1}{\left(s-4M_{Zn}^{2}\right)}\ln\left(\frac{s-4M_{Zn}^{2}+m_{hq}^{2}}{m_{hq}^{2}}\right)\right \rbrace \right. \nonumber \\
&\left.~~~~~~~~~~~~~~~~~~~~~~~~~~~~-\sum_{q={\rm even}}^{\infty}\left(m_{h}^{2}I^{nnq} + 2 M_{\Phi n}M_{\Phi q}I^{\prime qnn}\right)\frac{c1_{nnq}}{\left(s-4M_{Zn}^{2}\right)}\ln\left(\frac{s-4M_{Zn}^{2}+m_{hq}^{2}}{m_{hq}^{2}}\right)\right],&\\
&a_{0}(h^{n}A^{n}\rightarrow h^{n}A^{n})  = - \frac{1}{16\pi}\frac{1}{v^{2} M_{Zn}^{2}}\left[\left(m_{h}^{2}M_{\Phi n}^{2}I^{n}+2 M_{Z}^{4}I^{\prime n}\right)+M_{\Phi n}^{2}\left(m_{h}^{2}-M_{Z}^{2}\right)^{2}\left \lbrace \frac{1}{s-M_{Z}^{2}}\right.\right. \nonumber \\
&\left.~~~~~~~~~~~~~~~~~~~~~~~~~~\left.-\frac{s}{(s-m_{hn}^{2}-M_{Zn}^{2})^{2}}\ln X2_{n}\right \rbrace -3m_{h}^{2}s\frac{\left(m_{h}^{2}M_{\Phi n}^{2}+2 M_{Z}^{2}M_{Zn}^{2}\right)}{(s-m_{hn}^{2}-M_{Zn}^{2})^{2}}\ln X3{n}\right. \nonumber \\
&\left.~~~~~~~~~~~~~~~~~~~~~~~~~~ + \sum _{q={\rm even}}^{\infty}\frac{1}{M_{Zq}^{2}}\{M_{Z}^{2}c3_{nnq}^{2}+c4_{nnq}^{2}\}\left \lbrace \frac{1}{s-M_{Zq}^{2}}-\frac{s}{(s-m_{hn}^{2}-M_{Zn}^{2})^{2}}\ln Y2_{nq}\right \rbrace \right. \nonumber \\
&\left.~~~~~~~~~~~~~~~~~~~~~~~~~~ -\sum _{q={\rm even}}^{\infty}\frac{3m_{h}^{2}I^{nnq}c1_{nnq}s}{(s-m_{hn}^{2}-M_{Zn}^{2})^{2}}\ln Y3{nq}\right],&\\
&a_{0}(h^{n}G_{Z}^{n}\rightarrow h^{n}G_{Z}^{n})  = - \frac{1}{16\pi}\frac{M_{Z}^{2}}{v^{2} M_{Zn}^{2}}\left[\left(m_{h}^{2}I^{n}+2 M_{\Phi n}^{2}I^{\prime n}\right)+m_{hn}^{4}\left \lbrace \frac{1}{s-M_{Z}^{2}}\right.\right. \nonumber \\
&\left.~~~~~~~~~~~~~~~~~~~~~~~~~~~~~\left.-\frac{s}{(s-m_{hn}^{2}-M_{Zn}^{2})^{2}}\ln X2_{n}\right \rbrace-\frac{3m_{h}^{4}s}{(s-m_{hn}^{2}-M_{Zn}^{2})^{2}}\ln X3_{n}\right. \nonumber \\
&\left.~~~~~~~~~~~~~~~~~~~~~~~~~~~~~ +\sum _{q={\rm even}}^{\infty}\frac{1}{M_{Zq}^{2}}\{M_{Z}^{2}c5_{nnq}^{2}+c6_{nnq}^{2}\}\left \lbrace \frac{1}{s-M_{Zq}^{2}}-\frac{s}{(s-m_{hn}^{2}-M_{Zn}^{2})^{2}}\ln Y2_{nq}\right \rbrace \right. \nonumber \\
&\left.~~~~~~~~~~~~~~~~~~~~~~~~~~~~~-\sum _{q={\rm even}}^{\infty}\left(m_{h}^{2}I^{nnq}+ 2 M_{\Phi n}M_{\Phi q}I^{\prime qnn}\right)\frac{3m_{h}^{2}I^{nnq}s}{(s-m_{hn}^{2}-M_{Zn}^{2})^{2}}\ln Y3_{nq}\right],&\\
&a_{0}(h^{n}G_{Z}^{n}\rightarrow H^{n \pm}G^{n \mp})  = - \frac{1}{16\pi}\frac{M_{Z} M_{W}}{v^{2} M_{Zn}}\left[M_{\Phi n}(1-\cos2\theta_{W})I^{\prime n}+\frac{M_{\Phi n} m_{h}^{2}}{(s-M_{Z}^{2})}\right. \nonumber \\
&\left.~~~~~~~~~~~~~~~~~~~~~~~~~~~~~~~-\frac{2M_{\Phi n}m_{hn}^{2}\cos2\theta_{W}\sqrt{s}}{\sqrt{s-4M_{Wn}^{2}}(s-m_{hn}^{2}-M_{Zn}^{2})}\ln X4_{n}+\sum _{q={\rm even}}^{\infty}\frac{(-M_{Z}^{2}c10_{nnq} c5_{nnq}+c9_{nnq} c6_{nnq})}{M_{Zq}^{2}(s-M_{Zq}^{2})} \right. \nonumber \\
&\left.~~~~~~~~~~~~~~~~~~~~~~~~~~~~~~~+ \sum _{q={\rm even}}^{\infty}\frac{2\sqrt{s}}{{\sqrt{s-4M_{Wn}^{2}}(s-m_{hn}^{2}-M_{Zn}^{2})}}\right. \nonumber \\
&\left.~~~~~~~~~~~~~~~~~~~~~~~~~~~~~~~\times\left \lbrace \frac{M_{\Phi n}M_{\Phi q}}{M_{Wn}^{2} M_{Wq}}I^{\prime qnn}c8_{nnq}-\frac{c11_{nnq}c12_{nnq}}{M_{Wn}^{2} M_{Wq}^{2}}-\frac{M_{W}^{2}}{M_{Wq}^{2}}c5_{nnq}c10_{nnq}\right \rbrace \ln Y4_{nq}\right], &\\
&a_{0}(H^{n \pm}h^{n} \rightarrow H^{n \pm}h^{n})  = - \frac{1}{16\pi}\frac{1}{v^{2} M_{Wn}^{2}}\left[\left(m_{h}^{2}M_{\Phi n}^{2}I^{n}+2 M_{W}^{4}I^{\prime n}\right)+ M_{\Phi n}^{2}\left(m_{h}^{2}-M_{W}^{2}\right)^{2}\left \lbrace \frac{1}{s-M_{W}^{2}}\right.\right. \nonumber \\
&\left.~~~~~~~~~~~~~~~~~~~~~~~~~~~~~~~\left.-\frac{s}{(s-m_{hn}^{2}-M_{Wn}^{2})^{2}}\ln X7_{n}\right \rbrace -3m_{h}^{2}\frac{\left(m_{h}^{2}M_{\Phi n}^{2}+2 M_{W}^{2}M_{Wn}^{2}\right)s}{(s-m_{hn}^{2}-M_{Wn}^{2})^{2}}\ln X8_{n}\right. \nonumber \\
&\left.~~~~~~~~~~~~~~~~~~~~~~~~~~~~~~~+\sum _{q={\rm even}}^{\infty}\frac{\left(M_{W}^{2}c13_{nnq}^{2}+c8_{nnq}^{2}\right)}{M_{Wq}^{2}}\left \lbrace\frac{1}{s-M_{Wq}^{2}}-\frac{s}{(s-m_{hn}^{2}-M_{Wn}^{2})^{2}}\ln Y7_{nq}\right \rbrace \right. \nonumber \\
&\left.~~~~~~~~~~~~~~~~~~~~~~~~~~~~~~~-\sum _{q={\rm even}}^{\infty}3m_{h}^{2}I^{nnq}\frac{c2_{nnq}s}{(s-m_{hn}^{2}-M_{Wn}^{2})^{2}}\ln Y8_{nq}\right],&\\
&a_{0}(G^{n \pm}h^{n} \rightarrow G^{n \pm}h^{n})  = - \frac{1}{16\pi}\frac{M_{W}^{2}}{v^{2} M_{Wn}^{2}}\left[\left(m_{h}^{2}I^{n}+2 M_{\Phi n}^{2} I^{\prime n}\right)+ M_{\Phi n}^{2}\left(m_{h}^{2}-M_{W}^{2}\right)^{2}\left \lbrace \frac{1}{s-M_{W}^{2}}\right.\right. \nonumber \\
&\left.~~~~~~~~~~~~~~~~~~~~~~~~~~~~~~~\left.-\frac{s}{(s-m_{hn}^{2}-M_{Wn}^{2})^{2}}\ln X7_{n}\right \rbrace-3m_{h}^{4}\frac{s}{(s-m_{hn}^{2}-M_{Wn}^{2})^{2}}\ln X8_{n}\right. \nonumber \\
&\left.~~~~~~~~~~~~~~~~~~~~~~~~~~~~~~~+\sum _{q={\rm even}}^{\infty}\frac{\left(M_{W}^{2}c5_{nnq}^{2}+c12_{nnq}^{2}\right)}{M_{Wq}^{2}}\left \lbrace \frac{1}{s-M_{Wq}^{2}}-\frac{s}{(s-m_{hn}^{2}-M_{Wn}^{2})^{2}}\ln Y7_{nq}\right \rbrace \right. \nonumber \\
&\left.~~~~~~~~~~~~~~~~~~~~~~~~~~~~~~~-\sum _{q={\rm even}}^{\infty}\left(m_{h}^{2}I^{nnq}+2M_{\Phi n}M_{\Phi q}I^{\prime qnn}\right)\frac{3m_{h}^{2}I^{nnq}s}{(s-m_{hn}^{2}-M_{Wn}^{2})^{2}}\ln Y8_{nq}\right],&\\
&a_{0}(H^{n \pm}G_{Z}^{n} \rightarrow H^{n \pm}G_{Z}^{n})  = - \frac{1}{16\pi}\frac{M_{Z}^{2}}{v^{2} M_{Zn}^{2} M_{Wn}^{2}}\left[q2_{n} + (M_{\Phi n}^{2}M_{Wn}^{4}{\cos}^{2}2\theta_{W})\left \lbrace \frac{1}{s-M_{W}^{2}}\right.\right. \nonumber \\
&\left.~~~~~~~~~~~~~~~~~~~~~~~~~~~~~~~~~\left.-\frac{s}{(s-M_{Zn}^{2}-M_{Wn}^{2})^{2}}\ln X9_{n}\right \rbrace -m_{h}^{2}\left(m_{h}^{2}M_{\Phi n}^{2}+2 M_{W}^{2}M_{Wn}^{2}\right)\right. \nonumber \\
&\left.~~~~~~~~~~~~~~~~~~~~~~~~~~~~~~~~~\times \frac{s}{(s-M_{Zn}^{2}-M_{Wn}^{2})^{2}}\ln X10_{n} + \sum _{q={\rm even}}^{\infty}\frac{1}{M_{Wq}^{2}}\left(M_{W}^{2}M_{Wn}^{4}c14_{nnq}^{2}+c11_{nnq}^{2}\right)\right. \nonumber \\
&\left.~~~~~~~~~~~~~~~~~~~~~~~~~~~~~~~\times \left \lbrace \frac{1}{s-M_{Wq}^{2}}-\frac{s}{(s-M_{Zn}^{2}-M_{Wn}^{2})^{2}}\ln Y9_{nq}\right \rbrace\right. \nonumber \\
&\left.~~~~~~~~~~~~~~~~~~~~~~~~~~~~~~~-\sum _{q={\rm even}}^{\infty}\left(m_{h}^{2}I^{nnq}+2M_{\Phi n}M_{\Phi q}I^{\prime qnn}\right)\frac{s~c2_{nnq}}{(s-M_{Zn}^{2}-M_{Wn}^{2})^{2}}\ln Y10_{nq}\right],&\\
&a_{0}(H^{n \pm} A^{n} \rightarrow H^{n \pm}A^{n})  = - \frac{1}{16\pi}\frac{M_{Z}^{2}}{v^{2} M_{Zn}^{2} M_{Wn}^{2}}\left[M_{\Phi n}^{2}q3_{n}+(M_{\Phi n}^{2}+2M_{W}^{2})(M_{Z}^{2}-M_{W}^{2})^{2}\left \lbrace \frac{1}{s-M_{W}^{2}}\right.\right. \nonumber \\
&\left.~~~~~~~~~~~~~~~~~~~~~~~~~~~~~~~~-\left.\frac{s}{(s-M_{Zn}^{2}-M_{Wn}^{2})^{2}}\ln X9_{n}\right \rbrace  \right. \nonumber \\
&\left.~~~~~~~~~~~~~~~~~~~~~~~~~~~~~~~~-\left(m_{h}^{2}M_{\Phi n}^{2}+2 M_{Z}^{2}M_{Zn}^{2}\right)\left(m_{h}^{2}M_{\Phi n}^{2}+2 M_{W}^{2}M_{Wn}^{2}\right)\frac{s}{(s-M_{Zn}^{2}-M_{Wn}^{2})^{2}}\ln X10_{n} \right. \nonumber \\
&\left.~~~~~~~~~~~~~~~~~~~~~~~~~~~~~~~~+\sum _{q={\rm even}}^{\infty}\frac{1}{M_{Wq}^{2}}\left(M_{W}^{2}c16_{nnq}^{2}+c15_{nnq}^{2}\right)\left \lbrace \frac{1}{s-M_{Wq}^{2}}-\frac{s}{(s-M_{Zn}^{2}-M_{Wn}^{2})^{2}}\ln Y9_{nq}\right \rbrace  \right. \nonumber \\
&\left.~~~~~~~~~~~~~~~~~~~~~~~~~~~~~~~~-\sum _{q={\rm even}}^{\infty}c1_{nnq}c2_{nnq}\frac{s}{(s-M_{Zn}^{2}-M_{Wn}^{2})^{2}}\ln Y10_{nq}\right],&\\
&a_{0}(G^{n \pm} A^{n} \rightarrow G^{n \pm}A^{n})  = - \frac{1}{16\pi}\frac{M_{W}^{2}}{v^{2} M_{Zn}^{2} M_{Wn}^{2}}\left[q4_{n}+(M_{\Phi n}^{2} M_{Zn}^{4})\left \lbrace \frac{1}{s-M_{W}^{2}}-\frac{s}{(s-M_{Zn}^{2}-M_{Wn}^{2})^{2}}\ln X9_{n}\right \rbrace \right. \nonumber \\
&\left.~~~~~~~~~~~~~~~~~~~~~~~~~~~~~~~-m_{h}^{2}\left(m_{h}^{2}M_{\Phi n}^{2}+2 M_{Z}^{2}M_{Zn}^{2}\right)\frac{s}{(s-M_{Zn}^{2}-M_{Wn}^{2})^{2}}\ln X10_{n}\right. \nonumber \\
&\left.~~~~~~~~~~~~~~~~~~~~~~~~~~~~~~~+\sum _{q={\rm even}}^{\infty}\frac{1}{M_{Wq}^{2}}\left(M_{W}^{2}M_{Zn}^{4}c18_{nnq}^{2}+c17_{nnq}^{2}\right)\right. \nonumber \\
&\left.~~~~~~~~~~~~~~~~~~~~~~~~~~~~~~~~\times\left \lbrace \frac{1}{s-M_{Wq}^{2}}-\frac{s}{(s-M_{Zn}^{2}-M_{Wn}^{2})^{2}}\ln Y9_{nq}\right \rbrace \right. \nonumber \\
&\left.~~~~~~~~~~~~~~~~~~~~~~~~~~~~~~~-\sum _{q={\rm even}}^{\infty}\left(m_{h}^{2}I^{nnq}+2M_{\Phi n}M_{\Phi q}I^{\prime qnn}\right)c1_{nnq}\frac{s}{(s-M_{Zn}^{2}-M_{Wn}^{2})^{2}}\ln Y10_{nq}\right],&\\
&a_{0}(A^{n} G_{Z}^{n} \rightarrow G^{n \pm}H^{n \mp})  = - \frac{1}{16\pi}\frac{M_{W}M_{Z}}{v^{2} M_{Zn}^{2} M_{Wn}^{2}}\left[q5+(m_{h}^{2}-M_{Zn}^{2})(m_{h}^{2}-M_{Wn}^{2})\frac{M_{\Phi n}^{2}}{s-m_{h}^{2}}\right. \nonumber \\
&\left.~~~~~~~~~~~~~~~~~~~~~~~~~~~~~~~~-\frac{2 M_{\Phi n}^{2}\cos2\theta_{W} M_{Zn}^{2} M_{Wn}^{2}}{\sqrt{\left(s-4M_{Zn}^{2}\right)\left(s-4M_{Wn}^{2}\right)}}\ln X6_{n}+\sum _{q={\rm even}}^{\infty}\frac{c7_{nnq} c19_{nnq}}{s-m_{hq}^{2}} \right. \nonumber \\
&\left.~~~~~~~~~~~~~~~~~~~~~~~~~~~~~~~~+ \sum _{q={\rm even}}^{\infty}\left(M_{W}^{2}M_{Wn}^{2}M_{Zn}^{2}c14_{nnq}c18_{nnq}+c17_{nnq}c11_{nnq}\right.\right. \nonumber \\
&\left.~~~~~~~~~~~~~~~~~~~~~~~~~~~~~~~~\left.-M_{Wq}^{2}(1-\cos2\theta_{W})M_{\Phi n}I^{\prime qnn}c15_{nnq}\right)\frac{2}{M_{Wq}^{2}\sqrt{\left(s-4M_{Zn}^{2}\right)\left(s-4M_{Wn}^{2}\right)}}\ln Y6_{nq}\right].&
\end{flalign}
In the above expressions all the symbols $ci_{nnq}$s and $qj_{n}$s where $i : 1 \rightarrow 19$ and $j : 1 \rightarrow 5$, are given explicitly in APPENDIX C; and the explicit expressions of the symbols of $Xk_{n}$s and $Yk_{nq}$s having $k : 1 \rightarrow 10$ are given in APPENDIX D. The overlap intergrals $I^{nnq}, I^{n}, I^{\prime nnq}, I^{\prime qnn}, I^{\prime n}$ are obtained from Eqs.~\ref{Is}-\ref{In1}.
\begin{center}
{\bf \large APPENDIX B}
\end{center}
\section*{Feynman Rules:}
In the following Feynman rules, $n ~{\rm or}~ q$ stands for KK-mode and 0 for SM.\\
\subsection*{$0nn$ Coupling $\left(n: {\rm even ~ or ~odd} \right)$}
\setlength{\mathindent}{0pt}
\begin{flalign*}
&h^{0} h^{n} h^{n} :- 3i\left(\frac{m_{h}^{2}}{v}\right), &\\
&h^{0} A^{n} A^{n} :- \frac{i}{v}\left(\frac{m_{h}^{2}M_{\Phi n}^{2} + 2 M_{Z}^{2}M_{Zn}^{2}}{M_{Zn}^{2}}\right), &\\
&h^{0} G_{Z}^{n} G_{Z}^{n} :- \frac{i}{v}\left(\frac{m_{h}^{2}M_{Z}^{2}}{M_{Zn}^{2}}\right), &\\
&h^{0} A^{n} G_{Z}^{n} :- i\frac{M_{Z} M_{\Phi n}}{v}\left(\frac{m_{h}^{2}-M_{Zn}^{2}}{M_{Zn}^{2}}\right), &\\
&\chi^{0} h^{n} A^{n} :- \frac{i}{v}M_{\Phi n}\left(\frac{m_{h}^{2}-M_{Z}^{2}}{M_{Zn}}\right), &\\
&\chi^{0} h^{n} G_{Z}^{n} :- i\frac{M_{Z}}{v}\left(\frac{m_{hn}^{2}}{M_{Zn}}\right), &\\
&h^{0} H^{n+} H^{n-} :- \frac{i}{v}\left(\frac{m_{h}^{2}M_{\Phi n}^{2} + 2 M_{W}^{2}M_{Wn}^{2}}{M_{Wn}^{2}}\right), &\\
&h^{0} G^{n+} G^{n-} :- i\frac{m_{h}^{2}}{v}\left(\frac{M_{W}^{2}}{M_{Wn}^{2}}\right), &\\
&h^{0} G^{n\pm} H^{n\mp} :\pm \frac{M_{\Phi n}M_{W}}{v}\left(\frac{m_{h}^{2}-M_{Wn}^{2}}{M_{Wn}^{2}}\right), &\\
&\phi^{0 \mp} h^{n} G^{n \pm} :\pm \frac{M_{W}}{v}\left(\frac{m_{hn}^{2}}{M_{Wn}}\right), &\\
&\phi^{0 \mp} h^{n} H^{n \pm} :-i\frac{M_{\Phi n}}{v}\left(\frac{m_{h}^{2}-M_{W}^{2}}{M_{Wn}}\right), &\\
&\phi^{0 \mp} G_{Z}^{n} H^{n \pm} :\mp \frac{M_{\Phi n}M_{Z}}{v}\cos2\theta_{W}\left(\frac{M_{Wn}}{M_{Zn}}\right), &\\
&\phi^{0 \mp} A^{n} G^{n \pm} : i\frac{M_{\Phi n}M_{W}}{v}\left(\frac{M_{Zn}}{M_{Wn}}\right), &\\
&\phi^{0 \pm} A^{n} H^{n \mp} : \pm \frac{\left(M_{\Phi n}^{2}+2 M_{W}^{2}\right)\left(M_{Z}^{2}-M_{W}^{2}\right)}{v M_{Zn} M_{Wn}}, &\\
&\chi^{0} H^{n \mp} G^{n \pm} : -i\frac{M_{\Phi n}M_{W}}{v}, &\\
\end{flalign*}
\subsection*{$nnq$ Coupling $\left(n: {\rm even ~ or ~odd}; q: {\rm even} \right)$}
\setlength{\mathindent}{0pt}
\begin{flalign*}
&h^{n} h^{n} h^{q} :- 3i\left(\frac{m_{h}^{2}}{v}\right)I^{nnq}, &\\
&A^{n} A^{n} h^{q}:-\frac{i}{v  M_{Zn}^{2}}c1_{nnq},&\\
&H^{n+} H^{n-} h^{q}:-\frac{i}{v  M_{Wn}^{2}}c2_{nnq}, &\\
&h^{n} A^{n} G_{Z}^{q}:\frac{i M_{Z}}{v  M_{Zn} M_{Zq}}c3_{nnq}, &\\
&h^{n} A^{n} A^{q}:-\frac{i}{v  M_{Zn} M_{Zq}}c4_{nnq}, &\\
&G_{Z}^{n} G_{Z}^{n} h^{q}:-\frac{i M_{Z}^{2}}{v M_{Zn}^{2}}\left(m_{h}^{2}I^{nnq}+ 2 M_{\Phi n}M_{\Phi q} I^{\prime qnn} \right), &\\
&G_{Z}^{n} h^{n} G_{Z}^{q}:-\frac{i M_{Z}^{2}}{v M_{Zn} M_{Zq}}c5_{nnq}, &\\
&h^{n} G_{Z}^{n} A^{q}:\frac{i M_{Z}}{v  M_{Zn} M_{Zq}}c6_{nnq}, &\\
&A^{n} G_{Z}^{n} h^{q}:-i\frac{M_{Z}}{v  M_{Zn}^{2}}c7_{nnq}, &\\
&H^{n \pm} H^{q \mp} h^{n}:-\frac{i}{v M_{W n} M_{W q}}c8_{nnq}, &\\
&G^{n +} G^{n -} h^{q}:-i\frac{M_{W}^{2}}{v M_{W n}^{2}}\left(m_{h}^{2}I^{nnq}+2M_{\Phi n}M_{\Phi q}I^{\prime qnn}\right),&\\
&G^{n \mp} G^{q \pm} h^{n}:-i\frac{M_{W}^{2}}{v M_{W n} M_{W q}}c5_{nnq},&\\
&H^{n \pm} G^{n \mp} A^{q}:\frac{i M_{W}}{v M_{Zq}}c9_{nnq}, &\\
&H^{n \pm} G^{n \mp} G_{Z}^{q}:\frac{iM_{Z}M_{W}}{v M_{Z q}}c10_{nnq}, &\\
&H^{n \mp} H^{q \pm} G_{Z}^{n}:\mp \frac{M_{Z}}{v M_{Zn} M_{W n} M_{W q}}c11_{nnq}, &\\
&G^{n \mp} H^{q \pm} h^{n}:\pm \frac{M_{W}}{v M_{W n} M_{W q}}c12_{nnq}, &\\
&H^{n \pm} G^{q \mp} h^{n}:\pm \frac{M_{W}}{v M_{W n} M_{W q}}c13_{nnq},&\\
&H^{n \pm} G^{q \mp} G_{Z}^{n}:-i\frac{M_{W}M_{Z}M_{W n}}{v M_{Z n} M_{W q}}c14_{nnq},&\\
&H^{n \pm} H^{q \mp} A^{n}:\pm \frac{1}{v M_{Z n} M_{W n} M_{W q}}c15_{nnq},&\\
&H^{n \pm} G^{q \mp} A^{n}: i\frac{M_{W}}{v M_{Z n} M_{W n} M_{W q}}c16_{nnq},&\\
&G^{n \pm} H^{q \mp} A^{n}: i\frac{M_{W}}{v M_{Z n} M_{W n} M_{W q}}c17_{nnq},&\\
&G^{n \pm} G^{q \mp} A^{n}:\pm \frac{M_{W}^{2}M_{Z n}}{v M_{W n} M_{W q}}c18_{nnq},&\\
&G^{n \pm} H^{n \mp} h^{q}: \pm \frac{M_{W}}{v M_{W n}^{2}}c19_{nnq},&\\
&G^{n \pm} H^{q \mp} G_{Z}^{n}:-i\frac{\left(1-\cos2\theta_{W}\right)M_{W}M_{Z}M_{\Phi n}M_{W q}}{v M_{W n} M_{Z n}}I^{\prime qnn}. &\\
\end{flalign*}
\subsection*{$nnn$ Coupling $\left(n: {\rm even}\right)$}
\setlength{\mathindent}{0pt}
\begin{flalign*}
&h^{n} h^{n} h^{n}:- 3i\left(\frac{m_{h}^{2}}{v}\right)I^{3n}, &\\
&A^{n} A^{n} h^{n}:-\frac{i}{v  M_{Zn}^{2}}c1_{3n},&\\
&H^{n+} H^{n-} h^{n}:-\frac{i}{v  M_{Wn}^{2}}c2_{3n}, &\\
&h^{n} A^{n} G_{Z}^{n}:\frac{i M_{Z} M_{\Phi n}}{v  M_{Zn}^{2}}c3_{3n}, &\\
&G_{Z}^{n} G_{Z}^{n} h^{n}:-\frac{i M_{Z}^{2}}{v M_{Zn}^{2}}c4_{3n}, &\\
&G^{n +} G^{n -} h^{n}:-i\frac{M_{W}^{2}}{v M_{W n}^{2}}c4_{3n},&\\
&G^{n \pm} H^{n \mp} A^{n}: i\frac{M_{W}}{v M_{Z n}}c5_{3n},&\\
&G^{n \pm} H^{n \mp} G_{Z}^{n}:-i\frac{M_{W}M_{Z}}{v M_{Z n}}c6_{3n}, &\\
&G^{n \pm} H^{n \mp} h^{n}: \mp \frac{M_{W} M_{\Phi n}}{v M_{W n}^{2}}c7_{3n}.&
\end{flalign*}
The overlap integrals $I^{3n} {\rm and}~I^{'3n}$ can be calculated from Eqs.~\ref{Is} and \ref{Is1} or \ref{Is2} respectively.
\subsection*{Quartic Coupling $\left(nnnn; n: {\rm even~or~odd}\right)$}
\setlength{\mathindent}{0pt}
\begin{flalign*}
&h^{n} h^{n} h^{n} h^{n}:- 3i\left(\frac{m_{h}^{2}}{v^2}\right)I^{n}, &\\
&h^{n} h^{n} A^{n} A^{n}:-\frac{i}{v^{2}  M_{Zn}^{2}}\left(m_{h}^{2}M_{\Phi n}^{2}I^{n}+2 M_{Z}^{4}I^{\prime n}\right), &\\
&A^{n} A^{n} A^{n} A^{n}:-\frac{3i}{v^{2}  M_{Zn}^{4}}\left(m_{h}^{2}M_{\Phi n}^{4}I^{n}+4 M_{Z}^{4}M_{\Phi n}^{2}I^{\prime n}\right),&\\
&h^{n} h^{n} G_{Z}^{n} G_{Z}^{n}:-\frac{i M_{Z}^{2}}{v^{2} M_{Zn}^{2}}\left(m_{h}^{2}I^{n}+2M_{\Phi n}^{2}I^{\prime n}\right), &\\
&A^{n} A^{n} G_{Z}^{n} G_{Z}^{n}:-\frac{i M_{Z}^{2}}{v^{2} M_{Zn}^{4}}q1_{n}, &\\
&H^{n+} H^{n-} H^{n+} H^{n-}:-\frac{2i}{v^{2}  M_{Wn}^{4}}\left(m_{h}^{2}M_{\Phi n}^{4}I^{n}+4 M_{W}^{4}M_{\Phi n}^{2}I^{\prime n}\right),&\\
&h^{n} G_{Z}^{n} H^{n \pm} G^{n \mp} : -i\frac{M_{W}M_{Z}M_{\Phi n}}{v^{2} M_{Z n}}\left(1-\cos2\theta_{W}\right)I^{\prime n}, &\\
&h^{n} h^{n} H^{n+} H^{n-}:-\frac{i}{v^{2}  M_{Wn}^{2}}\left(m_{h}^{2}M_{\Phi n}^{2}I^{n}+2 M_{W}^{4}I^{\prime n}\right),&\\
&h^{n} h^{n} G^{n+} G^{n-}:-\frac{iM_{W}^{2}}{v^{2}M_{Wn}^{2}}\left(m_{h}^{2}I^{n}+2 M_{\Phi n}^{2}I^{\prime n}\right),&\\
&G_{Z}^{n} G_{Z}^{n} H^{n+} H^{n-}: -\frac{iM_{Z}^{2}}{v^{2}M_{Wn}^{2}M_{Zn}^{2}}q2_{n},&\\
&A^{n} A^{n} H^{n+} H^{n-}: -\frac{iM_{\Phi n}^{2}}{v^{2}M_{Wn}^{2}M_{Zn}^{2}}q3_{n},&\\
&A^{n} A^{n} G^{n+} G^{n-}: -\frac{iM_{W}^{2}}{v^{2}M_{Wn}^{2}M_{Zn}^{2}}q4_{n},&\\
&A^{n} G_{Z}^{n} G^{n \pm} H^{n\mp}: \pm \frac{M_{W}M_{Z}}{v^{2}M_{Wn}^{2}M_{Zn}^{2}}q5_{n}.&
\end{flalign*}
\begin{center}
{\bf \large APPENDIX C}
\end{center}
\setlength{\mathindent}{0pt}
\begin{flalign*}
&c1_{nnq} : m_{h}^{2}M_{\Phi n}^{2}I^{nnq}+2 M_{Z}^{2} M_{Zn}^{2} I^{\prime nnq}- 2 M_{Z}^{2}M_{\Phi n}M_{\Phi q} I^{\prime qnn},& \\
&c2_{nnq} : m_{h}^{2}M_{\Phi n}^{2}I^{nnq}+2 M_{W}^{2} M_{Wn}^{2} I^{\prime nnq}- 2 M_{W}^{2}M_{\Phi n}M_{\Phi q} I^{\prime qnn},& \\
&c3_{nnq} : -m_{h}^{2}M_{\Phi n} I^{nnq}+M_{Z}^{2}M_{\Phi q} I^{\prime qnn}+ M_{Z}^{2}M_{\Phi n} I^{\prime nnq} ,& \\
&c4_{nnq} : m_{h}^{2}M_{\Phi n}M_{\Phi q} I^{nnq}+M_{Z}^{2}(2M_{Z}^{2}+M_{\Phi q}^{2}) I^{\prime qnn}- M_{Z}^{2}M_{\Phi n}M_{\Phi q} I^{\prime nnq}, & \\
&c5_{nnq} : m_{h}^{2}I^{nnq}+ M_{\Phi n}^{2}I^{\prime nnq}+ M_{\Phi n}M_{\Phi q} I^{\prime qnn}, & \\
&c6_{nnq} : -m_{h}^{2}M_{\Phi q} I^{nnq}+M_{\Phi n}(I^{\prime qnn}(2M_{Z}^{2}+M_{\Phi q}^{2})-M_{\Phi n}M_{\Phi q} I^{\prime nnq}), & \\
&c7_{nnq} : m_{h}^{2}M_{\Phi n} I^{nnq}-M_{Zn}^{2} M_{\Phi n}I^{\prime nnq}-(M_{Z}^{2}-M_{\Phi n}^{2})M_{\Phi q} I^{\prime qnn}, & \\
&c8_{nnq} : M_{\Phi n}M_{\Phi q}\left(m_{h}^{2}I^{nnq}-M_{W}^{2}I^{\prime nnq}\right)+M_{W}^{2}\left(2M_{W}^{2}+M_{\Phi q}^{2}\right)I^{\prime qnn}, &\\
&c9_{nnq} : M_{Z q}^{2}I^{\prime qnn}-\left(M_{\Phi n}M_{\Phi q}I^{\prime nnq}+ \cos2\theta_{W}M_{Z}^{2} I^{\prime qnn}\right), &\\
&c10_{nnq} : -M_{\Phi n}I^{\prime nnq}+\cos2\theta_{W}M_{\Phi q}I^{\prime qnn}, &\\
&c11_{nnq} :M_{W}^{2}(M_{\Phi q}^{2}-M_{\Phi n}^{2})I^{\prime qnn}+M_{\Phi n}\cos2\theta_{W}(I^{\prime qnn}M_{W q}^{2}M_{\Phi n}-I^{\prime nnq}M_{W n}^{2}M_{\Phi q}), &\\
&c12_{nnq} : -m_{h}^{2}M_{\Phi q}I^{nnq}+M_{\Phi n}M_{W q}^{2}I^{\prime qnn}+M_{\Phi n}\left(M_{W}^{2}I^{\prime qnn}-M_{\Phi n}M_{\Phi q}I^{\prime nnq}\right), &\\
&c13_{nnq} : -m_{h}^{2}M_{\Phi n}I^{nnq}+M_{\Phi n}(M_{W}^{2}I^{\prime nnq}-M_{\Phi q}M_{\Phi n}I^{\prime qnn})+M_{\Phi q}M_{W n}^{2}I^{\prime qnn}, &\\
&c14_{nnq} : M_{\Phi q}I^{\prime qnn}-\cos2\theta_{W}M_{\Phi n}I^{\prime nnq}, &\\
&c15_{nnq} : M_{W}^{2}M_{Z n}^{2}\left(M_{\Phi n}I^{\prime qnn}-M_{\Phi q}I^{\prime nnq}\right) + I^{\prime qnn}M_{\Phi n}M_{W}^{2}\left(M_{\Phi q}^{2}-M_{\Phi n}^{2}\right)&\nonumber \\
&~~~~~~~~~~-\cos2\theta_{W}M_{Z}^{2}\left(I^{\prime qnn}M_{W q}^{2}M_{\Phi n}-I^{\prime nnq}M_{W n}^{2}M_{\Phi q}\right), &\\
&c16_{nnq} : M_{\Phi n}M_{\Phi q}(M_{Z}^{2}-M_{W}^{2})I^{\prime qnn}-M_{Z}^{2}M_{W n}^{2}\cos2\theta_{W}I^{\prime nnq}+M_{Z n}^{2}M_{W}^{2}I^{\prime nnq}, & \\
&c17_{nnq} : M_{Z n}^{2}\left(M_{\Phi n}M_{\Phi q}I^{\prime nnq}+  M_{W}^{2}I^{\prime qnn}\right)-M_{W q}^{2}I^{\prime qnn}\left(M_{\Phi n}^{2}+M_{Z}^{2}\cos2\theta_{W}\right), & \\
&c18_{nnq} :M_{\Phi n}I^{\prime nnq}-M_{\Phi q}I^{\prime qnn}, & \\
&c19_{nnq} : m_{h}^{2}M_{\Phi n} I^{nnq}-M_{Wn}^{2} M_{\Phi n}I^{\prime nnq}+(M_{\Phi n}^{2}-M_{W}^{2})M_{\Phi q} I^{\prime qnn}, &\\
&c1_{3n} : m_{h}^{2}M_{\Phi n}^{2}I^{3n}+2 M_{Z}^{4}I^{\prime 3n},&\\
&c2_{3n} : m_{h}^{2}M_{\Phi n}^{2}I^{3n}+2 M_{W}^{4}I^{\prime 3n},&\\
&c3_{3n} : -m_{h}^{2} I^{3n}+ 2M_{Z}^{2}I^{\prime 3n}, & \\
&c4_{3n} : m_{h}^{2}I^{3n}+2M_{\Phi n}^{2}I^{\prime 3n}, & \\
&c5_{3n} : M_{Z}^{2}I^{\prime 3n}\left(1-\cos2\theta_{W}\right), & \\
&c6_{3n} : M_{\Phi n}I^{\prime 3n}\left(1-\cos2\theta_{W}\right), & \\
&c7_{3n} : -m_{h}^{2} I^{3n}+ 2M_{W}^{2}I^{\prime 3n}, & \\
&q1_{n} : 3m_{h}^{2}M_{\Phi n}^{2}I^{n}+2(M_{Z}^{4}-4M_{Z}^{2}M_{\Phi n}^{2}+M_{\Phi n}^{4})I^{\prime n}, &\\
&q2_{n} : m_{h}^{2}M_{\Phi n}^{2}I^{n}+\left(2M_{W}^{4}-4M_{W}^{2}M_{\Phi n}^{2}(1-\cos2\theta_{W})+M_{\Phi n}^{4}(1+\cos4\theta_{W})\right)I^{\prime n}, & \\
&q3_{n} : m_{h}^{2}M_{\Phi n}^{2}I^{n}+\left(2M_{W}^{4}+4M_{W}^{2}M_{Z}^{2}(1-\cos2\theta_{W})+M_{Z}^{4}(1+\cos4\theta_{W})\right)I^{\prime n},&\\
&q4_{n} : m_{h}^{2}M_{\Phi n}^{2}I^{n}+\left(M_{Z}^{4}(1+\cos4\theta_{W})-4M_{Z}^{2}M_{\Phi n}^{2}(1-\cos2\theta_{W})+ 2M_{\Phi n}^{4}\right)I^{\prime n}, & \\
&q5_{n} :m_{h}^{2}M_{\Phi n}^{2}I^{n}+\left((1-\cos2\theta_{W})\left(M_{W}^{2}(M_{Z}^{2}-M_{\Phi n}^{2})+M_{\Phi n}^{4}\right)-M_{Z}^{2}M_{\Phi n}^{2}(3+\cos4\theta_{W})\right)I^{\prime n}.&
\end{flalign*}
\begin{center}
{\bf \large APPENDIX D}
\end{center}
\setlength{\mathindent}{0pt}
\begin{flalign*}
&X1_{n} :\frac{s-2m_{h}^{2}-4M_{\Phi n}^{2}-\sqrt{\left(s-4m_{hn}^{2}\right)\left(s-4M_{Zn}^{2}\right)}}{2\sqrt{m_{h}^{4}+M_{Z}^{2}\left(s-4m_{hn}^{2}\right)}},& \\
&Y1_{nq} : \frac{s-2m_{h}^{2}-4M_{\Phi n}^{2}+2M_{\Phi q}^{2}-\sqrt{\left(s-4m_{hn}^{2}\right)\left(s-4M_{Zn}^{2}\right)}}{2\sqrt{{\left(m_{h}^{2}-M_{\Phi q}^{2}\right)}^{2}+M_{\Phi q}^{2}\left(s-4M_{\Phi n}^{2}\right)+M_{Z}^{2}\left(s-4m_{hn}^{2}\right)}},& \\
&X2_{n} : \frac{s\{2(m_{hn}^{2}+M_{Zn}^{2})-s\}-2(s M_{Z}^{2} +m_{hn}^{2}M_{Zn}^{2})}{m_{hn}^{4}+M_{Zn}^{4}-sM_{Z}^{2}}, &\\
&Y2_{nq} : \frac{s\{2(m_{hn}^{2}+M_{Zn}^{2})-s\}-2(s M_{Zq}^{2} +m_{hn}^{2}M_{Zn}^{2})}{m_{hn}^{4}+M_{Zn}^{4}-sM_{Zq}^{2}}, &\\
&X3_{n} : 1+ \frac{(s-m_{hn}^{2}-M_{Zn}^{2})^{2}}{s m_{h}^{2}}, &\\
&Y3_{nq} : 1+ \frac{(s-m_{hn}^{2}-M_{Zn}^{2})^{2}}{s m_{hq}^{2}}, &\\
&X4_{n} : \frac{\sqrt{s}\left(s-2M_{\Phi n}^{2}-m_{hn}^{2}-M_{Zn}^{2}\right)-\sqrt{s-4M_{Wn}^{2}}(s-m_{hn}^{2}-M_{Zn}^{2})}{2\sqrt{(s-m_{hn}^{2}-M_{Zn}^{2})\{sM_{W}^{2}-M_{Wn}^{2}(m_{hn}^{2}+M_{Zn}^{2})\}+sM_{\Phi n}^{4}}}, &\\
&Y4_{nq} : \frac{\sqrt{s}\left(s+2M_{\Phi q}^{2}-2M_{\Phi n}^{2}-m_{hn}^{2}-M_{Zn}^{2}\right)-\sqrt{s-4M_{Wn}^{2}}(s-m_{hn}^{2}-M_{Zn}^{2})}{2\sqrt{(s-m_{hn}^{2}-M_{Zn}^{2})\{sM_{Wq}^{2}-M_{Wn}^{2}(m_{hn}^{2}+M_{Zn}^{2})\}+s(M_{\Phi q}^{2}-M_{\Phi n}^{2})^{2}}}, &\\
&X5_{n} :\frac{s-2m_{h}^{2}-4M_{\Phi n}^{2}-\sqrt{\left(s-4m_{hn}^{2}\right)\left(s-4M_{Wn}^{2}\right)}}{2\sqrt{m_{h}^{4}+M_{W}^{2}\left(s-4m_{hn}^{2}\right)}},& \\
&Y5_{nq} : \frac{s-2m_{h}^{2}-4M_{\Phi n}^{2}+2M_{\Phi q}^{2}-\sqrt{\left(s-4m_{hn}^{2}\right)\left(s-4M_{Wn}^{2}\right)}}{2\sqrt{{\left(m_{h}^{2}-M_{\Phi q}^{2}\right)}^{2}+M_{\Phi q}^{2}\left(s-4M_{\Phi n}^{2}\right)+M_{W}^{2}\left(s-4m_{hn}^{2}\right)}},& \\
&X6_{n} :\frac{s-2M_{Z}^{2}-4M_{\Phi n}^{2}-\sqrt{\left(s-4M_{Zn}^{2}\right)\left(s-4M_{Wn}^{2}\right)}}{2\sqrt{M_{Z}^{4}+M_{W}^{2}\left(s-4M_{Zn}^{2}\right)}},& \\
&Y6_{nq} : \frac{s-2M_{Z}^{2}-4M_{\Phi n}^{2}+2M_{\Phi q}^{2}-\sqrt{\left(s-4M_{Zn}^{2}\right)\left(s-4M_{Wn}^{2}\right)}}{2\sqrt{{\left(M_{Z}^{2}-M_{\Phi q}^{2}\right)}^{2}+M_{\Phi q}^{2}\left(s-4M_{\Phi n}^{2}\right)+M_{W}^{2}\left(s-4M_{Zn}^{2}\right)}},& \\
&X7_{n} : \frac{s\{2(m_{hn}^{2}+M_{Wn}^{2})-s\}-2(s M_{W}^{2} +m_{hn}^{2}M_{Wn}^{2})}{m_{hn}^{4}+M_{Wn}^{4}-sM_{W}^{2}}, &\\
&Y7_{nq} : \frac{s\{2(m_{hn}^{2}+M_{Wn}^{2})-s\}-2(s M_{Wq}^{2} +m_{hn}^{2}M_{Wn}^{2})}{m_{hn}^{4}+M_{Wn}^{4}-sM_{Wq}^{2}}, &\\
&X8_{n} : 1+ \frac{(s-m_{hn}^{2}-M_{Wn}^{2})^{2}}{s m_{h}^{2}}, &\\
&Y8_{nq} : 1+ \frac{(s-m_{hn}^{2}-M_{Wn}^{2})^{2}}{s m_{hq}^{2}}, &\\
&X9_{n} : \frac{s\{2(M_{Zn}^{2}+M_{Wn}^{2})-s\}-2(s M_{W}^{2} +M_{Zn}^{2}M_{Wn}^{2})}{M_{Zn}^{4}+M_{Wn}^{4}-sM_{W}^{2}}, &\\
&Y9_{nq} : \frac{s\{2(M_{Zn}^{2}+M_{Wn}^{2})-s\}-2(s M_{Wq}^{2} +M_{Zn}^{2}M_{Wn}^{2})}{M_{Zn}^{4}+M_{Wn}^{4}-sM_{Wq}^{2}}, &\\
&X10_{n} : 1+ \frac{(s-M_{Zn}^{2}-M_{Wn}^{2})^{2}}{s m_{h}^{2}}, &\\
&Y10_{nq} : 1+ \frac{(s-M_{Zn}^{2}-M_{Wn}^{2})^{2}}{s m_{hq}^{2}}. &
\end{flalign*}
\begin{center}
{\bf \large APPENDIX E}
\end{center}
We now give the general form of the matrix elements explicitly. Here, the sum over KK-modes, symmetry factors and the factor $\frac{1}{\sqrt{2}}$ due to the presence of dibosonic states have been taken into account. While writing the elements, the ordering of neutral fields are important; e.g., the combinations $h^{n}h^{n}A^{n}A^{n}$ will differ from $h^{n}A^{n}h^{n}A^{n}$ by a factor of $\frac{1}{2}$ as $h^{n}h^{n}$ or $A^{n}A^{n}$ altogether implies the presence of dibosonic state. Another noteworthy point is the sum of the KK-numbers should be even, otherwise the elements will be zero due to the conservation of KK-parity. 
\subsection*{E.1 :}~ Elements of matrix $\mathcal{A}_{15\times 15}$ :
\setlength{\mathindent}{0pt}
\begin{flalign*}
&h^{0}h^{0}h^{0}h^{0} : -\frac{3}{2}\left(\frac{m_{h}^{2}}{v^{2}}\right),~~~~~~~~~~~~~~~~~~~~~~h^{0}h^{0}h^{n}h^{n} : -\frac{3}{2}\left(\frac{m_{h}^{2}}{v^{2}}\right),~~~~~~~~~~~~~~~~~~~~~~h^{0}h^{n}h^{0}h^{n} : -3\left(\frac{m_{h}^{2}}{v^{2}}\right),&\\
&h^{n}h^{n}h^{0}h^{m} : -\frac{3}{\sqrt{2}}\left(\frac{m_{h}^{2}}{v^{2}}\right)I^{nnm},~~~~~~~~~~~h^{n}h^{0}h^{n}h^{m} : -3\left(\frac{m_{h}^{2}}{v^{2}}\right)I^{nnm},~~~~~~~~~h^{0}h^{n}h^{m}h^{p} : -3\left(\frac{m_{h}^{2}}{v^{2}}\right)I^{nmp},&\\
&h^{0}h^{n}h^{n}h^{n} : -\frac{3}{\sqrt{2}}\left(\frac{m_{h}^{2}}{v^{2}}\right)I^{3n},~~~~~~~~~~~~~h^{n}h^{n}h^{n}h^{n} : -\frac{3}{2}\left(\frac{m_{h}^{2}}{v^{2}}\right)I^{n},~~~~~~~~~~h^{n}h^{n}h^{m}h^{m} : -\frac{3}{2}\left(\frac{m_{h}^{2}}{v^{2}}\right)I^{nnmm},&\\
&h^{n}h^{m}h^{n}h^{m} : -3\left(\frac{m_{h}^{2}}{v^{2}}\right)I^{nnmm},~~~~~~~~~h^{n}h^{n}h^{n}h^{m} : -\frac{3}{\sqrt{2}}\left(\frac{m_{h}^{2}}{v^{2}}\right)I^{nnnm},~~~h^{n}h^{n}h^{m}h^{p} : -\frac{3}{\sqrt{2}}\left(\frac{m_{h}^{2}}{v^{2}}\right)I^{nnmp},&\\
&h^{n}h^{m}h^{n}h^{p} : -3\left(\frac{m_{h}^{2}}{v^{2}}\right)I^{nnmp},~~~~~~~~~~~~h^{n}h^{m}h^{p}h^{q} : -3\left(\frac{m_{h}^{2}}{v^{2}}\right)I^{nmpq}.&
\end{flalign*}
\subsection*{E.2 :}~ Elements of matrix $\mathcal{D}_{10\times 10}$:
\setlength{\mathindent}{0pt}
\begin{flalign*}
&A^{n}A^{n}A^{n}A^{n} : -\frac{3}{2 v^{2} M_{Zn}^{4}}\left(m_{h}^{2} M_{\Phi n}^{4}I^{n} + 4 M_{Z}^{4} M_{\Phi n}^{2} I^{\prime n}\right),&\\
&A^{n}A^{n}A^{m}A^{m} : -\frac {\left \lbrace 3 m_{h}^{2} M_{\Phi n}^{2} M_{\Phi m}^{2} I^{nnmm} + 2 M_{Z}^{4} \left (M_{\Phi m}^{2} I_{1}^{nnmm} + M_{\Phi n}^{2} I_{1}^{mmnn} + 4 M_{\Phi n} M_{\Phi m} I_{1}^{nmnm} \right) \right \rbrace}{2 v^{2} M_{Zn}^{2} M_{Zm}^{2} },&\\
&A^{n}A^{m}A^{n}A^{m} : -\frac {\left \lbrace 3 m_{h}^{2} M_{\Phi n}^{2} M_{\Phi m}^{2} I^{nnmm} + 2 M_{Z}^{4} \left (M_{\Phi m}^{2} I_{1}^{nnmm} + M_{\Phi n}^{2} I_{1}^{mmnn} + 4 M_{\Phi n} M_{\Phi m} I_{1}^{nmnm} \right) \right \rbrace}{v^{2} M_{Zn}^{2} M_{Zm}^{2} },&\\
&A^{n}A^{n}A^{n}A^{m} : -\frac { 3 \left \lbrace m_{h}^{2} M_{\Phi n}^{3} M_{\Phi m} I^{nnnm} + 2 M_{Z}^{4} \left (M_{\Phi n} M_{\Phi m} I_{1}^{nnnm} + M_{\Phi n}^{2} I_{1}^{mnnn} \right) \right \rbrace}{\sqrt {2} v^{2} M_{Zn}^{3} M_{Zm}},&\\
&A^{n}A^{n}A^{m}A^{p} : -\frac{1}{\sqrt{2} v^{2} M_{Zn}^{2} M_{Zm} M_{Zp}} \left \lbrace 3 m_{h}^{2} M_{\Phi n}^{2} M_{\Phi m} M_{\Phi p} I^{nnmp}\right. &\\
&\left.~~~~~~~~~~~~~~~~~~ + 2 M_{Z}^{4} \left (M_{\Phi n}^{2} I_{1}^{mpnn} + M_{\Phi m}M_{\Phi p}  I_{1}^{nnmp} + 2 M_{\Phi n} \left( M_{\Phi m} I_{1}^{npmn}  + M_{\Phi p} I_{1}^{mnnp} \right)\right) \right \rbrace,&\\
&A^{n}A^{m}A^{n}A^{p} : -\frac{1}{v^{2} M_{Zn}^{2} M_{Zm} M_{Zp}} \left \lbrace 3 m_{h}^{2} M_{\Phi n}^{2} M_{\Phi m} M_{\Phi p} I^{nnmp}\right. &\\
&\left.~~~~~~~~~~~~~~~~~~ + 2 M_{Z}^{4} \left (M_{\Phi n}^{2} I_{1}^{mpnn} + M_{\Phi m}M_{\Phi p}  I_{1}^{nnmp} + 2 M_{\Phi n} \left( M_{\Phi m} I_{1}^{npmn}  + M_{\Phi p} I_{1}^{mnnp} \right)\right) \right \rbrace,&\\
&A^{n}A^{m}A^{p}A^{q} : -\frac{1}{v^{2} M_{Zn} M_{Zm} M_{Zp} M_{Zq}} \left[3 m_{h}^{2} M_{\Phi n} M_{\Phi m} M_{\Phi p} M_{\Phi q} I^{nmpq}\right. &\\
&\left.~~~~~~~~~~~~~~~~~~ + 2 M_{Z}^{4} \left \lbrace M_{\Phi q} \left(M_{\Phi p} I_{1}^{nmpq} + M_{\Phi m} I_{1}^{npmq}\right) + M_{\Phi n} \left( M_{\Phi p} I_{1}^{mqnp}  + M_{\Phi q} I_{1}^{mpnq} \right) \right.\right.&\\
&\left.\left.~~~~~~~~~~~~~~~~~~ +  M_{\Phi m} \left(M_{\Phi p} I_{1}^{nqmp} + M_{\Phi n} I_{1}^{pqnm}\right)\right \rbrace \right] .&\\
\end{flalign*}
\subsection*{E.3 :} ~ Elements of matrix $\mathcal{F}_{25\times 25}$ :
\setlength{\mathindent}{0pt}
\begin{flalign*}
&\phi^{+}\phi^{-}\phi^{+}\phi^{-} : -2\frac{m_{h}^{2}}{v^{2}},&\\
&\phi^{+}H^{n-}\phi^{+}H^{n-}/\phi^{-}H^{n+}\phi^{-}H^{n+} : -2\frac{m_{h}^{2} M_{\Phi n}^{2}}{v^{2} M_{Wn}^{2}}, &\\
&\phi^{+}\phi^{-}H^{n+}H^{n-} : -\frac{2}{v^{2} M_{Wn}^{2}}\left ( m_{h}^{2} M_{\Phi n}^{2} + M_{W}^{4} \right),&\\
\setlength{\mathindent}{0pt}
&\left \lbrace \begin{aligned}\phi^{+}H^{n+}H^{n-}H^{m-}\\
                            \phi^{-}H^{n+}H^{n-}H^{m+}
               \end{aligned} \right. : -\frac{2\left \lbrace m_{h}^{2} M_{\Phi n}^{2}  M_{\Phi m} I^{nnm} + M_{W}^{4} \left(M_{\Phi m} I_{1}^{nnm} + M_{\Phi n} I_{1}^{mnn} \right)\right\rbrace} {v^{2} M_{Wn}^{2} M_{Wm}}, &\\
\setlength{\mathindent}{0pt}
&\left \lbrace \begin{aligned}\phi^{+}H^{n+}H^{n-}H^{n-}\\
                            \phi^{-}H^{n+}H^{n-}H^{n+}
               \end{aligned} \right.: -\frac{2}{v^{2} M_{Wn}^{3}}\left(m_{h}^{2} M_{\Phi n}^{3} I^{3n} + 2 M_{W}^{4} M_{\Phi n} I^{\prime 3n}  \right), &\\
\setlength{\mathindent}{0pt}
&\left \lbrace \begin{aligned}\phi^{+}H^{n-}H^{n-}H^{m+}\\
                            \phi^{-}H^{n+}H^{n+}H^{m-}
               \end{aligned} \right.: -\frac{2\left (m_{h}^{2} M_{\Phi n}^{2} M_{\Phi m} I^{nnm} + 2 M_{W}^{4} M_{\Phi n} I_{1}^{mnn} \right)}{v^{2} M_{Wn}^{2} M_{Wm}}, &\\
\setlength{\mathindent}{0pt}
&\left \lbrace \begin{aligned}\phi^{+}H^{n-}H^{p+}H^{q-}\\
                           \phi^{-}H^{n+}H^{p-}H^{q+}
               \end{aligned} \right.: -\frac{2\left \lbrace m_{h}^{2} M_{\Phi n} M_{\Phi m} M_{\Phi p} I^{nmp} + M_{W}^{4} \left(M_{\Phi n} I_{1}^{mpn} + M_{\Phi p} I_{1}^{nmp} \right)\right\rbrace} {v^{2} M_{Wn} M_{Wm} M_{Wp}}, &\\
&H^{n+}H^{n-}H^{n+}H^{n-} : -\frac{2}{v^{2} M_{Wn}^{4}}\left ( m_{h}^{2} M_{\Phi n}^{4} I^{n} + 4 M_{W}^{4} M_{\Phi n}^{2} I^{\prime n}  \right),&\\
&H^{n+}H^{n-}H^{m+}H^{m-} : -\frac{2\left \lbrace m_{h}^{2} M_{\Phi n}^{2}  M_{\Phi m}^{2} I^{nnmm}  + M_{W}^{4}\left( M_{\Phi m}^{2} I_{1}^{nnmm} + M_{\Phi n}^{2} I_{1}^{mmnn} + 2 M_{\Phi n} M_{\Phi m} I_{1}^{nmnm} \right)\right \rbrace}{v^{2} M_{Wn}^{2} M_{Wm}^{2}}, &\\
&H^{n+}H^{m-}H^{n+}H^{m-} : -\frac{2}{v^{2} M_{Wn}^{2} M_{Wm}^{2}}\left(m_{h}^{2} M_{\Phi n}^{2}  M_{\Phi m}^{2} I^{nnmm} + 4 M_{W}^{4} M_{\Phi n} M_{\Phi m} I_{1}^{nmnm} \right), &\\
\setlength{\mathindent}{0pt}
&\left \lbrace \begin{aligned}H^{n+}H^{m-}H^{n+}H^{p-}\\
                            H^{n-}H^{m+}H^{n-}H^{p+}
               \end{aligned} \right.: -\frac{2\left \lbrace m_{h}^{2} M_{\Phi n}^{2}  M_{\Phi m} M_{\Phi p}  I^{nnmp} + 2 M_{W}^{4} M_{\Phi n}\left(M_{\Phi m} I_{1}^{npnm} + M_{\Phi p} I_{1}^{nmnp} \right)\right\rbrace} {v^{2} M_{Wn}^{2} M_{Wm} M_{Wp}}, &\\
\setlength{\mathindent}{0pt}
&\left \lbrace \begin{aligned}H^{n+}H^{n-}H^{n+}H^{m-}\\
                            H^{n-}H^{n+}H^{n-}H^{m+}
               \end{aligned} \right.: -\frac{2\left \lbrace m_{h}^{2} M_{\Phi n}^{3}  M_{\Phi m} I^{nnnm} + 2 M_{W}^{4} M_{\Phi n}\left(M_{\Phi m} I_{1}^{nnnm} + M_{\Phi n} I_{1}^{mnnn} \right)\right\rbrace} {v^{2} M_{Wn}^{3} M_{Wm}}, &\\
\setlength{\mathindent}{0pt}
&\left \lbrace \begin{aligned}H^{n+}H^{n-}H^{m+}H^{p-}\\
                            H^{n+}H^{n-}H^{m-}H^{p+}
               \end{aligned} \right.: -\frac{2}{v^{2} M_{Wn}^{2} M_{Wm} M_{Wp}}\left[ m_{h}^{2} M_{\Phi n}^{2}  M_{\Phi m} M_{\Phi p}  I^{nnmp} \right. &\\
&\left.~~~~~~~~~~~~~~~~~~~~~~~~~~~~~~ +  M_{W}^{4} \left \lbrace M_{\Phi n}\left(M_{\Phi m} I_{1}^{npnm} + M_{\Phi p} I_{1}^{nmnp} \right)+ M_{\Phi n}^{2} I_{1}^{mpnn} + M_{\Phi m} M_{\Phi p} I_{1}^{nnmp} \right\rbrace \right], &\\
&H^{n+}H^{m-}H^{p+}H^{q-} : -\frac{2}{v^{2} M_{Wn} M_{Wm} M_{Wp}  M_{Wq}}\left[ m_{h}^{2} M_{\Phi n} M_{\Phi m} M_{\Phi p} M_{\Phi q}  I^{nmpq} \right. &\\
&\left.~~~~~~~~~~~~~~~~~~~~~~~~~~ +  M_{W}^{4} \left \lbrace M_{\Phi n}\left(M_{\Phi m} I_{1}^{pqnm} + M_{\Phi q} I_{1}^{mpnq} \right)+
M_{\Phi p}\left(M_{\Phi m} I_{1}^{nqmp} + M_{\Phi q} I_{1}^{nmpq} \right) \right\rbrace \right]. &\\
\end{flalign*}
\subsection*{E.4 :} ~ Elements of matrix $\mathcal{B}_{15\times 10}$ :
\setlength{\mathindent}{0pt}
\begin{flalign*}
&h^{0}h^{0}A^{n}A^{n} : -\frac{1}{2}\frac{\left(m_{h}^{2}M_{\Phi n}^{2} + 2 M_{Z}^{4} \right)}{v^{2} M_{Zn}^{2}}, &\\
&h^{0}h^{n}A^{m}A^{m} : -\frac{1}{\sqrt{2}}\frac{\left(m_{h}^{2}M_{\Phi m}^{2} I^{nmm} + 2 M_{Z}^{4} I_{1}^{mmn} \right)}{v^{2} M_{Zm}^{2}}, &\\
&h^{0}h^{n}A^{m}A^{p} : -\frac{\left(m_{h}^{2}M_{\Phi m}M_{\Phi p} I^{nmp} + 2 M_{Z}^{4} I_{1}^{mpn} \right)}{v^{2} M_{Zm} M_{Zp}}, &\\
&h^{n}h^{n}A^{n}A^{n} : -\frac{1}{2}\frac{\left(m_{h}^{2} M_{\Phi n}^{2} I^{n} + 2 M_{Z}^{4} I^{\prime n} \right)}{v^{2} M_{Zn}^{2}}, &\\
&h^{n}h^{n}A^{m}A^{m} : -\frac{1}{2}\frac{\left(m_{h}^{2} M_{\Phi m}^{2} I^{nnmm} + 2 M_{Z}^{4} I_{1}^{mmnn} \right)}{v^{2} M_{Zm}^{2}}, &\\
&h^{n}h^{n}A^{m}A^{p} : -\frac{1}{\sqrt{2}}\frac{\left(m_{h}^{2} M_{\Phi m} M_{\Phi p} I^{nnmp} + 2 M_{Z}^{4} I_{1}^{mpnn} \right)}{v^{2} M_{Zm} M_{Zp}}, &\\
&A^{n}A^{n}h^{m}h^{p} : -\frac{1}{\sqrt{2}}\frac{\left(m_{h}^{2} M_{\Phi n}^{2} I^{mpnn} + 2 M_{Z}^{4} I_{1}^{nnmp} \right)}{v^{2} M_{Zn}^{2}}, &\\
&h^{n}h^{m}A^{n}A^{m} : -\frac{1}{\sqrt{2}}\frac{\left(m_{h}^{2} M_{\Phi n} M_{\Phi m} I^{nmnm} + 2 M_{Z}^{4} I_{1}^{nmnm} \right)}{v^{2} M_{Zn} M_{Zm}}, &\\
&h^{n}h^{m}A^{p}A^{q} : -\frac{\left(m_{h}^{2} M_{\Phi p} M_{\Phi q} I^{nmpq} + 2 M_{Z}^{4} I_{1}^{pqnm} \right)}{v^{2} M_{Zp} M_{Zq}}. &\\
\end{flalign*}
\subsection*{E.5 :}  ~ Elements of matrix $\mathcal{E}_{10\times 25}$ :
\setlength{\mathindent}{0pt}
\begin{flalign*}
&\phi^{+}\phi^{-}A^{n}A^{n} : -\frac{1}{\sqrt{2}}\frac{\left(m_{h}^{2}M_{\Phi n}^{2} + 2 M_{Z}^{4} \cos^{2}\theta_{W} \right)}{v^{2} M_{Zn}^{2}}, &\\
&\phi^{\pm}H^{m\mp}A^{n}A^{n} : -\frac{1}{\sqrt{2} v^{2} M_{Zn}^{2} M_{Wm}}\left[ m_{h}^{2} M_{\Phi n}^{2} M_{\Phi m} I^{nnm} \right. &\\
&\left.~~~~~~~~~~~~~~~~~~~~~ +  2 M_{Z}^{2}\left \lbrace \cos^{2}\theta_{W} M_{\Phi m} M_{Z}^{2} I_{1}^{nnm} +
M_{\Phi n} M_{W}^{2} I_{1}^{mnn} \left(1-\cos 2\theta_{W} \right) \right\rbrace \right], &\\
&\phi^{\pm}H^{n\mp}A^{n}A^{n} : -\frac{1}{\sqrt{2} v^{2} M_{Zn}^{2} M_{Wn}}\left[ m_{h}^{2} M_{\Phi n}^{3} I^{3n} \right. &\\
&\left.~~~~~~~~~~~~~~~~~~~~~ +  2 M_{Z}^{2} M_{\Phi n} I^{\prime 3n} \left \lbrace \cos^{2} 2\theta_{W} M_{Z}^{2}+
 M_{W}^{2} \left(1-\cos 2\theta_{W} \right) \right\rbrace \right], &\\
&\phi^{\pm}H^{n\mp}A^{n}A^{m} : -\frac{1}{v^{2} M_{Zn} M_{Zm} M_{Wn}}\left[ m_{h}^{2} M_{\Phi n}^{2} M_{\Phi m} I^{nnm} + 2 M_{Z}^{4} \cos^{2}2\theta_{W} M_{\Phi n} I_{1}^{mnn} \right. &\\
&\left.~~~~~~~~~~~~~~~~~~~~~ + M_{W}^{2} M_{Z}^{2} \left(1-\cos 2\theta_{W} \right) \left( M_{\Phi m} I_{1}^{nnm} + M_{\Phi n} I_{1}^{mnn}\right)  \right], &\\
&\phi^{\pm}H^{p\mp}A^{n}A^{m} : -\frac{1}{v^{2} M_{Zn} M_{Zm} M_{Wp}}\left[ m_{h}^{2} M_{\Phi n} M_{\Phi m} M_{\Phi p} I^{nmp} + 2 M_{Z}^{4} \cos^{2}2\theta_{W} M_{\Phi p} I_{1}^{nmp} \right. &\\
&\left.~~~~~~~~~~~~~~~~~~~~~ + M_{W}^{2} M_{Z}^{2} \left(1-\cos 2\theta_{W} \right) \left( M_{\Phi m} I_{1}^{pnm} + M_{\Phi n} I_{1}^{pmn}\right) \right], &\\
&A^{n}A^{n}H^{n+}H^{n-} : -\frac{M_{\Phi n}^{2}}{\sqrt{2} v^{2} M_{Wn}^{2} M_{Zn}^{2}} q3_{n}, &\\
&A^{n}A^{n}H^{m+}H^{m-} : -\frac{1}{\sqrt{2} v^{2} M_{Zn}^{2} M_{Wm}^{2}}\left[ m_{h}^{2} M_{\Phi n}^{2} M_{\Phi m}^{2} I^{nnmm} + 2 M_{Z}^{4} \cos^{2}2\theta_{W} M_{\Phi m}^{2} I_{1}^{nnmm} \right. &\\
&\left.~~~~~~~~~~~~~~~~~~~~~ + 4 M_{W}^{2} M_{Z}^{2} M_{\Phi n} M_{\Phi m} \left(1-\cos 2\theta_{W} \right)I_{1}^{nmnm} +  2 M_{W}^{4} M_{\Phi n}^{2} I_{1}^{mmnn}\right], &\\
&A^{n}A^{m}H^{p+}H^{p-} : -\frac{1}{v^{2} M_{Zn}  M_{Zm} M_{Wp}^{2}}\left[ m_{h}^{2} M_{\Phi n} M_{\Phi m} M_{\Phi p}^{2} I^{nmpp} + 2 M_{Z}^{4} \cos^{2}2\theta_{W} M_{\Phi p}^{2} I_{1}^{nmpp} \right. &\\
&\left.~~~~~~~~~~~~~~~~~~~~~ + 2 M_{W}^{2} M_{Z}^{2} M_{\Phi p} \left(1-\cos 2\theta_{W} \right) \left(M_{\Phi n} I_{1}^{pmnp} + M_{\Phi m} I_{1}^{nppm}\right)+  2 M_{W}^{4} M_{\Phi n} M_{\Phi m}I_{1}^{ppnm}\right], &\\
&A^{n}A^{m}H^{n+}H^{n-} : -\frac{1}{v^{2} M_{Zn}  M_{Zm} M_{Wn}^{2}}\left[ m_{h}^{2} M_{\Phi n}^{3} M_{\Phi m} I^{nnnm} + 2 M_{Z}^{4} \cos^{2}2\theta_{W} M_{\Phi n}^{2} I_{1}^{mnnn} \right. &\\
&\left.~~~~~~~~~~~~~~~~~~~~~ + 2 M_{W}^{2} M_{Z}^{2} M_{\Phi n} \left(1-\cos 2\theta_{W} \right) \left(M_{\Phi n} I_{1}^{mnnn} + M_{\Phi m} I_{1}^{nnnm}\right)+  2 M_{W}^{4} M_{\Phi n} M_{\Phi m}I_{1}^{nnnm}\right], &\\
&A^{n}A^{n}H^{m\pm}H^{p\mp} : -\frac{1}{\sqrt{2} v^{2} M_{Zn}^{2}  M_{Wm} M_{Wp}}\left[ m_{h}^{2} M_{\Phi n}^{2} M_{\Phi m} M_{\Phi p} I^{nnmp} + 2 M_{Z}^{4} \cos^{2}2\theta_{W} M_{\Phi m} M_{\Phi p} I_{1}^{nnmp} \right. &\\
&\left.~~~~~~~~~~~~~~~~~~~~~ + 2 M_{W}^{2} M_{Z}^{2} M_{\Phi n} \left(1-\cos 2\theta_{W} \right) \left(M_{\Phi m} I_{1}^{npmn} + M_{\Phi p} I_{1}^{mnnp}\right)+  2 M_{W}^{4} M_{\Phi n}^{2} I_{1}^{mpnn}\right], &\\
&A^{n}A^{n}H^{m\pm}H^{n\mp} : -\frac{1}{\sqrt{2} v^{2} M_{Zn}^{2} M_{Wm}  M_{Wn}}\left[ m_{h}^{2} M_{\Phi n}^{3} M_{\Phi m} I^{nnnm} + 2 M_{Z}^{4} \cos^{2}2\theta_{W} M_{\Phi n} M_{\Phi m} I_{1}^{nnnm} \right. &\\
&\left.~~~~~~~~~~~~~~~~~~~~~ + 2 M_{W}^{2} M_{Z}^{2} M_{\Phi n} \left(1-\cos 2\theta_{W} \right) \left(M_{\Phi n} I_{1}^{mnnn} + M_{\Phi m} I_{1}^{nnnm}\right)+  2 M_{W}^{4} M_{\Phi n}^{2} I_{1}^{mnnn}\right], &\\
&A^{n}A^{m}H^{p\pm}H^{q\mp} : -\frac{1}{v^{2} M_{Zn} M_{Zm} M_{Wp} M_{Wq}}\left[ m_{h}^{2} M_{\Phi n} M_{\Phi m} M_{\Phi p} M_{\Phi q} I^{nmpq} + 2 M_{Z}^{4} \cos^{2}2\theta_{W} M_{\Phi p} M_{\Phi q} I_{1}^{nmpq} \right. &\\
&\left.~~~~~~~~~~~~~~~~~~~~~ +2 M_{W}^{4} M_{\Phi n} M_{\Phi m}I_{1}^{pqnm} + M_{W}^{2} M_{Z}^{2} \left(1-\cos 2\theta_{W} \right) \left \lbrace M_{\Phi p}\left(M_{\Phi n} I_{1}^{mqnp} + M_{\Phi m} I_{1}^{nqmp}\right) \right. \right.&\\
&\left. \left. ~~~~~~~~~~~~~~~~~~~~~ + M_{\Phi q}\left(M_{\Phi n} I_{1}^{mpnp} + M_{\Phi m} I_{1}^{npmq}\right)\right\rbrace\right], &\\
&A^{n}A^{m}H^{n\pm}H^{m\mp} : -\frac{1}{v^{2} M_{Zn} M_{Zm} M_{Wn} M_{Wm}}\left[ m_{h}^{2} M_{\Phi n}^{2} M_{\Phi m}^{2} I^{nmnm} + 2 M_{Z}^{4} \cos^{2}2\theta_{W} M_{\Phi n} M_{\Phi m} I_{1}^{nmnm} \right. &\\
&\left.~~~~~~~~~~~~~~~~~~~~~ +2 M_{W}^{4} M_{\Phi n} M_{\Phi m}I_{1}^{nmnm} + M_{W}^{2} M_{Z}^{2} \left(1-\cos 2\theta_{W} \right) \left \lbrace M_{\Phi n}\left(M_{\Phi n} I_{1}^{mmnn} + M_{\Phi m} I_{1}^{nnmm}\right) \right. \right.&\\
&\left. \left. ~~~~~~~~~~~~~~~~~~~~~ + M_{\Phi m}\left(M_{\Phi n} I_{1}^{mnnm} + M_{\Phi m} I_{1}^{nnmm}\right)\right\rbrace\right], &\\
&A^{n}A^{m}H^{n\pm}H^{p\mp} : -\frac{1}{v^{2} M_{Zn} M_{Zm} M_{Wn} M_{Wp}}\left[ m_{h}^{2} M_{\Phi n}^{2} M_{\Phi m} M_{\Phi p} I^{nmnp} + 2 M_{Z}^{4} \cos^{2}2\theta_{W} M_{\Phi n} M_{\Phi p} I_{1}^{nmnp} \right. &\\
&\left.~~~~~~~~~~~~~~~~~~~~~ +2 M_{W}^{4} M_{\Phi n} M_{\Phi m}I_{1}^{npnm} + M_{W}^{2} M_{Z}^{2} \left(1-\cos 2\theta_{W} \right) \left \lbrace M_{\Phi n}\left(M_{\Phi n} I_{1}^{mpnn} + M_{\Phi m} I_{1}^{npmn}\right) \right. \right.&\\
&\left. \left. ~~~~~~~~~~~~~~~~~~~~~ + M_{\Phi p}\left(M_{\Phi n} I_{1}^{mnnp} + M_{\Phi m} I_{1}^{nnmp}\right)\right\rbrace\right]. &
\end{flalign*}
\subsection*{E.6 :}  ~ Elements of matrix $\mathcal{C}_{15\times 25}$ :
\setlength{\mathindent}{0pt}
\begin{flalign*}
&h^{0}h^{0}H^{n+}H^{n-} : -\frac{1}{\sqrt{2} v^{2} M_{Wn}^{2}}\left(m_{h}^{2} M_{\Phi n}^{2} + 2 M_{W}^{4} \right),&\\
&h^{0}h^{n}H^{n\pm}\phi^{\mp} : -\frac{m_{h}^{2}}{v^{2}}\frac{M_{\Phi n}}{M_{Wn}},&\\
&h^{n}h^{n}H^{m \pm}\phi^{\mp} : - \frac{m_{h}^{2}}{\sqrt{2} v^{2}}\frac{M_{\Phi m}}{M_{Wm}}I^{nnm},&\\
&h^{n}h^{n}H^{n \pm}\phi^{\mp} : - \frac{m_{h}^{2}}{\sqrt{2} v^{2}}\frac{M_{\Phi n}}{M_{Wn}}I^{3n},&\\
&h^{n}h^{m}H^{p \pm}\phi^{\mp} : - \frac{m_{h}^{2}}{v^{2}}\frac{M_{\Phi p}}{M_{Wp}}I^{nmp},&\\
&h^{n}h^{m}H^{n \pm}\phi^{\mp} : - \frac{m_{h}^{2}}{v^{2}}\frac{M_{\Phi n}}{M_{Wn}}I^{nnm},&\\
&h^{0}h^{n}H^{n+}H^{n-} : -\frac{1}{v^{2} M_{Wn}^{2}}\left(m_{h}^{2} M_{\Phi n}^{2} I^{3n} + 2 M_{W}^{4} I^{\prime 3n} \right),&\\
&h^{0}h^{n}H^{m+}H^{m-} : -\frac{1}{v^{2} M_{Wm}^{2}}\left(m_{h}^{2} M_{\Phi m}^{2} I^{nmm} + 2 M_{W}^{4} I_{1}^{mmn} \right),&\\
&h^{0}h^{n}H^{m \pm}H^{p \mp} : -\frac{1}{v^{2} M_{Wm} M_{Wp}}\left(m_{h}^{2} M_{\Phi m} M_{\Phi p} I^{nmp} + 2 M_{W}^{4} I_{1}^{mpn} \right),&\\
&h^{0}h^{n}H^{n\pm}H^{m\mp} : -\frac{1}{v^{2} M_{Wn} M_{Wm}}\left(m_{h}^{2} M_{\Phi n} M_{\Phi m} I^{nnm} + 2 M_{W}^{4} I_{1}^{mnn} \right),&\\
&h^{n}h^{n}H^{n+}H^{n-} : -\frac{1}{\sqrt{2} v^{2} M_{Wn}^{2}}\left(m_{h}^{2} M_{\Phi n}^{2} I^{n} + 2 M_{W}^{4} I^{\prime n} \right),&\\
&h^{n}h^{n}H^{m+}H^{m-} : -\frac{1}{\sqrt{2} v^{2} M_{Wm}^{2}}\left(m_{h}^{2} M_{\Phi m}^{2} I^{nnmm} + 2 M_{W}^{4} I_{1}^{mmnn} \right),&\\
&h^{n}h^{m}H^{p+}H^{p-} : -\frac{1}{v^{2} M_{Wp}^{2}}\left(m_{h}^{2} M_{\Phi p}^{2} I^{nmpp} + 2 M_{W}^{4} I_{1}^{ppnm} \right),&\\
&h^{n}h^{m}H^{n+}H^{n-} : -\frac{1}{v^{2} M_{Wn}^{2}}\left(m_{h}^{2} M_{\Phi n}^{2} I^{nmnn} + 2 M_{W}^{4} I_{1}^{nnnm} \right),&\\
&h^{n}h^{n}H^{m\pm}H^{p\mp} : -\frac{1}{\sqrt{2} v^{2} M_{Wm} M_{Wp}}\left(m_{h}^{2} M_{\Phi m} M_{\Phi p} I^{nnmp} + 2 M_{W}^{4} I_{1}^{mpnn} \right),&\\
&h^{n}h^{n}H^{n\pm}H^{m\mp} : -\frac{1}{\sqrt{2} v^{2} M_{Wn} M_{Wm}}\left(m_{h}^{2} M_{\Phi n} M_{\Phi m} I^{nnnm} + 2 M_{W}^{4} I_{1}^{mnnn} \right),&\\
&h^{n}h^{m}H^{p\pm}H^{q\mp} : -\frac{1}{v^{2} M_{Wp} M_{Wq}}\left(m_{h}^{2} M_{\Phi p} M_{\Phi q} I^{nmpq} + 2 M_{W}^{4} I_{1}^{pqnm} \right).&
\end{flalign*}
\subsection*{E.7 :}~Matrix elemnts of $\mathcal{M}^{(2)}_{NC, 20\times 20 }$ :
\setlength{\mathindent}{0pt}
\begin{flalign*}
&h^{0}A^{n}h^{0}A^{n} : -\frac{\left(m_{h}^{2}M_{\Phi n}^{2} + 2 M_{Z}^{4} \right)}{v^{2} M_{Zn}^{2}}, &\\
&h^{0}A^{m}h^{n}A^{m} : -\frac{\left(m_{h}^{2}M_{\Phi m}^{2} I^{nmm} + 2 M_{Z}^{4} I_{1}^{mmn} \right)}{v^{2} M_{Zm}^{2}}, &\\
&h^{0}A^{m}h^{n}A^{p} : -\frac{\left(m_{h}^{2}M_{\Phi m}M_{\Phi p} I^{nmp} + 2 M_{Z}^{4} I_{1}^{mpn} \right)}{v^{2} M_{Zm} M_{Zp}}, &\\
&h^{n}A^{n}h^{n}A^{n} : -\frac{\left(m_{h}^{2} M_{\Phi n}^{2} I^{n} + 2 M_{Z}^{4} I^{\prime n} \right)}{v^{2} M_{Zn}^{2}}, &\\
&h^{n}A^{m}h^{n}A^{m} : -\frac{\left(m_{h}^{2} M_{\Phi m}^{2} I^{nnmm} + 2 M_{Z}^{4} I_{1}^{mmnn} \right)}{v^{2} M_{Zm}^{2}}, &\\
&h^{n}A^{m}h^{n}A^{p} : -\frac{\left(m_{h}^{2} M_{\Phi m} M_{\Phi p} I^{nnmp} + 2 M_{Z}^{4} I_{1}^{mpnn} \right)}{v^{2} M_{Zm} M_{Zp}}, &\\
&A^{n}h^{m}A^{n}h^{p} : -\frac{\left(m_{h}^{2} M_{\Phi n}^{2} I^{mpnn} + 2 M_{Z}^{4} I_{1}^{nnmp} \right)}{v^{2} M_{Zn}^{2}}, &\\
&h^{n}A^{n}h^{m}A^{m} : -\frac{\left(m_{h}^{2} M_{\Phi n} M_{\Phi m} I^{nmnm} + 2 M_{Z}^{4} I_{1}^{nmnm} \right)}{v^{2} M_{Zn} M_{Zm}}, &\\
&h^{n}A^{p}h^{m}A^{q} : -\frac{\left(m_{h}^{2} M_{\Phi p} M_{\Phi q} I^{nmpq} + 2 M_{Z}^{4} I_{1}^{pqnm} \right)}{v^{2} M_{Zp} M_{Zq}}. &\\
\end{flalign*}
\subsection*{E.8 :}~Matrix elemnts of $\mathcal{G}_{CC, 20\times 20 }$ :
\setlength{\mathindent}{0pt}
\begin{flalign*}
&\phi^{+}A^{n}\phi^{+}A^{n} : \sqrt{2}~\phi^{+}\phi^{-}A^{n}A^{n},~~~~~ \phi^{+}A^{n}H^{m+}A^{n} : \sqrt{2}~\phi^{\pm}H^{m\mp}A^{n}A^{n},~~~~~\phi^{+}A^{n}H^{n +}A^{n} : \sqrt{2}~\phi^{\pm}H^{n\mp}A^{n}A^{n},  &\\
&\phi^{+}A^{n}H^{n+}A^{m} : \phi^{\pm}H^{n\mp}A^{n}A^{m},~~~~~\phi^{+}A^{n}H^{p+}A^{m} : \phi^{\pm}H^{p\mp}A^{n}A^{m},~~~~~A^{n}H^{n+}A^{n}H^{n+} : \sqrt{2}~A^{n}A^{n}H^{n+}H^{n-},&\\
&A^{n}H^{m+}A^{n}H^{m+} : \sqrt{2}~A^{n}A^{n}H^{m+}H^{m-},~~~~~~~~~~ A^{n}H^{p+}A^{m}H^{p+} : A^{n}A^{m}H^{p+}H^{p-}, &\\
&A^{n}H^{n+}A^{m}H^{n+} : A^{n}A^{m}H^{n+}H^{n-},~~~~~~~~~~~~~~~~A^{n}H^{m +}A^{n}H^{p +} : \sqrt{2}~A^{n}A^{n}H^{m\pm}H^{p\mp}, &\\
&A^{n}H^{m +}A^{n}H^{n +} : \sqrt{2}~A^{n}A^{n}H^{m\pm}H^{n\mp},~~~~~~~~~~~A^{n}H^{p +}A^{m}H^{q +} : A^{n}A^{m}H^{p\pm}H^{q\mp},  &\\
&A^{n}H^{n +}A^{m}H^{m +} : A^{n}A^{m}H^{n\pm}H^{m\mp},~~~~~~~~~~~~~~~A^{n}H^{n +}A^{m}H^{p +} : A^{n}A^{m}H^{n\pm}H^{p\mp}. &
\end{flalign*}
\subsection*{E.9 :}~Matrix elemnts of $\mathcal{H}_{CC, 25\times 25 }$ :
\setlength{\mathindent}{0pt}
\begin{flalign*}
&h^{0}H^{n+}h^{0}H^{n+} : \sqrt{2}~h^{0}h^{0}H^{n+}H^{n-},~~~~~~\phi^{+}h^{0}H^{n+}h^{n} : h^{0}h^{n}H^{n\pm}\phi^{\mp},~~~~~~\phi^{+}h^{n}H^{m +}h^{n} : \sqrt{2}~h^{n}h^{n}H^{m \pm}\phi^{\mp},&\\
&\phi^{+}h^{n}H^{n +}h^{n} : \sqrt{2}~h^{n}h^{n}H^{n \pm}\phi^{\mp},~~~~~~~~~~\phi^{+}h^{n}H^{p +}h^{m} : h^{n}h^{m}H^{p \pm}\phi^{\mp},~~~~~~~~~~~\phi^{+}h^{n}H^{n +}h^{m} : h^{n}h^{m}H^{n \pm}\phi^{\mp},&\\
&h^{0}H^{n+}h^{n}H^{n+} : h^{0}h^{n}H^{n+}H^{n-},~~~~~~h^{0}H^{m+}h^{n}H^{m+} : h^{0}h^{n}H^{m+}H^{m-},~~~~~~h^{0}H^{m +}h^{n}H^{p +} : h^{0}h^{n}H^{m \pm}H^{p \mp},&\\
&h^{0}H^{n+}h^{n}H^{m+} : h^{0}h^{n}H^{n\pm}H^{m\mp},~~~h^{n}H^{n+}h^{n}H^{n+} : \sqrt{2}~h^{n}h^{n}H^{n+}H^{n-},~~~~h^{n}H^{p+}h^{m}H^{p+} : h^{n}h^{m}H^{p+}H^{p-},&\\
&h^{n}H^{m+}h^{n}H^{m+} : \sqrt{2}~h^{n}h^{n}H^{m+}H^{m-},~~h^{n}H^{n+}h^{m}H^{n+} : h^{n}h^{m}H^{n+}H^{n-},~~h^{n}H^{m+}h^{n}H^{p+} : h^{n}h^{n}H^{m\pm}H^{p\mp},&\\
&h^{n}H^{n+}h^{n}H^{m+} : \sqrt{2}~h^{n}h^{n}H^{n\pm}H^{m\mp},~~~~~~h^{n}H^{p+}h^{m}H^{q+} : h^{n}h^{m}H^{p\pm}H^{q\mp}.&\\
\end{flalign*}

%
%
%
%
%
%
%
%
\bibliographystyle{JHEP}
\bibliography{reference}
\end{document}